\documentclass[twocolumn,tighten]{aastex61}
\received{April 7, 2017}
\revised{July 10, 2017}
\accepted{\today}
\submitjournal{ApJ}

\usepackage{graphicx}
\usepackage{epstopdf}
\usepackage{epsfig}
\usepackage{mathtools}
\usepackage{natbib}
\usepackage[caption = false]{subfig}
\usepackage{booktabs}
\usepackage{hyperref}

\DeclareMathOperator{\Lagr}{\mathcal{L}}

\shorttitle{Hot H$_2$ in Protoplanetary Disks}
\shortauthors{Hoadley et al.}

\begin{document}

\title{Signatures of Hot Molecular Hydrogen Absorption from Protoplanetary Disks: I. Non-thermal Populations}

\correspondingauthor{Keri Hoadley}
\email{keri.hoadley@colorado.edu}

\author[0000-0002-8636-3309]{Keri Hoadley}
\affil{Laboratory for Atmospheric and Space Physics (LASP) \\
University of Colorado \\
Space Science Building (SPSC) \\
3665 Discovery Drive \\
Boulder, CO 80303, USA}

\author{Kevin France}
\affiliation{Laboratory for Atmospheric and Space Physics (LASP) \\
University of Colorado \\
Space Science Building (SPSC) \\
3665 Discovery Drive \\
Boulder, CO 80303, USA}
\affil{Center for Astrophysical and Space Astronomy (CASA) \\ 
University of Colorado \\
389 UCB \\
Boulder, CO 80309, USA}

\author{Nicole Arulanantham}
\affiliation{Laboratory for Atmospheric and Space Physics (LASP) \\
University of Colorado \\
Space Science Building (SPSC) \\
3665 Discovery Drive \\
Boulder, CO 80303, USA}

\author{R.O. Parke Loyd}
\affiliation{Laboratory for Atmospheric and Space Physics (LASP) \\
University of Colorado \\
Space Science Building (SPSC) \\
3665 Discovery Drive \\
Boulder, CO 80303, USA}

\author{Nicholas Kruczek}
\affiliation{Laboratory for Atmospheric and Space Physics (LASP) \\
University of Colorado \\
Space Science Building (SPSC) \\
3665 Discovery Drive \\
Boulder, CO 80303, USA}


\begin{abstract}
The environment around protoplanetary disks (PPDs) regulates processes which drive the chemical and structural evolution of circumstellar material. We perform a detailed empirical survey of warm molecular hydrogen (H$_2$) absorption observed against \ion{H}{1}-Ly$\alpha$ (Ly$\alpha$: $\lambda$ 1215.67 {\AA}) emission profiles for 22 PPDs, using archival \textit{Hubble Space Telescope} (\textit{HST}) ultraviolet (UV) spectra to identify H$_2$ absorption signatures and quantify the column densities of H$_2$ ground states in each sightline.  
We compare thermal equilibrium models of H$_2$ to the observed H$_2$ rovibrational level distributions. We find that, for the majority of targets, there is a clear deviation in high energy states (T$_{exc}$ $\gtrsim$ 20,000 K) away from thermal equilibrium populations (T(H$_2$) $\gtrsim$ 3500 K). \\
We create a metric to estimate the total column density of non-thermal H$_2$ (N(H$_2$)$_{\textnormal{nLTE}}$) and find that the total column densities of thermal (N(H$_2$)) and N(H$_2$)$_{\textnormal{nLTE}}$ correlate for transition disks and targets with detectable \ion{C}{4}- pumped H$_2$ fluorescence. We compare N(H$_2$) and N(H$_2$)$_{\textnormal{nLTE}}$ to circumstellar observables and find that N(H$_2$)$_{\textnormal{nLTE}}$ correlates with X-ray and FUV luminosities, but no correlations are observed with the luminosities of discrete emission features (e.g., Ly$\alpha$, \ion{C}{4}). Additionally, N(H$_2$) and N(H$_2$)$_{\textnormal{nLTE}}$ are too low to account for the H$_2$ fluorescence observed in PPDs, so we speculate that this H$_2$ may instead be associated with a diffuse, hot, atomic halo surrounding the planet-forming disk. We create a simple photon-pumping model for each target to test this hypothesis and find that Ly$\alpha$ efficiently pumps H$_2$ levels with T$_{exc}$ $\geq$ 10,000 K out of thermal equilibrium.

\end{abstract}

\keywords{protoplanetary disks --- molecular processes --- line: profiles  --- stars: variables: T Tauri, Herbig Ae/Be --- techniques: spectroscopic -- ultraviolet: general}


%
%
%
\section{Introduction}
Protoplanetary disks (PPDs) are thought to provide the raw materials that form protoplanets and drive planetary systems to their final architectures \citep{Lubow+99,Lubow+D'Angelo+06,Armitage+03,Brown+09,Woitke+09b,Ayliffe+Bate+10,Dullemond+Monnier+10,Beck+12}. 
The presence of significant amounts of gas in the disk is a defining quality of PPDs, where the earliest stages are assumed to have the canonical interstellar medium (ISM) gas-to-dust ratio $\sim$ 100:1 (e.g., \citealt{Frisch+99,Schneider+15}). 
The gas content in PPDs controls essential processes tied to the formation and evolution of planetary systems, including dust grain growth (through the coupling of gas and dust dynamics), angular momentum transport, and thermal and chemical balance of the disk as it evolves \citep{Weidenschilling+77,Alexander+Armitage+07,Woitke+09b,Youdin+11,Levison+15}. 
However, over timescales of a few Myr, PPDs lose their massive gas disk, evolving to gas-sparse debris disks (with gas-to-dust ratios $\sim$0:1; \citealt{Alexander+14,Gorti+15}). 
The dispersal of the gas-rich disk is likely driven by a number of different physical processes throughout the PPD lifetime, ranging from photoevaporation of gas through thermal winds (for example, an atomic wind: e.g., \citealt{Owen+10}; a fully-ionized wind: e.g., \citealt{Alexander+06}; and/or a slow molecular wind: see review by \citealt{Alexander+14}) or magnetohydrodynamic (MHD) winds (e.g. \citealt{Ferreira+06,Bai+16}), to giant planet formation accreting and clearing gas remaining in a dust gap \citep{Lin+Papaloizou+86,Dodson-Robinson+Salyk+11,Zhu+11,Dong+15,Owen+16}.  
Probing the physical mechanisms that drive the dispersal of gas from PPDs is critical for inferring when, where, and how planet-forming disks lose their massive gas reservoir. In turn, these properties inform us of the physical and chemical environment in which planets form throughout the PPD lifetime. 
%

Internal radiation from the proto-stellar source can play an important role in determining the chemical and physical state of the gas-rich PPD \citep{Kamp+Dullemond+04,Nomura+04,Nomura+07,Oberg+10,Bethell+Bergin+11}. Ultraviolet (UV) and X-ray radiation, which are created by hot gas accretion onto and activity in the protostellar atmosphere, can effectively enhance the populations of rovibrationally-excited molecules, which create pathways for molecular dissociation (e.g. \citealt{Glassgold+Najita+01,Bergin+04,Gorti+Hollenbach+04,Glassgold+04,Kamp+05,Dullemond+07,Gudel+07,Kastner+16}). High-energy radiation may also help heat and regulate chemical processes in the disk atmosphere, leading to the production of atomic and molecular by-products (e.g. \citealt{Salyk+08,Walsh+15,Adamkovics+16}). Hot molecules can be swept up into thermal winds over the disk lifetime \citep{Alexander+06,Gorti+Hollenbach+09,Owen+10,Owen+16}, leading to the dispersal of the disk from the inside-out.

Molecular hydrogen (H$_2$) has been measured to be 10$^4$ times more abundant than other molecules (e.g., CO) in the warm regions of PPDs \citep{France+14b}, and large quantities of H$_2$ in the disk allow the molecule to survive at hot temperatures (T(H$_2$) $\sim$ 1000 $-$ 5000 K), shielding against collisional- and photo-dissociation \citep{Williams+00,Beckwith+78,Beckwith+83}. The properties of H$_2$ make it a reliable diagnostic of the spatial and structural behavior of warm molecules probed in and around PPDs \citep{Ardila+02,Herczeg+04}, as it is expected to trace residual amounts of gas in disks throughout their evolution ($\Sigma_{H_2}$ $\sim$ 10$^{-6}$ g cm$^{-2}$; e.g., \citealt{France+12b}).

However, H$_2$ is notoriously difficult to observe in PPDs; cold H$_2$ (T(H$_2$) $\sim$ 10 K) does not radiate and, due to its lack of a permanent dipole \citep{Sternberg+89}, ro-vibrational transitions of H$_2$ in the IR are dominated by weak, quadrupole transitions. Therefore, it has been easier to trace other molecular constituents of the inner disk, such as CO and HD, to interpret the behavior of the underlying H$_2$ reservoir (e.g., \citealt{Salyk+11b,Brown+13,Banzatti+15,McClure+16}). Most IR studies of H$_2$ in PPDs have been detections of shocked (hot) H$_2$ in collimated jets or streams \citep{Bary+03,Beck+12,Arulanantham+16}. 

The far ultraviolet (FUV: $\lambda \lambda$ 912 $-$ 1700 {\AA}) offers the strongest transition probabilities for dipole-allowed electronic transitions of H$_2$ photo-excited by UV photons, specifically absorption avenues coincident with \ion{H}{1}-Ly$\alpha$ ($\lambda$ 1215.67 {\AA}) photons, which are generated near the protostellar surface \citep{France+12b,Schindhelm+12b} and make up $\sim$ 90\% of the FUV flux in a typical T Tauri system \citep{France+14}. Warm H$_2$ (T $\geq$ 1000 K) can absorb Ly$\alpha$ photons, exciting the molecule up to either the Lyman ($2p\sigma B$ $^{1}\Sigma^{+}_{u}$) or Werner ($2p\pi C$ $^{1}\Pi_{u}$) electronic bands. Because of the large dipole-allowed transition probabilities (A$_{ul}$ $\sim$ 10$^8$ s$^{-1}$; \citealt{Abgrall+93,Abgrall+93b}), H$_2$ in these electronic states will decay instantaneously in a fluorescent cascade down to one of many different rovibration levels in the ground electronic state ($X ^{1}\Sigma^{+}_{g}$; \citealt{Herczeg+02}). Each fluorescence transition results in the discrete emission of a FUV photon, whose frequency depends on the electronic-to-ground state transition. We observed hundreds of these features throughout the FUV with the \textit{Hubble Space Telescope} (\textit{HST}) from $\lambda \lambda$ 1150 $-$ 1700 {\AA} (see \citealt{Herczeg+02,France+12b}). This process predominantly favors regions where warm molecules reside in disks \citep{Nomura+Millar+05,Nomura+07,Adamkovics+16}. The characterization of H$_2$ emission from PPDs has provided complimentary results to high-resolution IR-CO surveys probing PPD evolution (e.g. \citealt{Brown+13,Banzatti+15}).
 
We can also observe the excitation leg of the fluorescence process via H$_2$ absorption lines incident on the broad Ly$\alpha$ emission line in PPD systems. Several studies have looked to characterize and relate the H$_2$ absorption features within protostellar Ly$\alpha$ wings to fluorescent populations tied to the behavior of the inner disk material. \citet{Yang+11} detected the first signatures of Ly$\alpha$-H$_2$ absorption in DF Tau and V4046 Sgr. 
They found that, for V4046 Sgr, which hosts a cicumbinary disk with a relatively face-on inclination angle (i$_{disk}$ $\sim$ 35$^{\circ}$), the H$_2$ would have to be pumped near the accretion shock to explain how H$_2$ absorption features are detectable in the sightline. \citet{France+12a} performed an extensive study on warm molecules in the disk environment of AA Tau and were the first to empirically derive H$_2$ column densities from absorption features within the Ly$\alpha$ red stellar wing. The lower energy states of H$_2$ could be described by a warm thermal population (T(H$_2$) $\sim$ 2500 K $\pm$ 1000 K) consistent with H$_2$ fluorescence emission from the inner disk. They noticed that, for high excitation temperature states of H$_2$ (T$_{exc}$ $\geq$ 20,000 K), column densities deviated significantly from thermal distributions, providing the first hint that there may be additional excitation mechanisms in the disk atmosphere pumping H$_2$ out of local thermodynamic equilibrium (LTE). 

The behavior of these non-thermal states may provide clues about the mechanisms that drive molecules out of LTE and, potentially, the dispersal of gas from planet-forming disks. In the first paper of this study, we perform a quantitative, empirical survey of H$_2$ absorption observed against the Ly$\alpha$ stellar emission profiles of 22 PPD hosts. We aim to characterize the physical state of the gas in each sightline and learn how various stellar and disk mechanisms contribute to the excitation of non-thermal H$_2$ states. In Section 2, we present the archival observations used to perform this study. In Section 3, we describe the methodology of extracting H$_2$ absorption features from each Ly$\alpha$ emission profile and quantifying the column densities of each H$_2$ rovibrational level. In Section 4, we present results from fitting thermal models to the column density rotation diagrams for each target and what those results reveal about the non-thermal density distributions of H$_2$ in each sightline. In Section 5, we compare our results to observed disk and stellar properties, which probe different excitation mechanisms that may help explain excesses in non-thermal populations of H$_2$. We take all the evidence provided by this empirical study to infer the most likely location of H$_2$ absorption in the disk atmosphere. Finally, in Section 6, we conclude our paper with our major findings and future work that may help clarify open questions left unresolved by this study. In a follow-up study (Paper II), we will consider where the H$_2$ populations originate in the circumstellar environment. Additional plots and details about our absorption and thermal models are provided in the Appendix.

%
%
%
\section{Targets and Observations}

\begin{deluxetable*}{l c c c c c c c l}
\tabletypesize{\footnotesize}
\tablecaption{Target Properties \label{tab1}}
\tablewidth{0pt}
\tablehead{
	\colhead{Target} & \colhead{Spectral} & \colhead{Disk}	& \colhead{Distance} & \colhead{L$_{\star}$} & \colhead{M$_{\star}$} & \colhead{$\dot{M}$} 			& \colhead{i$_{disk}$}  & Ref.\tablenotemark{b} \\
	     \colhead{}   & \colhead{Type}    & \colhead{Type\tablenotemark{a}}  & \colhead{(pc)}     & \colhead{(L$_{\sun}$)} & \colhead{(M$_{\sun}$)}  & \colhead{($\times$ 10$^{-8}$ $M_{\sun}$ yr$^{-1}$)} 	& \colhead{($^{\circ}$)} & \colhead{}	
}	
\startdata
	AA Tau & K7 & P & 140 & 0.71 & 0.80 & 0.33 & 75 & 2, 4, 7, 12, 16, 52, 53 \\	
	AB Aur & A0 & T & 140 & 46.8 & 2.40 & 1.80 & 22 & 19, 39, 49, 50, 52, 53 \\
	AK Sco & F5 & P & 103 & 7.59 & 1.35 & 0.09 & 68 & 18, 20, 34, 57 \\
	BP Tau & K7 & P & 140 & 0.925 & 0.73 & 2.88 & 30 & 7, 12, 38, 52, 53 \\
	CS Cha & K6 & T & 160 & 1.32 & 1.05 & 1.20 & 60 & 21, 35, 40, 54 \\
	DE Tau & M0 & T & 140 & 0.87 & 0.59 & 2.64 & 35 & 7, 10, 12, 52, 53 \\
	DF Tau A & M2 & P & 140 & 1.97 & 0.19 & 17.70 & 85 & 7, 10, 52, 53 \\
	DM Tau & M1.5 & T & 140 & 0.24 &  0.50 & 0.29 & 35 & 16, 29, 32, 52, 53 \\
	GM Aur & K5.5 & T & 140 & 0.74 & 1.20 & 0.96 & 55 & 7, 16, 32, 52, 53 \\
	HD 104237 & A7.5 & T & 116 & 34.7 & 2.50 & 3.50 & 18 & 19, 23, 31, 45 \\
	HD 135344B & F3 & T & 140 & 8.13 & 1.60 & 0.54 & 11 & 19, 22, 31, 42, 57 \\
	HN Tau A & K5 & P & 140 & 0.19 & 0.85 & 0.13 & 40 & 6, 7, 12, 52, 53 \\
	LkCa 15 & K3 & T & 140 & 0.72 & 0.85 & 0.13 & 49 & 12, 29, 32, 52, 53 \\
	RECX-11 & K4 & P & 97 & 0.59 & 0.80 & 0.03 & 70 & 13, 24, 47, 55 \\
	RECX-15 & M2 & P & 97 & 0.08 & 0.40 & 0.10 & 60 & 13, 14, 15, 55 \\
	RU Lup & K7 & T & 121 & 0.42 & 0.80 & 3.00 & 24 & 25, 30, 36, 41, 56 \\
	RW Aur A & K4 & P & 140 & 2.3 & 1.40 & 3.16 & 77 & 5, 9, 11, 12, 17, 52, 53 \\
	SU Aur & G1 & T & 140 & 9.6 & 2.30 & 0.45 & 62 & 1, 3, 8, 11, 12, 52, 53 \\
	SZ 102 & K0 & T & 200 & 0.01 & 0.75 & 0.08 & 90 & 26, 37, 43, 48 \\
	TW Hya	& K6 & T & 54 & 0.17 & 0.60 & 0.02 & 4 & 27, 30, 42, 51, 56 \\
	UX Tau A & K2 & T & 140 & 3.5 & 1.30 & 1.00 & 35 & 12, 32, 52, 53 \\
	V4046 Sgr & K5 & T & 83 & 0.5+0.3 & 0.86+0.69 & 1.30 & 34 & 28, 33, 44, 46 \\
\enddata
%
\tablenotetext{a}{Disk Type is defined by either the detection of small dust grain depletion in the inner disk regions, resulting in disk holes or gaps, or the degree of dust settling in the disk, or both; PPDs can be categorized using an observable n$_{13-31}$, which is defined by the slope in the spectral energy distribution (SED) flux between 13 $\mu$m and 31 $\mu$m \citep{Furlan+09}: P = primordial (n$_{13-31}$ $<$ 0); T = transitional (n$_{13-31}$ $>$ 0).}
\tablenotetext{b}{\textbf{References:} (1) \citet{Akeson+02}, (2) \citet{Andrews+Williams+07}, (3) \citet{Bertout+88}, (4) \citet{Bouvier+99}, (5) \citet{Eisner+07}, (6) \citet{France+11}, (7) \citet{Gullbring+98}, (8) \citet{Gullbring+00}, (9) \citet{Hartigan+95}, (10) \citet{Johns-Krull+Valenti+01}, (11) \citet{Johns-Krull+00}, (12) \citet{Kraus+Hillenbrand+09}, (13) \citet{Lawson+04}, (14) \citet{Luhman+04}, (15) \citet{RamsayHowat+Greaves+07}, (16) \citet{Ricci+10}, (17) \citet{White+Ghez+01}, (18) \citet{vanderAncker+98}, (19) \citet{vanBoekel+05}, (20) \citet{Alencar+03}, (21) \citet{Lawson+96}, (22) \citet{Lyo+11}, (23) \citet{Feigelson+03}, (24) \citet{Lawson+01}, (25) \citet{Herczeg+05}, (26) \citet{Comeron+Fernandez+10}, (27) \citet{Webb+99}, (28) \citet{Quast+00}, (29) \citet{Hartmann+98}, (30) \citet{Herczeg+Hillenbrand+08}, (31) \citet{GarciaLopez+06}, (32) \citet{Andrews+11}, (33) \citet{France+12a}, (34) \citet{GomezdeCastro+09}, (35) \citet{Espaillat+07a}, (36) \citet{Stempels+07}, (37) \citet{Comeron+03}, (38) \citet{Simon+00}, (39) \citet{Tang+12}, (40) \citet{Espaillat+11}, (41) \citet{Stempels+12}, (42) \citet{Pontoppidan+08}, (43) \citet{Coffey+04}, (44) \citet{Rodriguez+10}, (45) \citet{Grady+04}, (46) \citet{Rosenfeld+12}, (47) \citet{Ingleby+11}, (48) \citet{Hughes+94}, (49) \citet{Hashimoto+11}, (50) \citet{Donehew+Brittain+11}, (51) \citet{Rosenfeld+12b}, (52) \citet{Bertout+99}, (53) \citet{Loinard+07}, (54) \citet{Luhman+04}, (55) \citet{Mamajek+99}, (56) \citet{vanLeeuwen+07}, (57) \citet{Grady+09}.}
\end{deluxetable*}

%
Our target list is derived from \citet{McJunkin+14}, who analyzed the reddening of the HI-Ly$\alpha$ profiles of 31 young stellar systems to create a comprehensive list of interstellar dust extinction estimates along each sight line.Of the 31 original targets, 22 of the protostellar targets showed signs of H$_2$ absorption in the Ly$\alpha$ profiles. All of these observations have been described previously in studies of H$_2$ (e.g. \citealt{France+12b,Hoadley+15}), hot gas (e.g. \citealt{Ardila+13}), and UV radiation (e.g. \citealt{France+14}). Several of the targets are known binaries or multiples (DF Tau: \citealt{Ghez+93}; HN Tau, RW Aur, and UX Tau: \citealt{Correia+06}; AK Sco and HD 104237 are spectroscopic binaries: \citealt{GomezdeCastro+09, Bohm+04}; and V4046 Sgr is a short-period binary, which acts as a point source for most applications: \citealt{Quast+00}), and only the primary stellar component is observed within the aperture. The majority of the targets are observed within the Taurus-Auriga and $\eta$ Chamaeleontis star-forming regions, with distances of 140 and 97 pc, respectively. Young stars observed in these star-forming regions have ages ranging a few Myr, while field pre-main sequence stars (e.g. TW Hya, V4046 Sgr) have ages range between 10 $-$ 30 Myr. The majority of these targets have age ranges comparable to the depletion timescale of gas and circumstellar dust via accretion processes \citep{Hernandez+07,Fedele+10}, making them ideal candidates for understanding the abundance and physical state of H$_2$ at a variety of PPD evolutionary stages. Table~\ref{tab1} presents relevant stellar and disk properties. All observations of the stellar Ly$\alpha$ profiles were taken either with the Cosmic Origins Spectrograph (COS) or Space Telescope Imaging Spectrograph (STIS) aboard the \textit{Hubble Space Telescope} (\textit{HST}).

\begin{figure}[htp]
\centering
	\includegraphics[angle=90,width=0.475\textwidth]{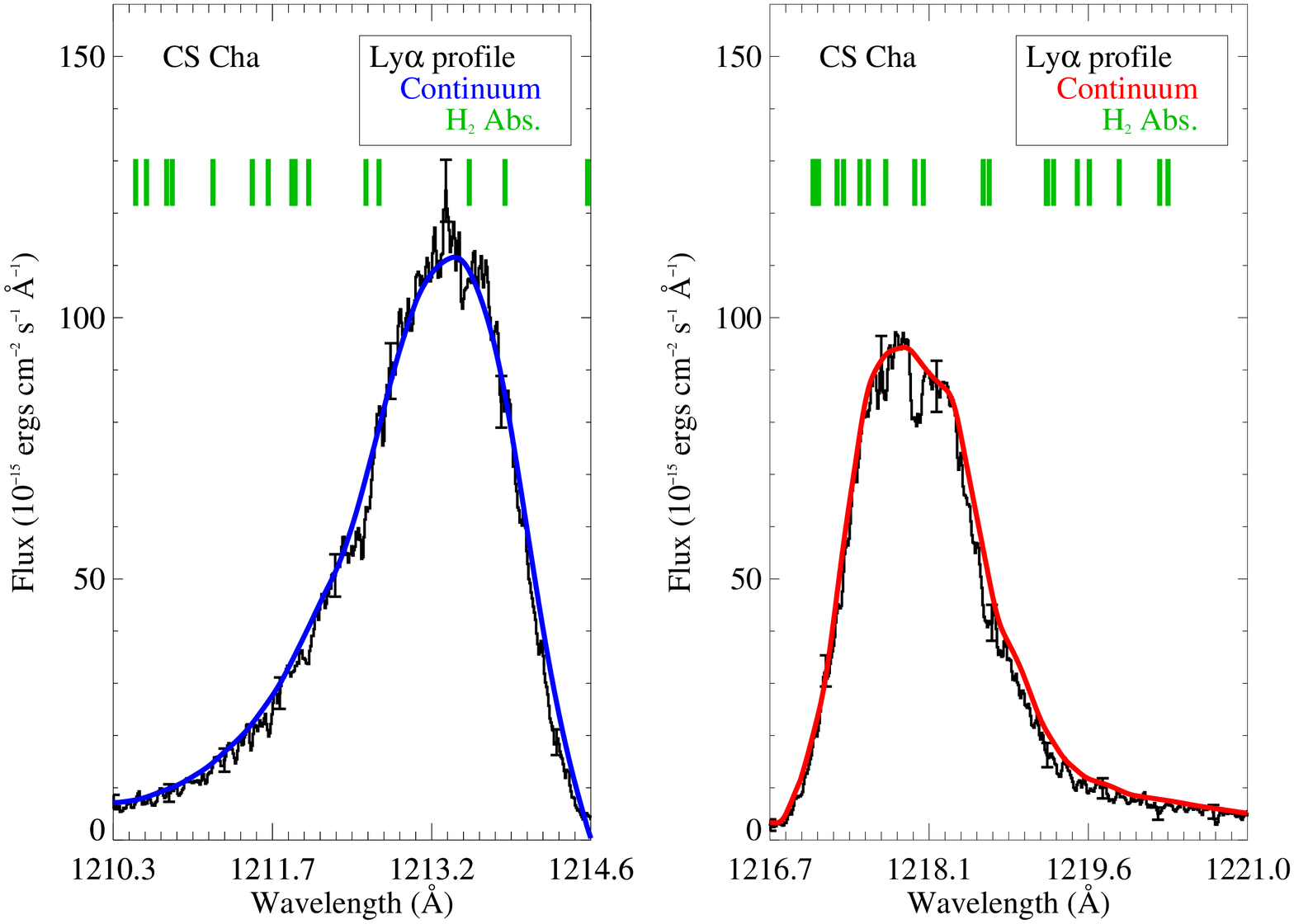}
\caption{The stellar Ly$\alpha$ wings for one target (CS Cha) in our survey. Within the wings themselves, absorption signatures can be seen. The mean flux array over the Ly$\alpha$ profile is selected to minimize contamination from the absorption features and trace the shape of the Ly$\alpha$ wings. The mean flux array is smoothed (and shown in blue over the blue Ly$\alpha$ wing and red over the red Ly$\alpha$ wing), and the observed line profile is divided by the mean flux array to create relative absorption spectra across the Ly$\alpha$ profile. H$_2$ absorption transitions are identified with green hashes and properties about each transition are shown in Table~\ref{tab3}. \label{fig1}}
\end{figure}

\subsection{COS Observations}

Each PPD spectrum collected with \textit{HST}/COS was taken either during the Disk, Accretion, and Outflows (DAO) of Tau Guest Observing (GO) program (PID 11616; PI: G. Herczeg) or COS Guaranteed Time Observing (PIDs 11533 and 12036; PI: J. Green). Each spectrum was observed with the medium-resolution far-UV modes of the spectrograph (G130M and G160M ($\Delta v$ $\approx$ 18 km s$^{-1}$ at Ly$\alpha$); \citealt{Green+12}). Multiple central wavelength positions were included to minimize fixed-pattern noise. The COS data were processed using the COS calibration pipeline (CALCOS) and were aligned and co-added with the procedure described by \citet{Danforth+10}. By design, COS is a slitless spectrograph, allowing the full 2\arcsec.5 field of view through the instrument. This means the instrument is exposed to strong contamination from geocoronal Ly$\alpha$ (Ly$\alpha_{\oplus}$). To mitigate this contamination, we mask the central $\sim$ 2 {\AA} of the Ly$\alpha$ spectra. 

\subsection{STIS Observations}

Several targets either exceeded the COS bright-object limit or had archival STIS observations available with the desired far-UV bandpass and resolution (AB Aur, HD 104237, TW Hya). The archival data were obtained with the STIS medium-resolution grating mode (G140M ($\Delta v$ $\approx$ 30 km s$^{-1}$ between 1150 $-$ 1700 {\AA}): \citet{Kimble+98,Woodgate+98}), while the COS-bright objects were observed with the echelle medium-resolution mode (E140M ($\Delta v$ $\approx$ 7 km s$^{-1}$ between 1150 $-$ 1700 {\AA})). The STIS echelle spectra were processed using echelle calibration software developed for the STIS StarCAT catalog \citep{Ayres+10}. Unlike COS, STIS has a small slit aperture (0\arcsec.2 $\times$ 0\arcsec.2), so the Ly$\alpha_{\oplus}$ signal is weaker; nonetheless, we remove the inner region of the Ly$\alpha$ profile ($\sim$ 0.5 $-$ 2 {\AA}) for consistency among all the data. 

\begin{figure}[htp]
\centering
\includegraphics[angle=90,width=0.475\textwidth]{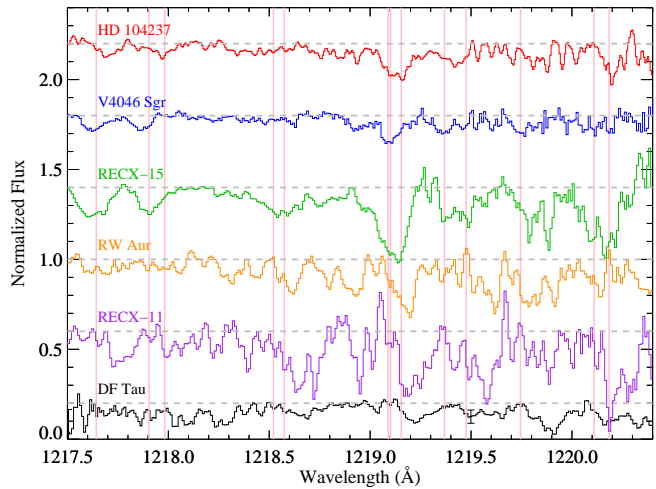}
	\caption{The normalized absorption spectra for 6 targets in our survey, ordered by increasing disk inclination angle ($i_{disk}$) from the top-down (11$^{\circ}$, 34$^{\circ}$, $\sim$60$^{\circ}$, $\sim$70$^{\circ}$, 77$^{\circ}$, and 85$^{\circ}$, respectively) from $\lambda \lambda$ 1217.5 $-$ 1220.5 {\AA}. Each target is shown in a different color and offset from $I_{Ly \alpha}$ $\approx$ 1.0, which is shown with the dashed gray horizontal line. The laboratory wavelengths of H$_2$ absorption features considered in this study are show with solid pink vertical lines. For each target, absorption profiles are expected to be red-shifted by $v$ sin$i_{disk}$ $\approx$ 2, 0, 10, 15, 19, and 16 km s$^{-1}$, respectively \citep{Nguyen+12,Woitke+13,Quast+00}, which correspond to $\Delta \lambda$ $\sim$ 0.01, 0.00, 0.04, 0.06, 0.08, and 0.07 {\AA}.   \label{fig2}}
\end{figure}

\section{Ly$\alpha$ Normalization and Absorption Line Spectroscopy}\label{sec:lya_norm}

\begin{figure*}[htp!]
\centering
		\subfloat{\includegraphics[angle=90,width=0.475\textwidth]{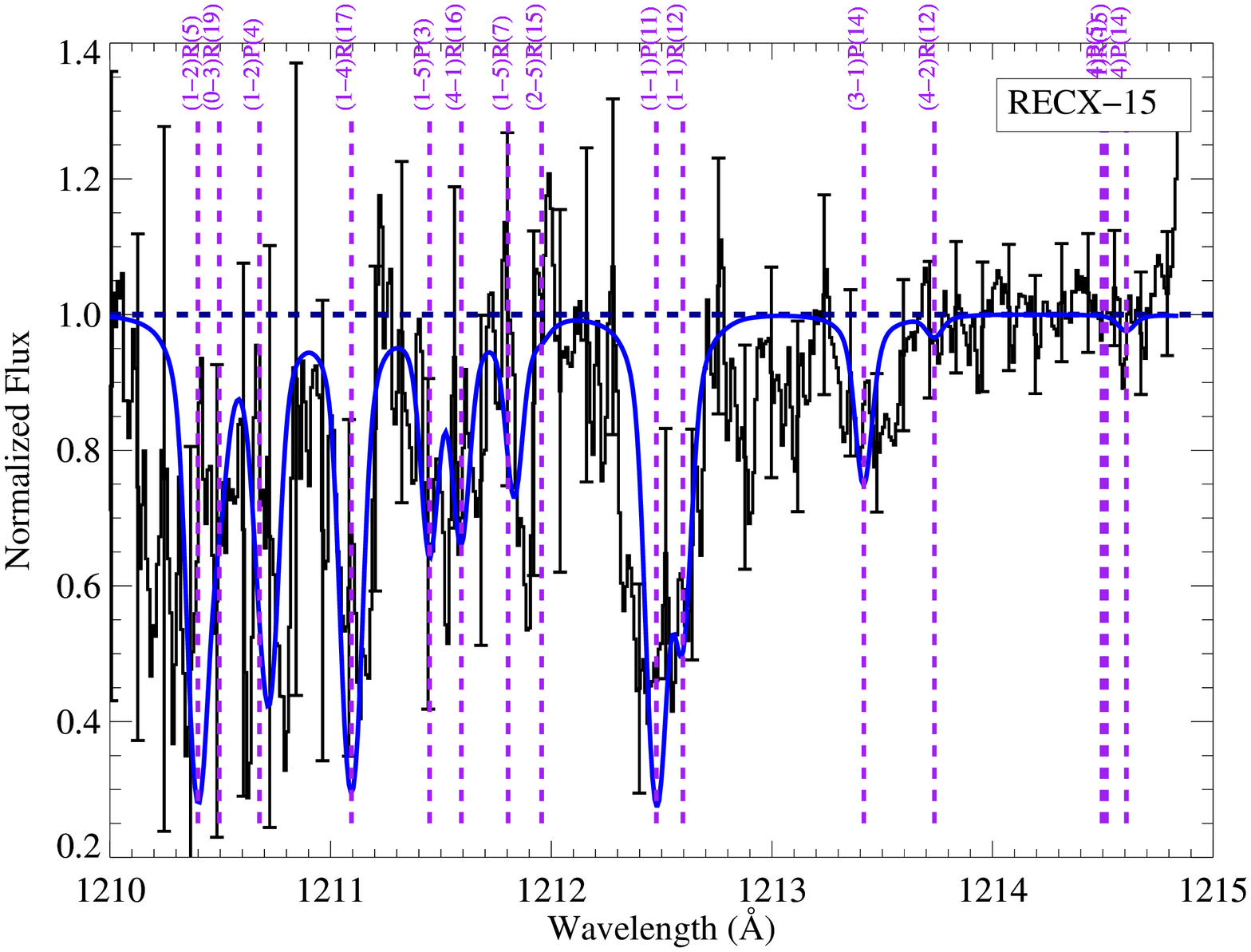}}
    \subfloat{\includegraphics[angle=90,width=0.475\textwidth]{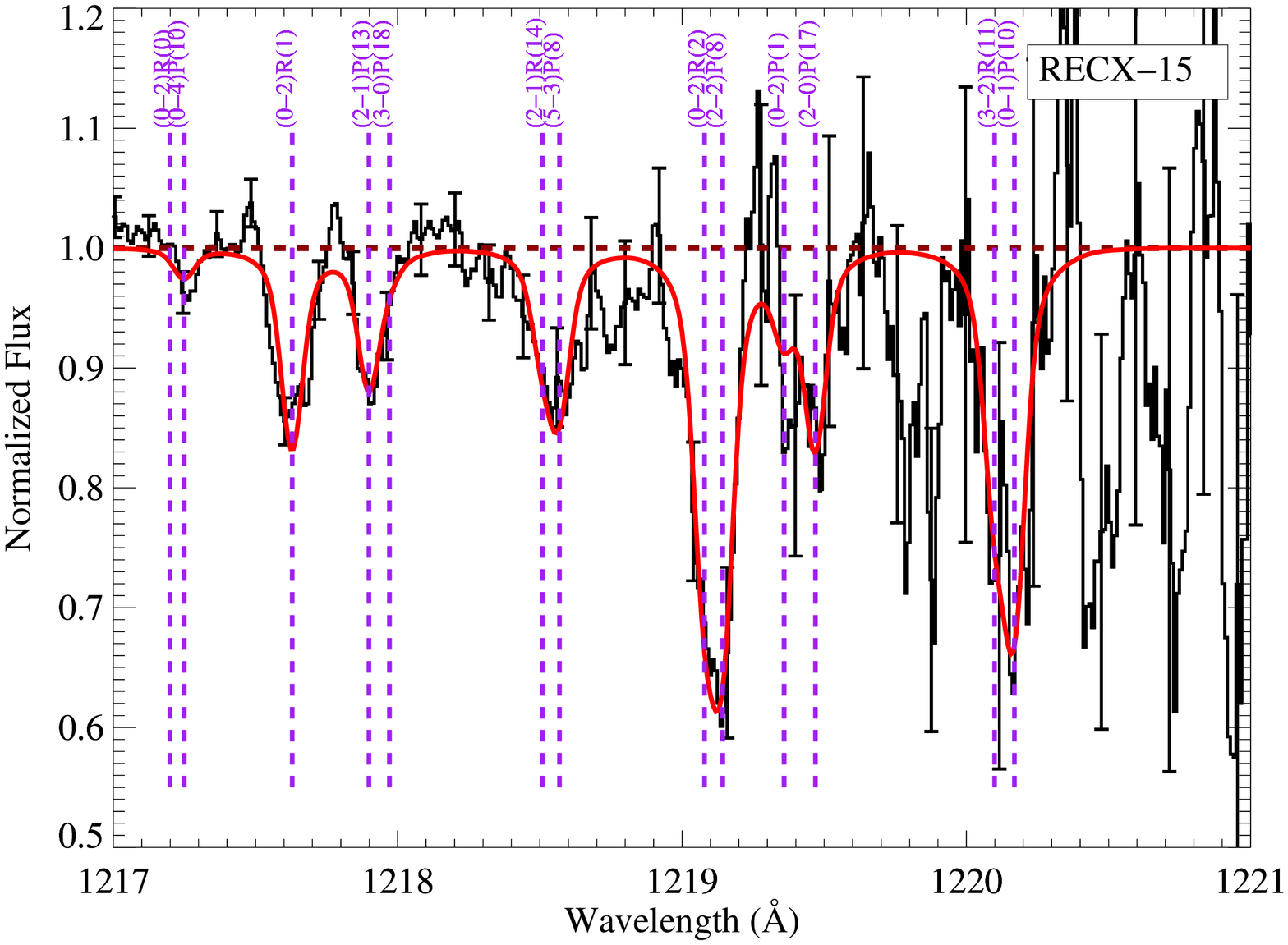}}
	\caption{The best-fit H$_2$ absorption spectrum for RECX-15, assuming $b_{H_2}$ $=$ 5 km s$^{-1}$ is shown in blue (\emph{left}) and red (\emph{right}) over the data (black). Prominent absorption features are labeled in each figure. Table~\ref{tab3} lists all H$_2$ features considered for the fit, and Table~\ref{tab4} presents best-fit thermal model parameters, given the distribution of rovibrational column densities derived from these absorption line fits.  \label{fig2a}}
\end{figure*}

We identify absorption signatures of H$_2$ in each sightline by creating transmission spectra of the stellar Ly$\alpha$ profiles of each PPD host. We treat each Ly$\alpha$ profile as a ``continuum'' source and normalize the emission feature, such that $I_{Ly \alpha}$ $\approx$ 1.0. 
We create a grid of 5 $-$ 10 unique spectral bins from $\lambda \lambda$ 1216.5 $-$ 1221.5 {\AA} (or $\lambda \lambda$ 1210.0 $-$ 1215.0 {\AA} for the blue wing component), which are each selected by hand to avoid molecular absorption features. Each grid bin is defined over 0.35 {\AA}, to both smooth the Ly$\alpha$ emission feature and avoid washing out the H$_2$ absorption features. Within each grid, we measure the mean and standard deviation along the Ly$\alpha$ profile and store them in binned flux and error arrays. We smooth each flux array with a boxcar function of size 0.5 {\AA} over the Ly$\alpha$ bandpass and normalize the Ly$\alpha$ profile with this smoothed grid. An example of the smoothed grid array over the Ly$\alpha$ profile for one of our survey targets is shown in Figure~\ref{fig1}, and all Ly$\alpha$ profiles are presented in Appendix~\ref{app:fig2}.

Figure~\ref{fig2} presents the normalized Ly$\alpha$ spectra for 6 targets, shown in order of inclination angle (edge-on targets on the bottom, and face-on targets towards the top). The effective ``continuum'' levels of the normalized Ly$\alpha$ flux profiles are indicated by the gray dashed lines of each spectrum, and relative flux minima with full width half maximum (FWHM) greater than the spectral resolution of the data are interpreted as absorption features. We highlight where H$_2$ absorption features are expected to reside in the spectrum with solid pink lines. For the edge-on targets (DF Tau, RECX-11, RW Aur), we see the absorption features appear systematically red-shifted. For face-on targets (V4046 Sgr and HD 104237), the position of the absorption features matches the expected laboratory wavelength of H$_2$. The observed red-shift in H$_2$ absorption is expected to fall within corrections made for the radial velocity ($v$ sin$i_{disk}$) of each target and the uncertainty in the COS wavelength solution ($\Delta v$ $\sim$ 15 km s$^{-1}$). Additionally, there are several absorption features seen in more than one target that do not coincide with marked H$_2$ features, most notably around 1218.35 {\AA} and 1219.80 {\AA}. As a first-order check that all H$_2$ and additional absorption features are not artifacts of instrument systematic errors (e.g., gain sag signature from the COS MAMA detector), we compare the Ly$\alpha$ normalized absorption profiles from two observing modes ($\lambda$1291 and $\lambda$1327) of the G130M grating for RECX-15 and find that absorption features appear in both observing modes, giving confidence that these features are real. We will attempt to identify unknown features and verify that these features are real, performing the same check on two different observing modes of COS, in Paper II.

\begin{figure}[htp!]
\centering
    \includegraphics[angle=90,width=0.475\textwidth]{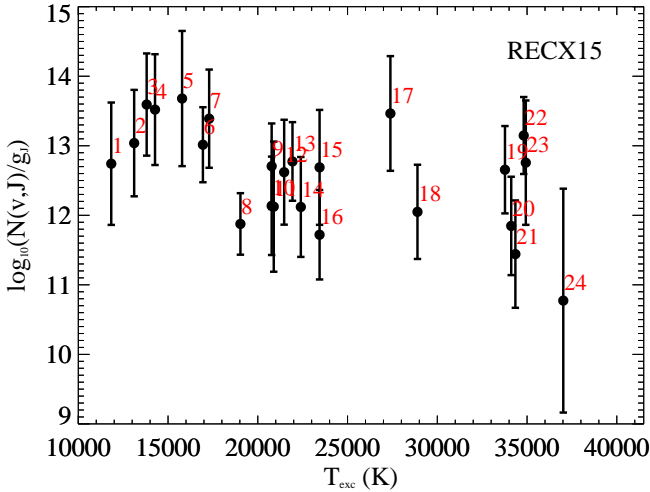}
	\caption{The rotation diagram produced for H$_2$ ground state rovibration levels probed in the protostellar Ly$\alpha$ profile of RECX-15. 
	The column density in each rovibration state is determined from the H$_2$ absorption line fits shown in Figure~\ref{fig2a}. Each label number in the plot corresponds with the following H$_2$ transitions: 1. (0-2)R(1); 2. (1-2)P(4); 3. (1-2)R(5); 4. (0-1)P(10); 5. (1-1)P(11); 6. (2-2)P(8); 7. (1-1)R(12); 8. (2-1)P(13); 9. (3-1)P(14); 10. (2-1)R(14); 11. (3-2)R(11); 12. (2-0)P(17); 13. (5-3)P(8); 14. (4-2)R(12); 15. (4-1)R(16); 16. (3-0)P(18); 17. (1-5)P(3); 18. (0-4)P(10); 19. (2-5)P(11); 20. (0-3)R(19); 21. (1-4)P(14); 22. (1-4)R(17); 23. (2-4)P(18); 24. (2-5)R(15).}  \label{fig2aa}
\end{figure}

We create a multi-component H$_2$ fitting routine to measure the column density in the absorption lines probed within the red and blue stellar wings of Ly$\alpha$, pumped either into the Lyman or Werner electronic band system. We create intrinsic line profiles from the molecular transition properties (listed in Table~\ref{tab3}) to infer the individual column densities probed in each observed rovibrational [$v$,$J$] level, as well as the average H$_2$ population properties (T(H$_2$), $b_{H_2}$, N(H$_2$)). The modeled b-value is fixed in all synthetic absorption spectra to replicate the thermal width of a warm bulk population of H$_2$ (T(H$_2$) $\geq$ 2500 K) in the absence of turbulent velocity broadening. Each line profile is co-added in optical depth space, and a transmission curve is created, which is convolved with either the COS LSF \citep{Kriss+11} or a Gaussian LSF at the STIS resolving power, prior to comparison with the observed Ly$\alpha$ spectra. Each best-fit, multi-absorption feature H$_2$ model is then determined using the \texttt{MPFIT} routine \citep{Markwardt+09}. 
Initial conditions for each transmission curve were first determined by manually fitting each H$_2$ spectrum. To remove bias introduced by the choice of initial conditions, a grid of initial parameters was searched for all sampled absorption spectra. The only parameter allowed to float continuously for all targets was the velocity shift of the line centers of the H$_2$ absorption features, $v_r$.

Figure~\ref{fig2a} shows the normalized H$_2$ absorption profiles in the blue and red Ly$\alpha$ emission profiles of RECX-15, with the best-fit synthetic H$_2$ absorption profiles overlaid in blue (left) and red (right) and labeled with the H$_2$ transition information. Figure~\ref{fig2aa} presents the resulting rotation diagram of H$_2$ ground state rovibrational in the sightline of RECX-15. All other synthetic H$_2$ absorption models and rotation diagrams are presented in Figure~\ref{app:fig3}.
The best-fit column densities and standard deviations are plotted in rotational diagrams against the rovibrational energy level (T$_{exc}$ = E$^{\prime\prime}$/k$_{B}$). Each H$_2$ level is statistically-weighed to correct for ortho- and para-H$_2$ species, such that g$_J$ = (2S + 1)(2J + 1), for S = 0 (para-H$_2$) and S = 1 (ortho-H$_2$).

\begin{figure*}[htp]
\centering
\subfloat{\includegraphics[angle=90,width=0.495\textwidth]{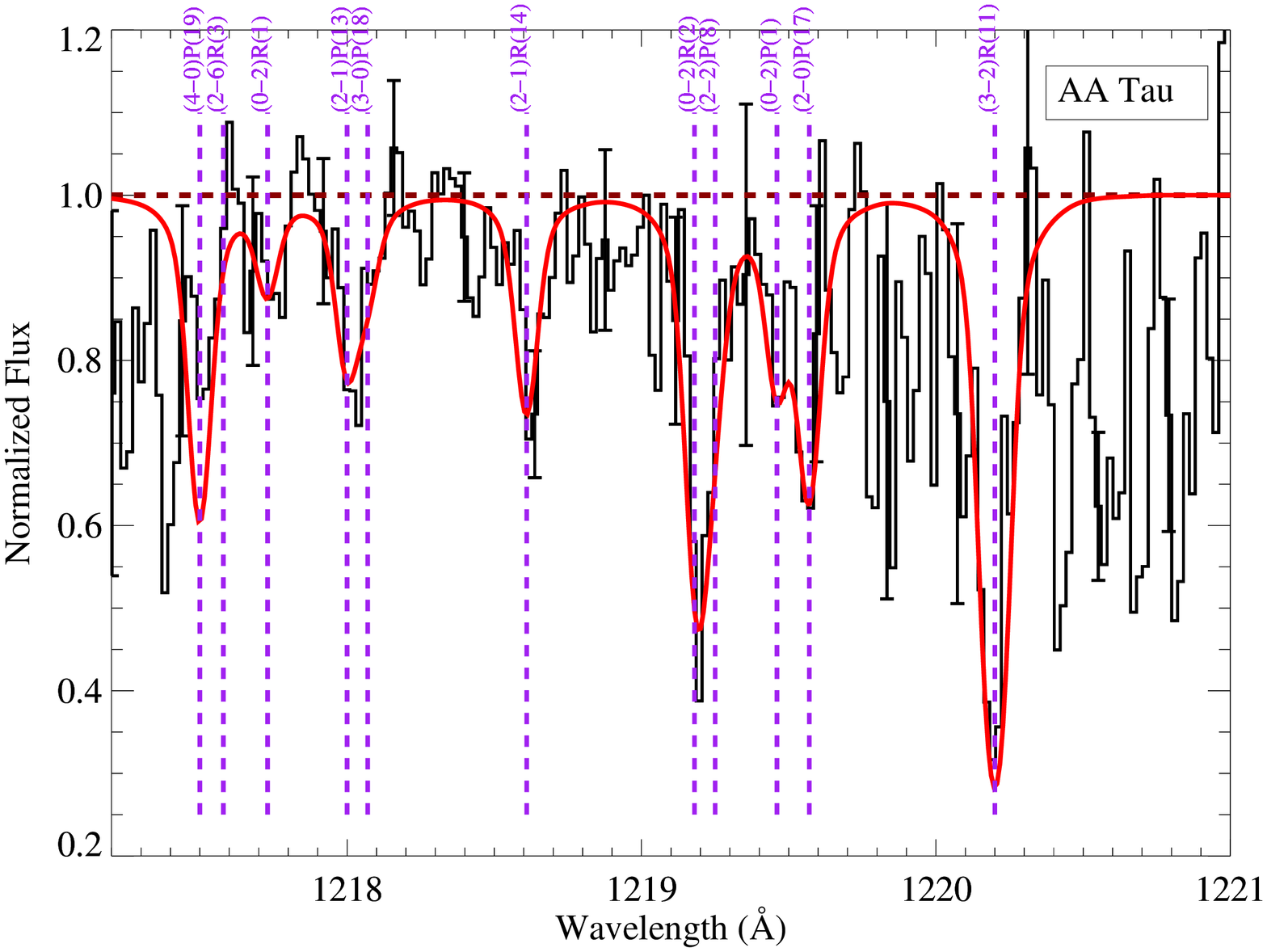}}
    \subfloat{\includegraphics[angle=90,width=0.43\textwidth]{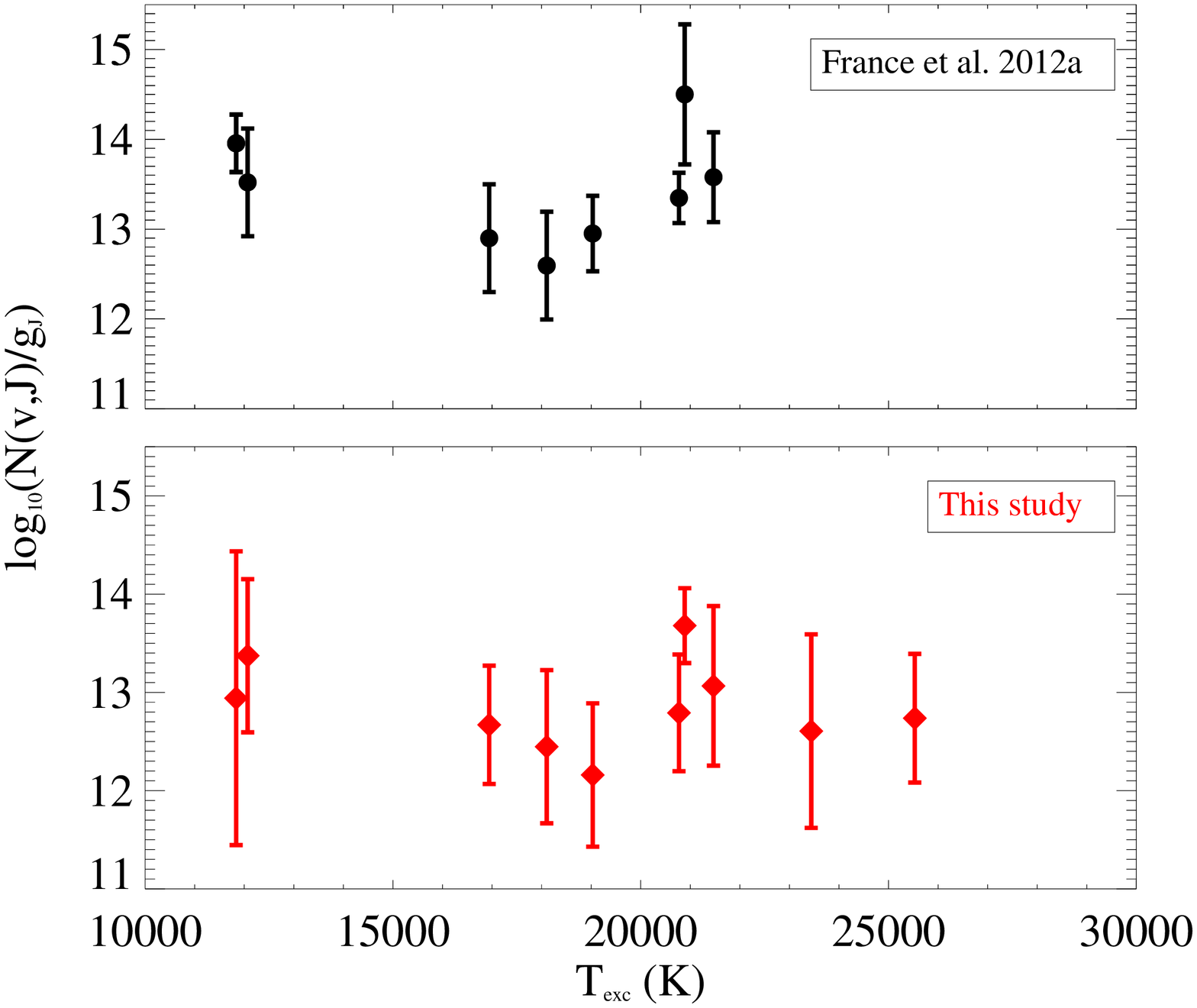}}
	\caption{(a) \emph{Left:} The best-fit synthetic H$_2$ absorption model (red) for AA Tau from 1217 $-$ 1221 {\AA} (black). Each transition is marked with dashed purple lines and identified with the progression ID.   
	\replaced{(b) \emph{Right:} The rotation diagram of H$_2$ populations for features modeled from 1217 $-$ 1221 {\AA} (purple). We compare our column density estimates to results from \citet{France+12a} (black). Both agree within the standard deviations determined from the absorption feature analysis.}{
	(b) \emph{Right:} We compare our column density estimates, determined from the synthetic H$_2$ optical depth modeling framework (red, bottom), to results from \citet{France+12a} (black, top) in H$_2$ rotation diagrams. Both agree within the standard deviations determined from the absorption feature analysis and show comparable temperature slopes.}   \label{fig2b}}
\end{figure*}

We check our methodology by comparing our results to \citet{France+12a}, who performed \replaced{the same}{a similar} procedure for the Ly$\alpha$ absorption spectrum of AA Tau. Figure~\ref{fig2b} (left) presents the H$_2$ absorption spectrum for the red Ly$\alpha$ spectrum of AA Tau, as performed in this study. \deleted{Additional details about the H$_2$ properties and initial conditions for each absorption model are provided in Appendix~\ref{app:h2abs}.} Figure~\ref{fig2b} (right) shows the H$_2$ rotation diagram for AA Tau determined in this study (purple) and \citet{France+12a} (black). The H$_2$ column densities in both studies agree within the error bars determined by the multi-component fit. Our study identified two additional H$_2$ absorption features not fit in \citet{France+12a} (H$_2$[0,19], pumped by $\lambda$1217.41 {\AA}, and H$_2$[6,3], pumped by $\lambda$1217.49 {\AA}). These transitions were important to capture, as observing and characterizing high energy H$_2$ ground states in PPD environments is a key motivation for this study.

%
%
%
%
%
\begin{deluxetable*}{l c c c c | l c c c c}
\tabletypesize{\footnotesize}
\tablecaption{Observed Ly$\alpha$-Pumped H$_2$ Transitions \label{tab3}}
\tablewidth{0pt}
\tablehead{
	\multicolumn{5}{c}{Blue Ly$\alpha$ Wing}	&	\multicolumn{5}{c}{Red Ly$\alpha$ Wing} \\
	\colhead{line ID\tablenotemark{a}} 	& \colhead{$\lambda_{pump}$} &	\colhead{f$_{osc}$\tablenotemark{b}}		& \colhead{E$^{\prime\prime}$\tablenotemark{c}} &	\colhead{A$_{ul}$} & \colhead{line ID\tablenotemark{a}} 	& \colhead{$\lambda_{pump}$} &	\colhead{f$_{osc}$\tablenotemark{b}}		& \colhead{E$^{\prime\prime}$\tablenotemark{c}} &	\colhead{A$_{ul}$}  \\
		\colhead{}		& \colhead{(\AA)}        	     &  \colhead{($\times$ 10$^{-3}$)}        & \colhead{(eV)}     & \colhead{($\times$ 10$^{8}$ s$^{-1}$)}  	& \colhead{}  & \colhead{(\AA)}        	     &  \colhead{($\times$ 10$^{-3}$)}        & \colhead{(eV)}     & \colhead{($\times$ 10$^{8}$ s$^{-1}$)}  	
}
\startdata
	B(1-2)R(5)	& 1210.352	& 36.3		& 1.19		& 1.4 	& C(1-5)R(5)	& 1216.988	& 7.1		& 2.46		& 0.39  \\
	C(0-3)R(19)	& 1210.449	& 25.4		& 2.94		& 1.1 	& C(1-5)R(9)	& 1216.997	& 19.7		& 2.76		& 0.80  \\
	B(1-2)P(4)	& 1210.631	& 29.1		& 1.13		& 1.7 	& B(3-3)R(2)	& 1217.031	& 1.24		& 1.50		& 0.04  \\
	C(2-5)P(11)	& 1210.682	& 30.1		& 2.91		& 1.5 	& B(3-3)P(1)	& 1217.038	& 1.28		& 1.48		& 0.17  \\
	C(1-4)R(17)	& 1211.048	& 37.2		& 3.00		& 1.6 	& B(0-2)R(0)	& 1217.205	& 44.0		& 1.00		& 0.66  \\
	C(1-5)P(3)	& 1211.402	& 7.5		& 2.36		& 0.48 	& C(0-4)Q(10)	& 1217.263	& 10.0		& 2.49		& 0.45  \\
	B(4-1)R(16) & 1211.546	& 25.7		& 2.02		& 1.1 	& B(4-0)P(19)	& 1217.410	& 9.28		& 2.20		& 0.44  \\
	C(1-5)R(7)	& 1211.758	& 24.2		& 2.57		& 0.97 	& C(2-6)R(3)	& 1217.488	& 36.4		& 2.73		& 1.30  \\
	C(2-4)P(18)	& 1211.787	& 15.2		& 3.01		& 0.73 	& B(0-2)R(1)	& 1217.643	& 28.9		& 1.02		& 0.78  \\
	C(2-5)R(15)	& 1211.910	& 32.8		& 3.19		& 1.4 	& B(2-1)P(13)	& 1217.904	& 19.2		& 1.64		& 0.93  \\
	B(1-1)P(11)	& 1212.426	& 13.3		& 1.36		& 0.66 	& B(3-0)P(18)	& 1217.982	& 6.64		& 2.02		& 0.32  \\
	B(1-1)R(12)	& 1212.543	& 10.9		& 1.49		& 0.46 	& B(2-1)R(14)	& 1218.521	& 18.1		& 1.79		& 0.76  \\
	B(3-1)P(14)	& 1213.356	& 20.6		& 1.79		& 1.00	& B(5-3)P(8)	& 1218.575	& 12.9		& 1.89		& 0.66  \\
	B(4-2)R(12)	& 1213.677	& 9.33		& 1.93		& 0.39 	& B(0-2)R(2)	& 1219.089	& 25.5		& 1.04		& 0.82  \\
	C(3-6)R(13)	& 1214.421	& 5.17		& 2.07		& 0.29 	& B(2-2)R(9)	& 1219.101	& 31.8		& 1.56		& 1.30  \\
	B(3-1)R(15)	& 1214.465	& 23.6		& 1.94		& 1.00	& B(2-2)P(8)	& 1219.154	& 21.4		& 1.46		& 1.10  \\
	C(1-4)P(14)	& 1214.566	& 28.3		& 2.96		& 1.40	 & B(0-2)P(1)	& 1219.368	& 14.9		& 1.02		& 2.00  \\
	B(4-3)P(5)	& 1214.781	& 9.90		& 1.65		& 0.55 	& B(2-0)P(17)	& 1219.476	& 3.98		& 1.85		& 0.19  \\
				&			&			&			&		& B(0-1)R(11)	& 1219.745	& 3.68		& 1.36		& 0.15 \\
				&			&			&			&		& B(3-2)R(11)	& 1220.110	& 21.3		& 1.80		& 0.88 \\
				&			&			&			&		& B(0-1)P(10)	& 1220.184	& 5.24		& 1.23		& 0.26 \\
\enddata
\tablenotetext{a}{Describes ground state-to-excited state transition, due to absorption of Ly$\alpha$ photon $\lambda_{pump}$. IDs beginning with ``B'' are excited to Lyman excitation level ($2p\sigma B ^{1}\Sigma^{+}_{u}$), and IDs beginning with ``C'' are excited to Werner excitation state ($2p\pi C ^{1}\Pi_{u}$).}
\tablenotetext{b}{The oscillator strength of the transition.}
\tablenotetext{c}{The energy level of ground state ($X ^{1}\Sigma^{+}_{g}$) H$_2$ before photo-excitation.}
\end{deluxetable*}

%
%
%
\section{Analysis \& Results} \label{sec:results}
We aim to characterize the behavior of the rovibrational H$_2$ populations identified in the PPD host Ly$\alpha$ spectra and estimate the total thermal and non-thermal column densities (N(H$_2$) and N(H$_2$)$_{\textnormal{nLTE}}$) of H$_2$ in each sightline. 

\begin{figure}[htp]
\centering
\includegraphics[angle=90,width=0.475\textwidth]{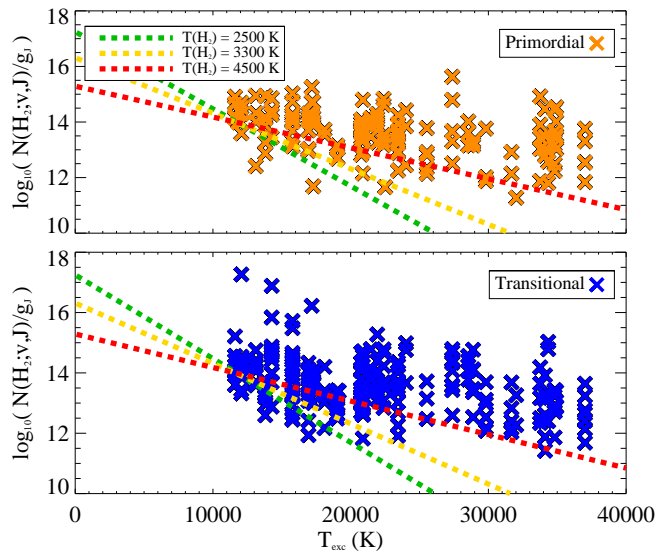}
	\caption{A relative comparison of H$_2$ rotation diagram behaviors. We normalize the rotation diagrams for H$_2$ distributions in all PPD sightlines to the [$v$ = 2,$J$ = 1] level. We split these relative spectra by disk evolution, where primordial targets are shown at the top (orange) and transitional targets are shown at the bottom (blue). We fit thermally-distributed H$_2$ through the [$v$ = 2,$J$ = 1] level for warm (T(H$_2$) = 2500 K; green) and hot (T(H$_2$) = 3500 K and 4500 K; yellow and red) H$_2$ populations. 
	\label{fig3aa} }
\end{figure}

Figure~\ref{fig3aa} presents the rotation diagrams for all targets in this survey. We split the sampled sightlines by PPD evolution phase, which we define by the behavior of excess infrared (dust) emission from 13 $-$ 31 $\mu$m \citep{Furlan+09}. Primordial PPDs are thought to be \added{``young''} disks with very little evidence of dust evolution and grain growth, meaning planet formation is either starting or in very early stages. Transitional disks are viewed as \added{``older''} disks where proto-planets have formed and are evolving, since the observed infrared dust distributions point to the build-up of larger dust grains. Transition disks also (typically) harbor one or more large dust cavities that indicate significant evolution of the disk material (e.g., see \citealt{Strom+89,Takeuchi+Artymowicz+01,Calvet+02,Calvet+05,Espaillat+07a}). 
To explore the behavior of H$_2$ populations simultaneously in all PPD sightlines, we normalize each H$_2$ rotation diagram to the \added{column density in the} [$v$ = 2,$J$ = 1] level. 
We include thermal models of warm/hot distributions of H$_2$ populations, drawn through the normalization rovibrational level [$v$ = 2,$J$ = 1], which range from the expected thermal populations of fluorescent H$_2$ in PPDs \citep{Herczeg+02,Herczeg+06,France+12b,Hoadley+15} to the dissociation limit of the molecule (red dashed line for T$_{diss}$ $\approx$ 4500 K; \citealt{Shull+82,Williams+00}). 

Despite the evolutionary differences in the dust distributions between the two PPD types, primordial and transitional PPD sightlines appear to show very similar H$_2$ rovibrational behaviors. Thermal distributions for T(H$_2$) $<$ 3300 K do not appear to describe the behavior of H$_2$ rovibration levels for T$_{exc}$ $>$ 23,000 K, but a thermal distribution of H$_2$ at or near the dissociation limit of the molecule does appear to be consistent with the lowest column densities of rovibrational H$_2$ at 23,000 K $<$ T$_{exc}$ $<$ 40,000 K. Still, we note that the majority of H$_2$ levels are significantly pumped, sometimes by as much as 4 dex, above the thermal distribution of H$_2$.

Additionally, we see a striking behavior in the distribution of H$_2$ rovibrational levels with T$_{exc}$ $>$ 20,000 K. At T$_{exc}$ $\sim$ 20,000 K, there is an abrupt upturn, or ``knee,'' away from the thermal distributions and an increase in rovibrational column density for higher excitation temperature states by $\gtrsim$ 1 dex. This ``knee'' appears to repeat around T$_{exc}$ $\sim$ 25,500 K and 31,000-32,000 K. This behavior, specifically between the ``knees'' at T$_{exc}$ $\sim$ 25,500 and 32,000 K, may be a result of under-sampling the distribution of highly-energetic H$_2$ with ground state energies in this range.

Non-thermal pumping mechanisms include many complex processes, which are challenging and computationally-expensive to model simultaneously; \citet{Nomura+Millar+05} and \citet{Nomura+07} show how many mechanisms, such as chemical processes (resulting in the destruction and formation of H$_2$), FUV/X-ray pumping, and dust grain formation and size distributions in PPD atmospheres \citep{Habart+04,Aikawa+Nomura+06,Nomura+06,Fleming+10}, affect the population ratios of H$_2$ and pump H$_2$ populations out of thermal equilibrium. However, \citet{Nomura+Millar+05} also show that small changes in any of these processes can have dramatic effects on the final structure of H$_2$ rovibrational levels. Since we do not sample the full suite of [$v$,$J$] ground states in this absorption line study, \replaced{it is not beneficial to}{we do not} attempt to model multiple, non-thermal mechanisms in the hope of reproducing the observed behavior of H$_2$ rovibration levels.

Instead, we compare the observed rovibration level distributions to thermal H$_2$ models. 
While thermal models alone will not explain the distributions and behaviors of H$_2$ in PPD sightlines, exploring various thermal distribution realizations will help place limits on the total thermal column density of H$_2$ in each PPD sightline.

We fit two thermal distributions to the rovibrational levels of each target: 
\begin{enumerate}
	\item Model 1: We fit purely thermal distributions of H$_2$ to all observed rovibrational states, regardless of excitation temperature. 
	\item Model 2: We fit purely thermal distributions of H$_2$ to only observed rovibrational states with T$_{exc}$ $\leq$ 17,500 K (E$^{\prime \prime}$ $\leq$ 1.5 eV). 
\end{enumerate}
We discuss the details of the molecular physics and energy equations used for Models 1 and 2 in Appendix~\ref{app:models}. 
Each model is optimized to the rotation diagram of each target through a Markov Chain Monte Carlo (MCMC) routine, performed with the Python \texttt{emcee} package \citep{Foreman+12}. The routine uses randomly-generated initial conditions and minimizes the likelihood function of the observed rovibrational column densities, given the range of model parameters. This process determines the best representative thermal model parameters (N(H$_2$),T(H$_2$)) to the data. Further details about the MCMC and parameter fits are discussed in Appendix~\ref{app:mcmc}.

\subsection{Thermal and Non-Thermal Column Densities} \label{sec:columns}

\begin{deluxetable*}{l c c c c c c c}
\tabletypesize{\small}
\tablecaption{Thermal H$_2$ Column Density \& Temperature Results \label{tab4}}
\tablewidth{0pt}
\tablehead{
						&				&  \multicolumn{2}{c}{Model 1}		 & \multicolumn{2}{c}{Model 2}   \\
\cmidrule(r{1em}){3-4} \cmidrule(r{1em}){5-6} 
	 \colhead{Target}  &	\colhead{$v_r$\tablenotemark{*}}	& \colhead{N(H$_2$)\tablenotemark{a}} &	\colhead{T(H$_2$)\tablenotemark{b}}  & \colhead{N(H$_2$)\tablenotemark{a}} & \colhead{T(H$_2$)\tablenotemark{b}} &  \colhead{N(H$_2$)$_{\textnormal{nLTE}}$\tablenotemark{a}} & \colhead{N(H$_2$[5,18])\tablenotemark{a,c}} 
}              
\startdata
	AA Tau & 20.1	&	16.27$^{+0.50}_{-0.34}$ & 4179$^{+585}_{-887}$ & 15.85$^{+0.11}_{-0.11}$ & 3578$^{+282}_{-221}$ & 16.40$^{+0.01}_{-0.01}$ & 10.35 \\
	AB Aur & -12.8	& 15.59$^{+0.31}_{-0.20}$ & 4488$^{+376}_{-704}$ & 15.34$^{+0.34}_{-0.26}$ & 3628$^{+744}_{-631}$ & 15.44$^{+0.01}_{-0.01}$ & - \\
	AK Sco &  -4.3	&	15.57$^{+0.17}_{-0.16}$ & 4880$^{+90}_{-190}$ & 15.52$^{+0.51}_{-0.29}$ & 3661$^{+872}_{-922}$ & 15.04$^{+0.05}_{-0.01}$ & -  \\
	BP Tau &  22.4	&	15.50$^{+0.21}_{-0.19}$ & 4855$^{+107}_{-220}$ & 15.11$^{+0.55}_{-0.31}$ & 3693$^{+868}_{-972}$ & 15.37$^{+0.01}_{-0.02}$ & 10.72 \\
	CS Cha &  13.6	&	15.82$^{+0.17}_{-0.16}$ & 4889$^{+83}_{-174}$ & 15.27$^{+0.57}_{-0.34}$ & 3536$^{+954}_{-962}$ & 15.52$^{+0.01}_{-0.02}$ & 9.92 \\
	DE Tau &  11.8	&	16.20$^{+0.50}_{-0.32}$ & 4082$^{+644}_{-927}$ & 16.08$^{+0.86}_{-0.50}$ & 3466$^{+1030}_{-1120}$ & 16.03$^{+0.01}_{-0.01}$ & - \\
	DF Tau A &  34.8	&	15.13$^{+0.29}_{-0.19}$ & 4375$^{+443}_{-695}$ & 14.98$^{+0.09}_{-0.09}$ & 3382$^{+188}_{-159}$ & 14.74$^{+0.01}_{-0.01}$ & 11.19 \\
	DM Tau &  40.6	&	16.02$^{+0.20}_{-0.18}$ & 4810$^{+140}_{-274}$ & 16.14$^{+0.75}_{-0.54}$ & 2900$^{+1170}_{-776}$ & 15.90$^{+0.01}_{-0.02}$ & 10.23 \\
	GM Aur &  25.4	&	15.84$^{+0.18}_{-0.17}$ & 4873$^{+95}_{-200}$ & 15.67$^{+0.68}_{-0.50}$ & 2966$^{+1096}_{-762}$ & 15.51$^{+0.01}_{-0.02}$ & - \\
	HD 104237 & 0.6	&	15.95$^{+0.27}_{-0.26}$ & 4831$^{+126}_{-264}$ & 15.16$^{+0.46}_{-0.28}$ & 3734$^{+830}_{-906}$ & 16.47$^{+0.01}_{-0.01}$ & - \\
	HD 135344 B & 6.7 &	15.60$^{+0.18}_{-0.17}$ & 4886$^{+86}_{-181}$ & 15.24$^{+0.42}_{-0.29}$ & 3544$^{+878}_{-770}$ & 15.26$^{+0.01}_{-0.02}$ & - \\
	HN Tau A &  24.2	&	16.92$^{+1.03}_{-0.64}$ & 3035$^{+1193}_{-966}$ & 16.85$^{+1.08}_{-0.72}$ & 2798$^{+1305}_{-912}$ & 14.63$^{+1.20}_{-0.20}$ & - \\
	LkCa15 &  35.0	&	17.77$^{+0.62}_{-0.51}$ & 4556$^{+324}_{-611}$ & 17.35$^{+0.11}_{-0.11}$ & 3516$^{+260}_{-200}$ & 17.64$^{+1.50}_{-0.20}$ & 10.01 \\
	RECX 11 &  24.5	&	15.84$^{+0.13}_{-0.13}$ & 4905$^{+71}_{-147}$ & 15.55$^{+0.24}_{-0.17}$ & 3939$^{+629}_{-611}$ & 15.64$^{+0.01}_{-0.01}$ & 9.98 \\
	RECX 15 & -2.7 	&	16.03$^{+0.21}_{-0.20}$ & 4858$^{+106}_{-219}$ & 15.47$^{+0.47}_{-0.27}$ & 3944$^{+729}_{-950}$ & 15.63$^{+0.01}_{-0.02}$ & 9.48 \\
	RU Lupi &  6.8	&	16.03$^{+0.21}_{-0.19}$ & 4765$^{+174}_{-336}$ & 15.38$^{+0.61}_{-0.34}$ & 3840$^{+807}_{-1106}$ & 15.66$^{+0.01}_{-0.02}$ & - \\
	RW Aur A &  12.4	&	16.23$^{+0.29}_{-0.27}$ & 4822$^{+133}_{-263}$ & 15.60$^{+0.56}_{-0.33}$ & 3729$^{+858}_{-1005}$ & 17.36$^{+0.01}_{-0.01}$ & - \\
	SU Aur &  36.0	&	16.21$^{+0.51}_{-0.38}$ & 4264$^{+525}_{-802}$ & 16.51$^{+3.48}_{-1.22}$ & 2574$^{+1654}_{-1565}$ & 15.31$^{+3.00}_{-0.20}$ & -  \\
	SZ 102 &  -20.7	&	15.43$^{+0.20}_{-0.15}$ & 4493$^{+362}_{-530}$ & 15.83$^{+0.32}_{-0.34}$ & 2785$^{+588}_{-366}$ & 15.26$^{+0.01}_{-0.01}$ & - \\
	TW Hya	&  19.6	&	15.40$^{+0.17}_{-0.16}$ & 4880$^{+89}_{-192}$ & 15.08$^{+0.54}_{-0.33}$ & 3483$^{+954}_{-887}$ & 15.19$^{+0.01}_{-0.02}$ & 11.31 \\
	UX Tau A &  33.0	&	16.76$^{+0.38}_{-0.34}$ & 4668$^{+244}_{-460}$ & 16.40$^{+1.32}_{-0.56}$ & 3129$^{+1283}_{-1383}$ & 16.38$^{+2.60}_{-0.20}$ & - \\
	V4046 Sgr &  -4.7	&	15.33$^{+0.15}_{-0.14}$ & 4894$^{+80}_{-164}$ & 15.05$^{+0.40}_{-0.25}$ & 3900$^{+740}_{-891}$ & 15.05$^{+0.01}_{-0.01}$ & 10.27  \\
	\hline 
	Avg. Model Results & 	&	15.97$^{+1.80}_{-0.84}$ & 4604$^{+301}_{-1570}$ & 15.70$^{+1.65}_{-0.72}$ & 3442$^{+500}_{-870}$ &		&		\\
\enddata
\tablenotetext{*}{The radial velocity of the system, derived from the synthetic H$_2$ optical depth curves, are expressed as km s$^{-1}$.}
\tablenotetext{a}{All column densities are to the power of 10 (log$_{10}$N(H$_2$)).}
\tablenotetext{b}{Thermal temperatures of the bulk H$_2$ populations (T(H$_2$)) are in Kelvin.}
\tablenotetext{c}{Estimated from the formalism outlined in \citet{Rosenthal+00} (Equation~\ref{eqn1}). We assume the H$_2$[5,18] population is optically thin.}
\end{deluxetable*}

\begin{figure}[htp]
\centering
\includegraphics[angle=270,width=0.475\textwidth]{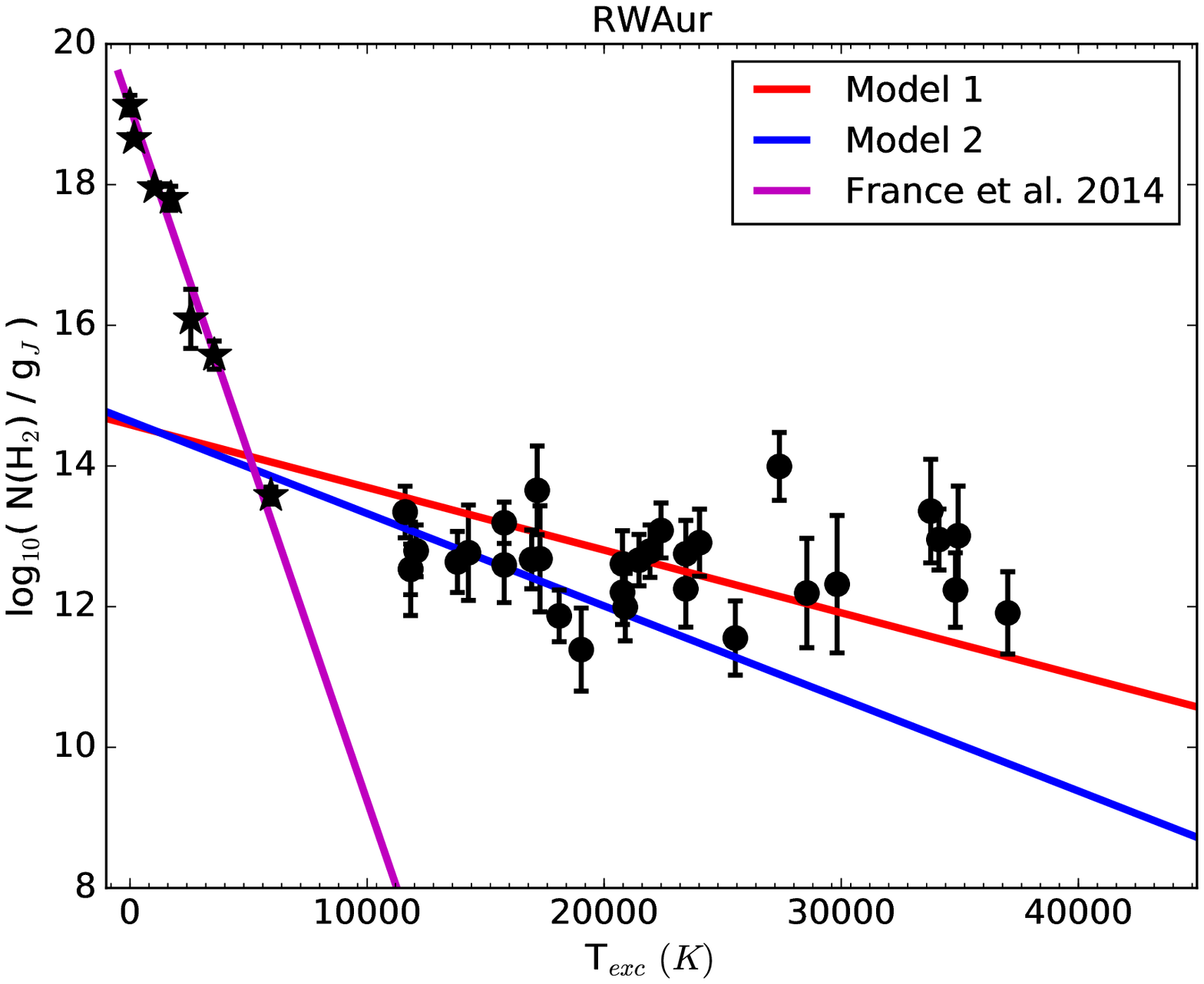}
	\caption{The rotation diagram for RW Aur, with rovibrational column densities derived in this study (black circles) and lower energy states \replaced{calculated}{measured} by \citet{France+14b} (black stars; $\lambda \lambda$ 1092.5 $-$ 1117 {\AA}). The red and blue solid lines represent thermal distributions of H$_2$ levels populated in Models 1 and 2, respectively. The magenta solid line shows the thermal distribution H$_2$ levels examined by \citet{France+14b}, with log$_{10}$( N(H$_2$) ) $=$ 19.90 cm$^{-2}$ and T(H$_2$) $=$ 440 K.
	\label{fig3a} }
\end{figure}

Each set of best-fit thermal model parameters is shown in Table~\ref{tab4}. 
Figure~\ref{fig3a} shows the rovibrational levels and thermal model realizations for RW Aur. We present data from this study (black circles) and lower excitation temperature states from \citet{France+14b} (black stars), which were detected against the FUV continuum between $\lambda \lambda$ 1092.5 $-$ 1117 {\AA}. RW Aur is the only target in our sample with both sets of H$_2$ data and provides \replaced{a great}{an} example for visualizing how higher excitation temperature ground states deviate from the warm thermal levels of H$_2$, which are likely probing the denser regions within the disk atmosphere (log$_{10}$(N(H$_2$)) $=$ 19.90 cm$^{-2}$ and T(H$_2$) $=$ 440 K: magenta; \citealt{France+14b}). Higher energy rovibrational H$_2$ levels appear to scatter out of thermal equilibrium and are described by higher \replaced{bulk}{effective} temperatures, as predicted by \citet{Nomura+Millar+05}. We present all H$_2$ rotation diagrams and thermal distribution realizations for each target in our survey in Appendix~\ref{app:fig4}.

Table~\ref{tab3} lists the average S/N of each Ly$\alpha$ emission profile as observed by either \textit{HST}/COS or \textit{HST}/STIS.  
We compute a Spearman rank coefficient between the best-fit thermal model N(H$_2$) and the Ly$\alpha$ wing S/N and find significant trends for both Model 1 ($\rho$ = -0.71, with a probability to exceed the null hypothesis that the data are drawn from random distributions (p\footnote{The strength of p is defined as follows: p $>$ 5\% (5.0$\times$10$^{-2}$) = no correlation; 1\% $<$ p $<$ 5\% = possible correlation; 0.1\% $<$ p $<$ 1\% = correlation; p $<$ 0.1\% = strong correlation.} = 7.0$\times$10$^{-3}$) and Models 2 ($\rho$ = -0.78, p = 5.6$\times$10$^{-2}$). 
However, when we exclude one low S/N data point from the correlation (LkCa 15) and re-calculate the Spearman rank coefficient for both model realizations, we see a more randomly-distributed set of modeled column density estimates (Model 1: $\rho$ = -0.22, p = 3.91$\times$10$^{-1}$ and Model 2: $\rho$ = -0.27, p = 1.92$\times$10$^{-1}$). Therefore, for the remainder of this study, we do not include LkCa 15 results in further analysis.

We use the results from Models 1 and 2 to estimate the total column density of thermally-distributed H$_2$ (N(H$_2$)) in each sight line. 
We choose to represent the thermal distributions of hot H$_2$ with the results from Model 2. T(H$_2$) from Model 2 represents a more realistic determination of the bulk temperature profiles of thermal H$_2$ (T(H$_2$) $\sim$ 2500 $-$ 3500 K) in each sightline, whereas Model 1 produces T(H$_2$) $\approx$ T$_{diss}$(H$_2$). In reality, there is very little difference between N(H$_2$) determined from Models 1 and 2; both model realizations predict similar N(H$_2$), though Model 2 results tend to under-predict N(H$_2$) when compared to Model 1 results, and thus provide a lower limit to the total thermal column density of hot H$_2$.

To approximate how much of the total observed H$_2$ column density is associated with excess H$_2$ populations in highly energetic (non-thermal) states, we define a metric for the total non-thermal column density of H$_2$ in highly excited levels (E$^{\prime\prime}$ $>$ 1.75 eV, or T$_{exc}$ $>$ 20,000 K), which we refer to as N(H$_2$)$_{\textnormal{nLTE}}$. N(H$_2$)$_{\textnormal{nLTE}}$ is calculated by integrating the residual between observed H$_2$ rovibration levels with T$_{exc}$ $>$ 20,000 K and the predicted populations of H$_2$ at that same rovibration level from the modeled thermal distributions, or N(H$_2$)$_{\textnormal{nLTE}}$ = $\Sigma_{[v,J]} ($N(H$_2$[$v$,$J$])$_{\textnormal{obs}}$ - N(H$_2$[$v$,$J$])$_{\textnormal{model}} )$. For consistency, we calculate N(H$_2$)$_{\textnormal{nLTE}}$ from all best-fit model realizations from both Models 1 and 2. We find we are able to produce approximately the same N(H$_2$)$_{\textnormal{nLTE}}$ estimate from N(H$_2$) of both Models 1 and 2. Associated error bars on N(H$_2$)$_{\textnormal{nLTE}}$ are estimated as the minimum and maximum deviations away from the median N(H$_2$)$_{\textnormal{nLTE}}$ for all Model 1 and Model 2 best-fit thermal parameters.
Table~\ref{tab4} includes our estimates of N(H$_2$)$_{\textnormal{nLTE}}$ for each target (for which we include LkCa 15, but we do not use in further analysis).

\subsection{\ion{C}{4}-Pumped H$_2$ Fluorescence}\label{sec:civh2}
Molecular hydrogen populations photo-excited by \ion{C}{4} photons ($\lambda$ 1548.20, 1550.77 {\AA}) are found in highly excited ground states ([3,25], [5,18], and [7,13]; E$^{\prime\prime}$ $\geq$ 3.8 eV, T$_{exc}$ $>$ 43,000 K) that are difficult to explain with thermally-generated H$_2$ populations alone.\deleted{ at temperatures probed in PPDs (T(H$_2$) $\sim$ 2000 - 3000 K; \citet{Herczeg+06,France+12b,Hoadley+15,McJunkin+16}).} These highly excited states are also unlikely to be directly populated by the fluorescence process. Electronic transitions are dipole-allowed, meaning \replaced{$J^{\prime \prime}$ = $\pm$1}{$J^{\prime} - J^{\prime \prime}$ = $\pm$1 or 0 (for Werner band lines with $J^{\prime \prime}$ $\neq$ 0)} between excited and ground state transitions. \replaced{Therefore, the}{The} decay from excited electronic to ground states can easily increase the ground electronic vibrational levels, but will not substantially change the ground electronic quantum rotational levels \citep{Herczeg+06}. Therefore, other physical \replaced{processes}{mechanisms}, such as collisional \citep{Bergin+04} and chemical \citep{Takahashi+99,Adamkovics+16} processes, must be responsible for populating these highly energetic levels of H$_2$. 

\begin{figure}[htp]
\centering
	\includegraphics[angle=90,width=0.475\textwidth]{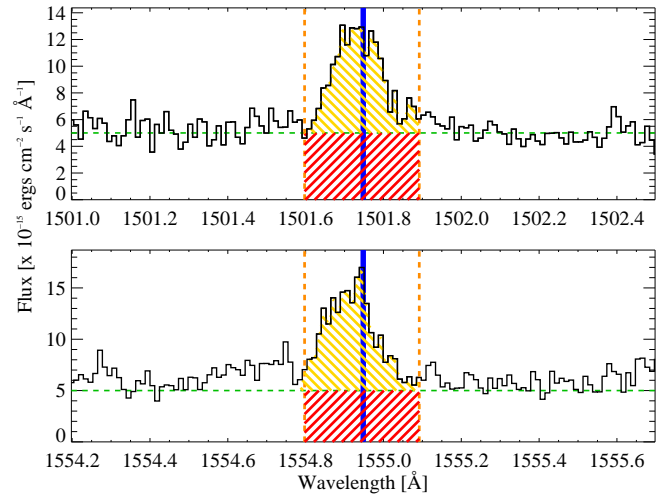}
	\caption{The presence of \ion{C}{4}-pumped H$_2$ emission from the B(0-5)P(18) progression. 
	We present \replaced{this process}{example fluorescence lines} for BP Tau, for emission lines at 1501.75 {\AA} ((0-5)R(16)) and 1554.95 {\AA} ((0-6)R(16)), indicated by the blue dotted lines. The green dashed line shows the continuum levels in each spectral region. The orange dashed lines mark off the region considered for each fluorescence line. The yellow hashed region represents the integrated flux F(\ion{C}{4}-H$_2$) within the orange region, while the red hashed region represents the integrated continuum flux in the same region.  \label{fig5}}
\end{figure}

Since we do not know which processes dominate the pumping of H$_2$ into these highly energetic upper rotational levels, we use the emission from \ion{C}{4}-pumped H$_2$ as a proxy for a variety of non-thermal processes that may excite H$_2$ to highly non-thermal states. We estimate the column density of H$_2$ populating these energetic levels from the total fluorescent emission produced by \ion{C}{4}-pumped H$_2$. We stipulate two conditions to verify whether the target exhibits \ion{C}{4}-pumped H$_2$ emission in the FUV spectrum: 1) each emission line must have an elevated flux level $\geq$ 1.5$\sigma$ above the continuum floor, and 2) at least two emission lines from the same progression must be present. Figure~\ref{fig5} demonstrates how we determine that the emission line exists above the FUV continuum. Only fluorescence from the B(0-5)P(18) 1548.15 {\AA} transition meets this criteria for \replaced{all targets}{targets which show signs of \ion{C}{4}-pumped H$_2$ emission} in our survey. The two brightest transitions from the B(0-5)P(18) 1548.15 {\AA} cascade, $\lambda$ 1501.75 {\AA} and $\lambda$ 1554.95 {\AA}, are free of blending from other atomic or molecular contaminants \citep{Herczeg+06}. Therefore, emission features observed at these wavelengths are detected fluorescence transitions, originating from the highly non-thermal H$_2$ state [5,18]. Of the 22 targets, 10 show statistically significant emission lines from \ion{C}{4}-pumped H$_2$ fluorescence.

To estimate the density of highly excited H$_2$, we use the flux from the two brightest emission features at $\lambda$ 1501.75 {\AA} and $\lambda$ 1554.95 {\AA}, after subtracting the UV continuum. We assume that H$_2$[5,18] is optically thin and estimate the total column density of this highly non-thermal ground state (N(\ion{C}{4}-H$_2$)) from the formalism described in \citet{Rosenthal+00}, 
\begin{equation}
\scalebox{0.9}{$
	N(C IV \rightarrow H_2[v^{\prime \prime},J^{\prime \prime}]) = \frac{4 \pi \lambda}{h c} \frac{F(C IV-H_2)([v^{\prime},J^{\prime}] \rightarrow [v^{\prime \prime},J^{\prime \prime}])}{A_{ul}([v^{\prime},J^{\prime}] \rightarrow [v^{\prime \prime},J^{\prime \prime}])}
	$}
	\label{eqn1}
\end{equation}

where N(\ion{C}{4}$\rightarrow$H$_2$[$v^{\prime \prime}$,$J^{\prime \prime}$]) is the column density of \ion{C}{4}-pumped H$_2$ that decays to ground state [$v^{\prime \prime}$,$J^{\prime \prime}$], $\lambda$ is the transition wavelength between electronic and ground states, F(\ion{C}{4}-H$_2$) is the integrated flux in the emission line produced by the transition between excited electronic level [$v^{\prime}$,$J^{\prime}$] and ground level [$v^{\prime \prime}$,$J^{\prime \prime}$], and $A_{ul}$ is the spontaneous decay \replaced{coefficient}{rate} for the transition. For each emission line, we calculate N(\ion{C}{4}-H$_2$) and take the average of the results from the two emission features as the estimate of N(\ion{C}{4}-H$_2$). Error bars on N(\ion{C}{4}-H$_2$) are taken as the residual between the N(\ion{C}{4}-H$_2$) and the column density derived from each emission feature at $\lambda$ 1501.75 {\AA} and $\lambda$ 1554.95 {\AA}. Derived N(\ion{C}{4}-H$_2$) values are listed in Table~\ref{tab4}. All column densities derived from the fluorescence emission from the B(0-5)P(18) progression are log$_{10}$(N(\ion{C}{4}-H$_2$)) $<$ 12.0 cm$^{-2}$, which is consistent with a thin layer of highly energetic H$_2$ \citep{Herczeg+06}.

%
%
%
\section{Discussion}

This study has focused on characterizing the column density of H$_2$ from observed distributions of rovibrational states derived from their respective absorption features embedded within the stellar Ly$\alpha$ wings of PPD hosts. We discovered that we systematically find higher excitation levels with larger column densities than expected from thermally-distributed, warm populations of H$_2$. 
The observed H$_2$ distributions of rovibrational states tells us that some mechanism or mechanisms in and/or around the circumstellar environment is/are affecting the equilibrium state of warm molecules in these sightlines. We aim to characterize the general behavior of thermal and non-thermal H$_2$ populations and column densities in PPD environments by comparing these quantities to stellar and circumstellar observables.

\begin{figure*}[htp]
\centering
	\subfloat{\includegraphics[angle=90,width=0.475\textwidth]{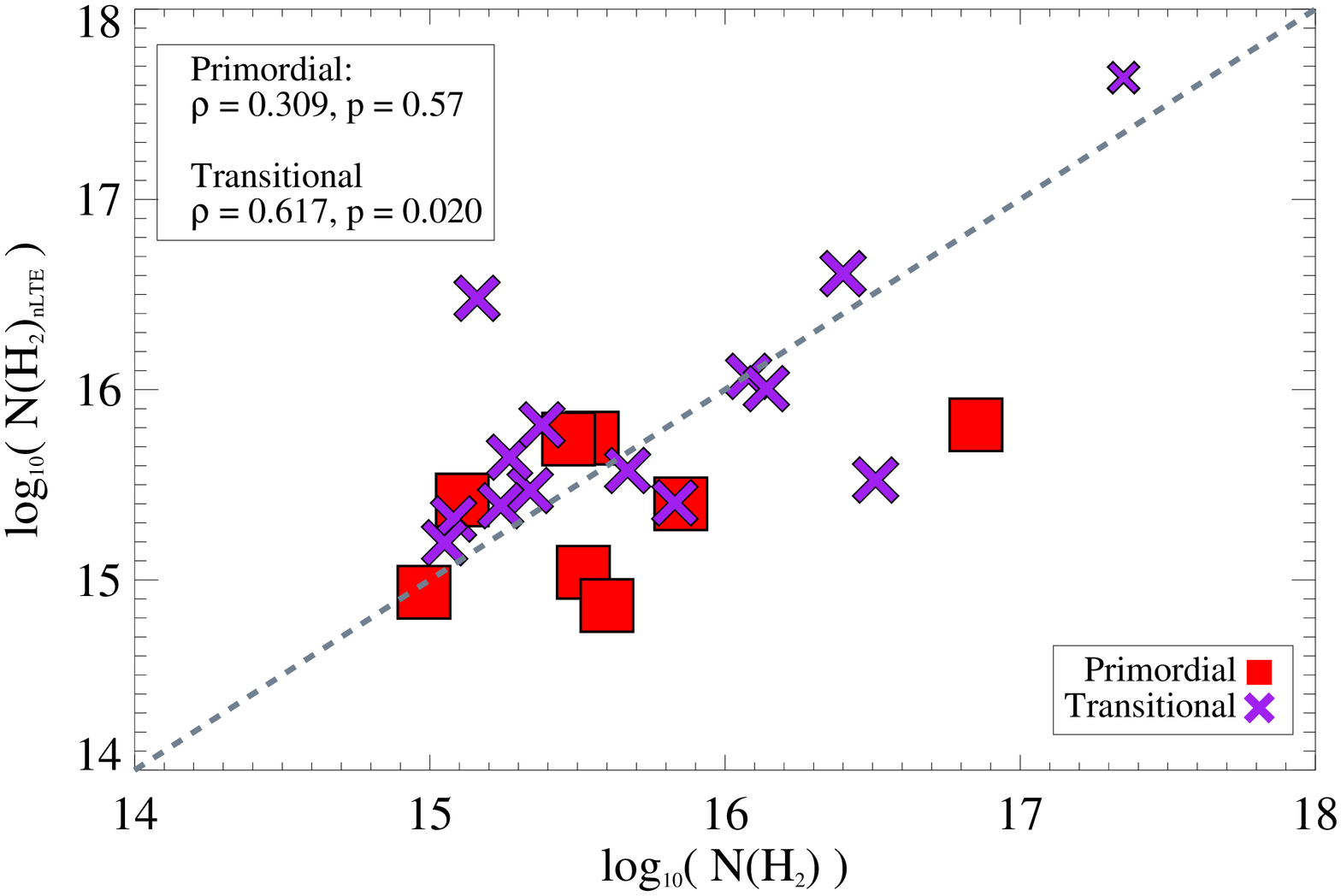}}
    \subfloat{\includegraphics[angle=90,width=0.475\textwidth]{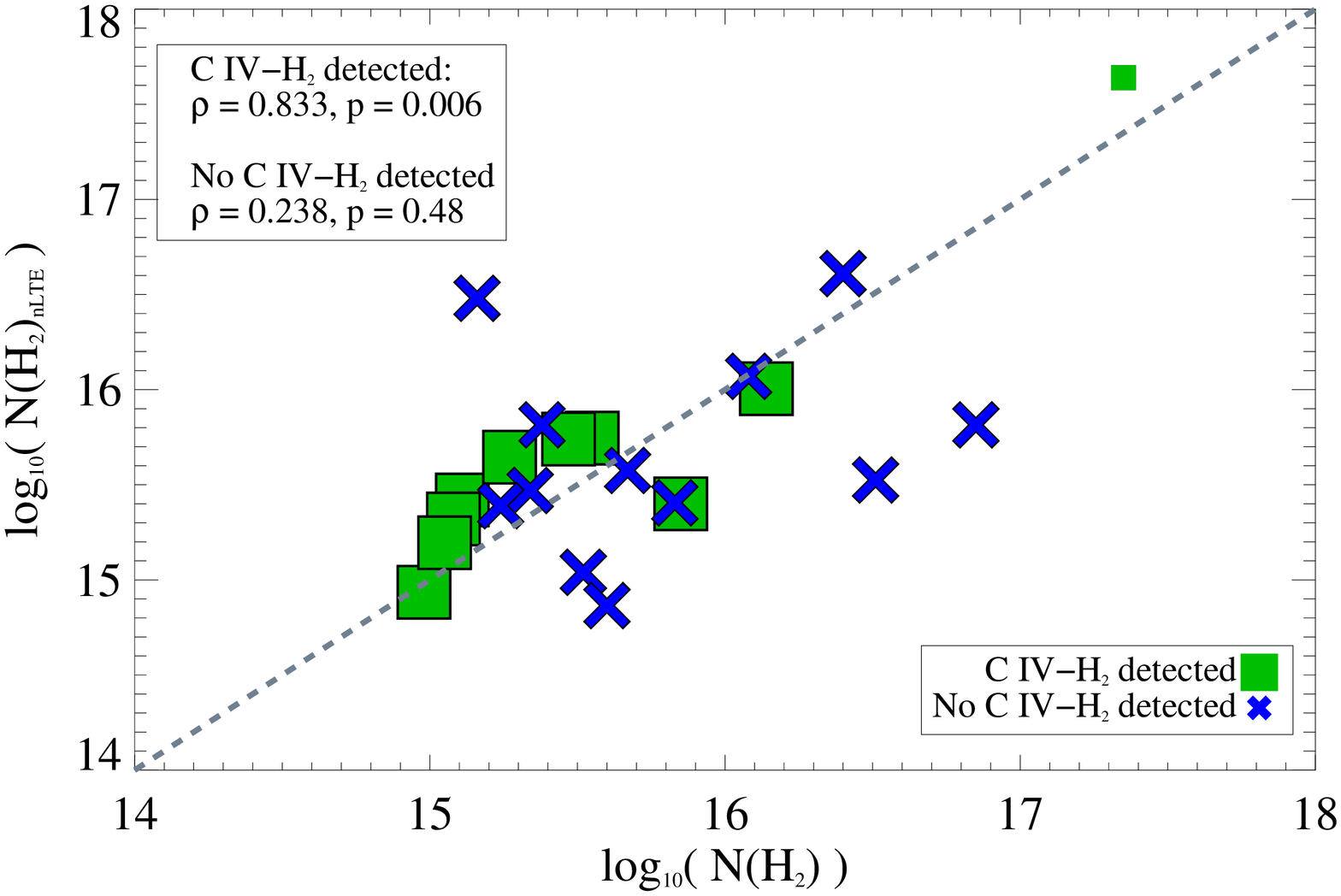}}
	\caption{We compare model-derived N(H$_2$) to N(H$_2$)$_{\textnormal{nLTE}}$ and separate populations by disk evolutionary phase (\emph{left}) and whether there is evidence of \ion{C}{4}-pumped H$_2$ fluorescence in the FUV spectrum (\emph{right}). Transitional disk targets and targets with detected \ion{C}{4}-pumped H$_2$ fluorescence (AA Tau, BP Tau, CS Cha, DF Tau, DM Tau, LkCa 15, RECX 11, RECX 15, TW Hya, and V4046 Sgr) appear to have direct correlations with N(H$_2$) $\sim$ N(H$_2$)$_{\textnormal{nLTE}}$.
	\label{fig4}}
\end{figure*}

First, we look for correlations between the modeled distributions of warm, thermal H$_2$ (T(H$_2$) $>$ 2500 K) and the populations of non-thermal H$_2$ states for the sampled PPD sightlines.
Figure~\ref{fig4} compares thermal, model-derived N(H$_2$) to the sum of the residuals in highly-energetic H$_2$ states, N(H$_2$)$_{\textnormal{nLTE}}$. Before noting the distributions of total column densities by categorization, we see that the general trend between the distributions of N(H$_2$) and N(H$_2$)$_{\textnormal{nLTE}}$ appear roughly related, with a Spearman rank coefficient which agrees with this assessment ($\rho$ = +0.54), but a PTE that suggests there is no strong indication of a trend between the two variables (p = 1.17$\times$10$^{-1}$). However, when we categorize targets by their disk evolution and whether \ion{C}{4}-pumped H$_2$ fluorescence is detected in their FUV spectra, we see much clearer trends that point to target distributions which have correlated N(H$_2$) and N(H$_2$)$_{\textnormal{nLTE}}$ populations. Transitional disks appear to predominantly straddle the N(H$_2$) = N(H$_2$)$_{\textnormal{nLTE}}$ equality line ($\rho$ = +0.62, p = 2.00$\times$10$^{-2}$), and targets which have detectable \ion{C}{4}-pumped H$_2$ fluorescence show the same behavior ($\rho$ = +0.83, p = 6.03$\times$10$^{-3}$). 
\deleted{The observed presence of \ion{C}{4}-pumped H$_2$ fluorescence is suggestive of H$_2$ populations that should not be populated if the H$_2$ are thermal and are therefore attributed to populations in existence because of non-thermal processes, such as H$_2$ formation \citep{Herczeg+02,Herczeg+06}.} 
Primordial disk targets appear to have more scattered distributions of N(H$_2$) and N(H$_2$)$_{\textnormal{nLTE}}$ ($\rho$ = +0.31, p = 5.69$\times$10$^{-1}$), as do targets with no detected \ion{C}{4}-pumped H$_2$ fluorescence ($\rho$ = +0.24, p = 4.82$\times$10$^{-1}$).

\subsection{H$_2$ Column Densities \& the Circumstellar Environment} \label{sec:nlte}

\deleted{We explore possible connections between circumstellar radiation (from the protostar, accretion shock, and disk molecular fluorescence) and derived N(H$_2$) and N(H$_2$)$_{\textnormal{nLTE}}$ from our thermal H$_2$ models. The physical evolution of PPDs is thought to be primarily driven by internal irradiation from the host protostar and planet formation \citep{Takahashi+99,Nomura+07,Dodson-Robinson+Salyk+11,Zhu+11,Owen+16}. \citet{Nomura+Millar+05} and \citet{Nomura+07} examined in great detail the expected effects of stellar UV and X-ray irradiation on the state of the molecular disk and discovered that excess UV/X-ray emission pumps H$_2$ to highly energetic, non-thermal ground levels. Consequently, we observe similar behavior in each empirical PPD H$_2$ rotation diagram developed through this work (e.g., Figure~\ref{fig3aa}). }
\deleted{We compare N(H$_2$) and N(H$_2$)$_{\textnormal{nLTE}}$ to observables that may be linked to excitation processes favoring higher-energy rovibrational H$_2$ levels in the PPD environments, including X-ray, FUV, Ly$\alpha$, \ion{C}{4}, H$_2$ fluorescence, and H$_2$ dissociation ``bump'' luminosities (L$_X$, L$_{FUV}$, L$_{Ly\alpha}$, L$_{C IV}$, L$_{H_2}$, and L$_{Bump}$); total flux from $\lambda \lambda$ 912 $-$ 1150 {\AA} (F$_{1110 \textnormal{\AA}}$); and the column density of highly energetic H$_2$ pumped by stellar \ion{C}{4} (N(\ion{C}{4}-H$_2$); see Section~\ref{sec:civh2}).}

\begin{figure*}[htp]
\centering
	\subfloat{\includegraphics[angle=90,width=0.475\textwidth]{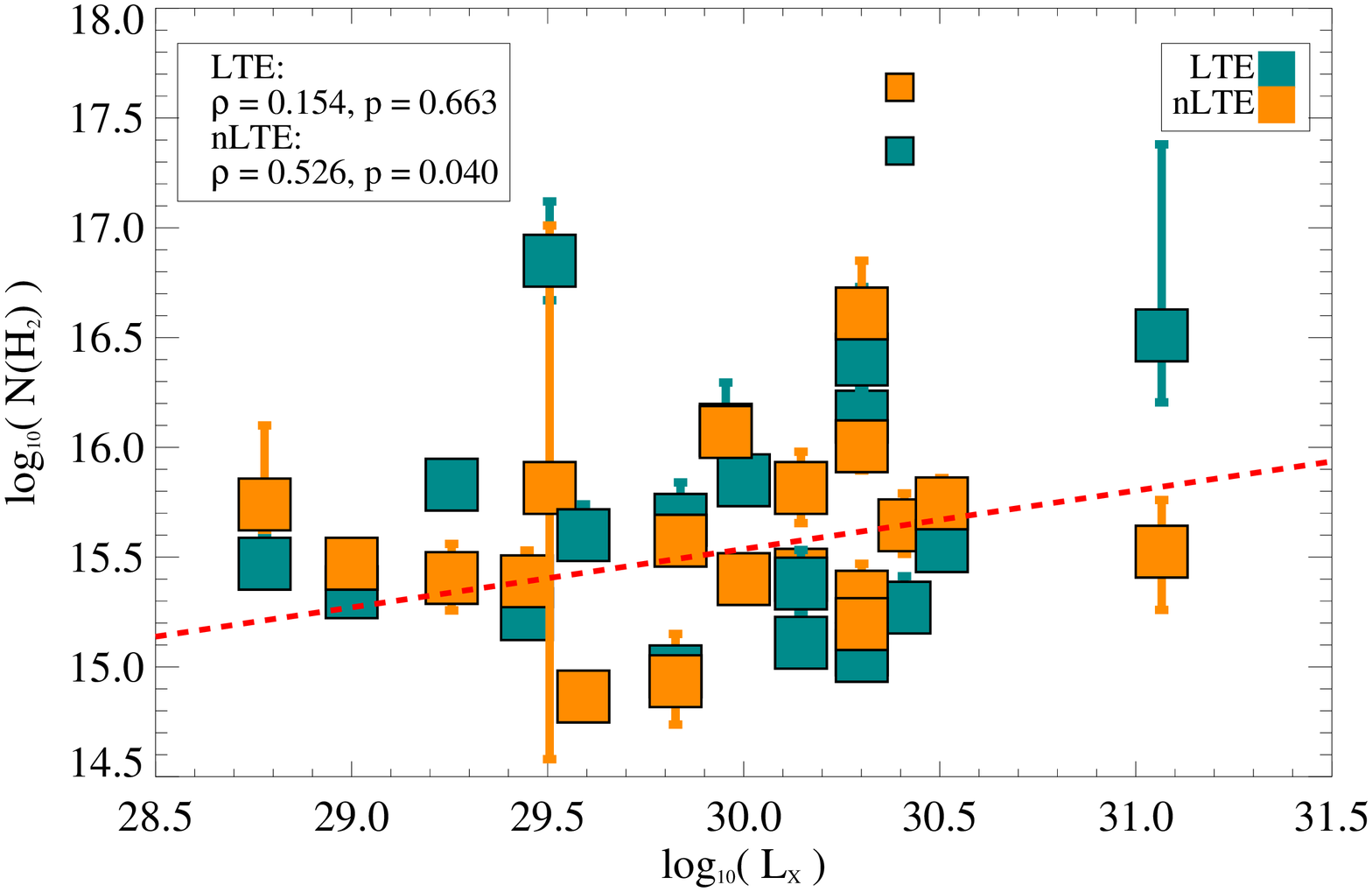}}
	\hfill
    \subfloat{\includegraphics[angle=90,width=0.47\textwidth]{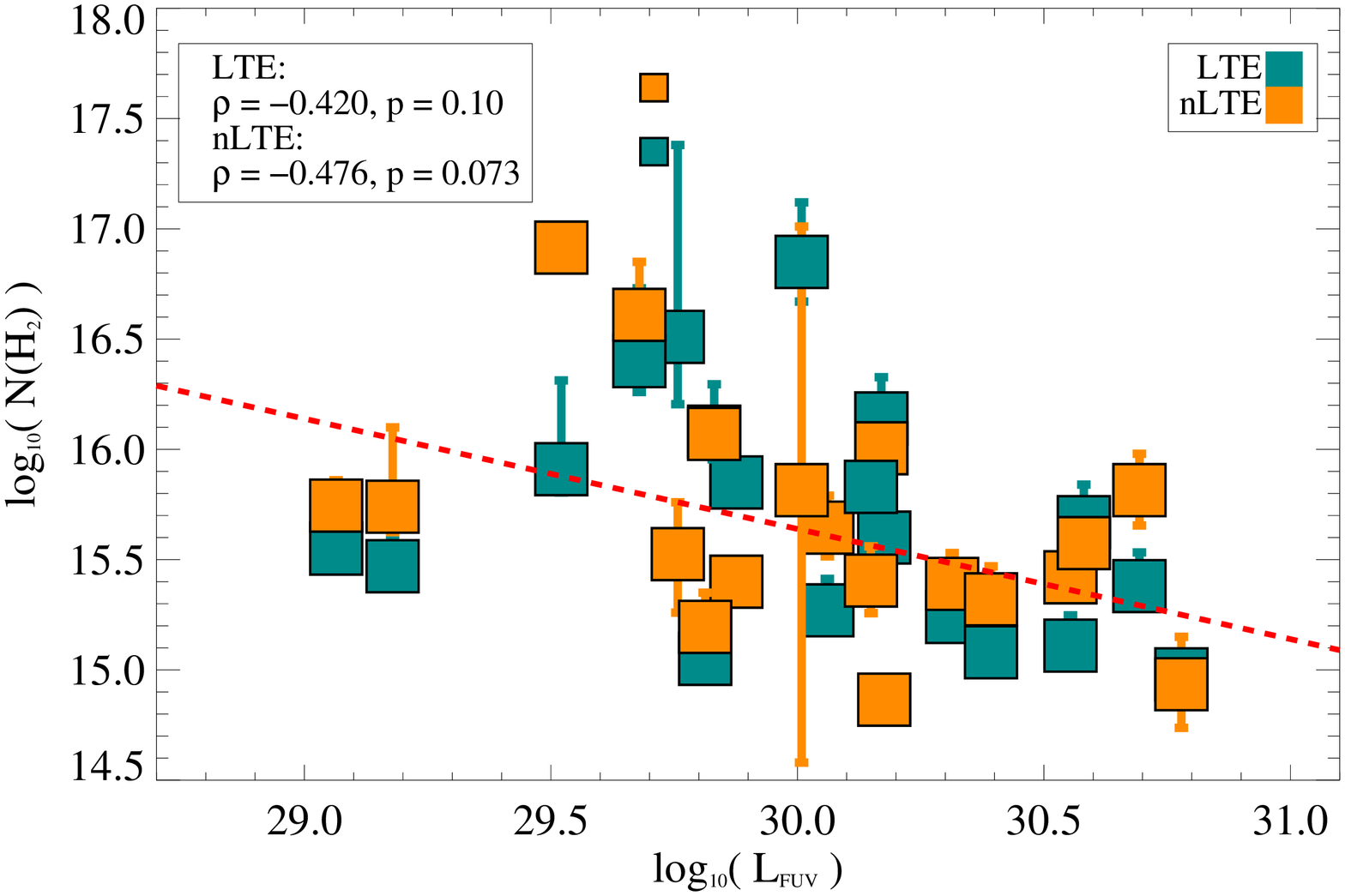}} \\
		\subfloat{\includegraphics[angle=90,width=0.475\textwidth]{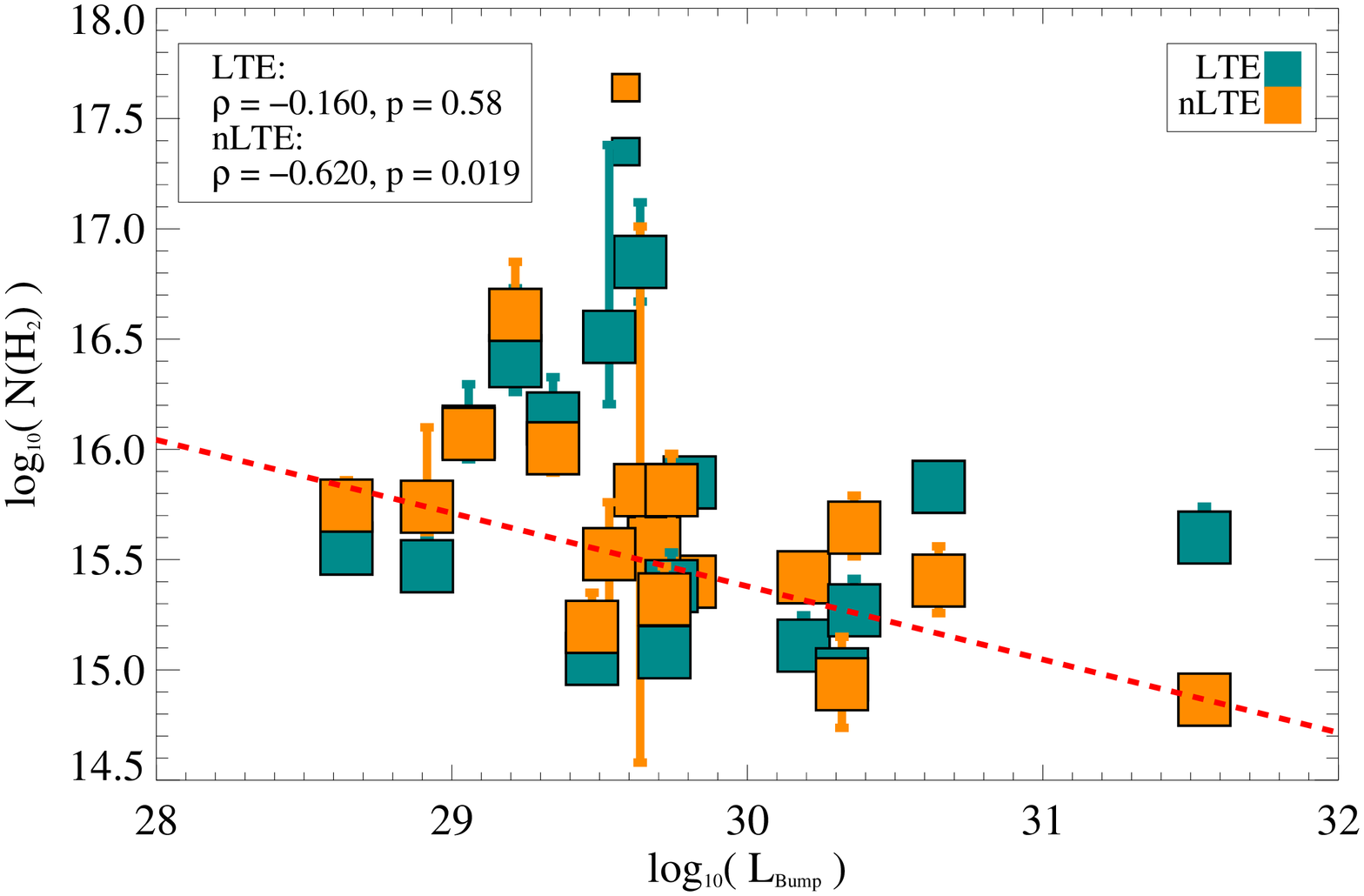}}
	\hfill
    \subfloat{\includegraphics[angle=90,width=0.475\textwidth]{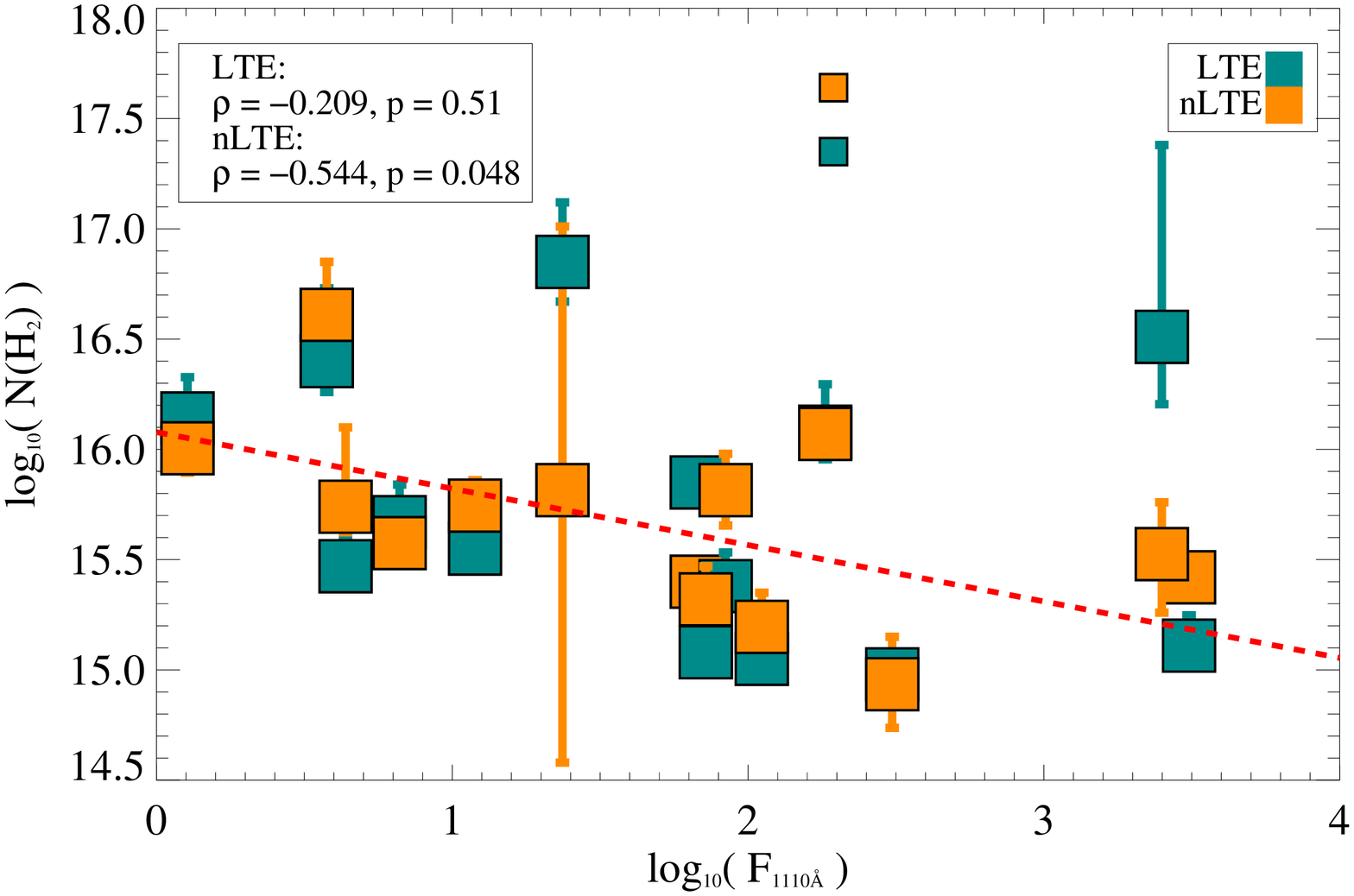}}
	\caption{We compare the total column densities of thermal and non-thermal H$_2$ to the total X-ray luminosity (\emph{top left}), the total FUV continuum luminosity (\emph{top right}), the total H$_2$ dissociation ``bump'' luminosity around $\lambda$ 1600 {\AA} (\emph{bottom left}), and the integrated flux from $\lambda \lambda$ 912 $-$ 1150 {\AA} \added{estimated at 1 AU from each protostar} (\emph{bottom right}). N(H$_2$) shows no significant correlations with any high-energy radiation observables, while N(H$_2$)$_{\textnormal{nLTE}}$ shows confident trends with L$_X$, L$_{Bump}$, and F$_{1110 \textnormal{\AA}}$. Both total column densities show a very loose trend with L$_{FUV}$. \added{LkCa 15 is included in each plot as smaller square symbols.} Outside of log space, the column density variables have units of cm$^{-2}$, the luminosity variables have units of erg s$^{-1}$, and the flux variables have units of erg cm$^{-2}$ s$^{-1}$.
	\label{fig8}}
\end{figure*}

We first consider the role of excess FUV and X-ray emission on the modeled thermal and non-thermal total column densities of H$_2$, to explore if the distributions of observed H$_2$ levels match the behaviors observed in \citet{Nomura+Millar+05} and \citet{Nomura+07}. We split the various excess emission into the following categories: the total X-ray luminosity (L$_X$; \citealt{France+17} and references therein), the total FUV continuum luminosity (L$_{FUV}$: $\lambda \lambda$ 1490 $-$ 1690 {\AA}, excluding any discrete or extended emission features; \citealt{France+14}), the total H$_2$ dissociation continuum around $\lambda$ 1600 {\AA} (L$_{Bump}$; \citealt{France+17}), and the total observed flux, corrected for ISM reddening, of FUV continuum+discrete emission features from $\lambda \lambda$ 912 $-$ 1150 {\AA} (F$_{1110 \textnormal{\AA}}$; \citealt{France+14}). Figure~\ref{fig8} shows the comparison of N(H$_2$) and N(H$_2$)$_{\textnormal{nLTE}}$ to these circumstellar observables. We note a correlation between L$_X$ and N(H$_2$)$_{\textnormal{nLTE}}$ ($\rho$ = +0.53, p = 4.00$\times$10$^{-2}$), but no correlation between N(H$_2$) and L$_X$ ($\rho$ = +0.15, p = 6.62$\times$10$^{-1}$). 
We observe an anti-correlation between N(H$_2$)$_{\textnormal{nLTE}}$ and L$_{Bump}$ ($\rho$ = -0.62, p = 1.90$\times$10$^{-2}$) and no strong trend between N(H$_2$) and L$_{Bump}$ ($\rho$ = -0.16, p = 5.83$\times$10$^{-1}$). We again find an anti-correlation between N(H$_2$)$_{\textnormal{nLTE}}$ and F$_{1110 \textnormal{\AA}}$ ($\rho$ = -0.54, p = 4.81$\times$10$^{-2}$), yet no indication of a trend between N(H$_2$) and F$_{1110 \textnormal{\AA}}$ ($\rho$ = -0.21, p = 5.14$\times$10$^{-1}$). Finally, both N(H$_2$) and N(H$_2$)$_{\textnormal{nLTE}}$ show suggestive anti-correlations with L$_{FUV}$, but they are not statistically significant (N(H$_2$): $\rho$ = -0.42, p = 1.02$\times$10$^{-1}$; N(H$_2$)$_{\textnormal{nLTE}}$: $\rho$ = -0.48, p = 7.30$\times$10$^{-2}$).

\begin{figure*}[htp]
\centering
	\subfloat{\includegraphics[angle=90,width=0.465\textwidth]{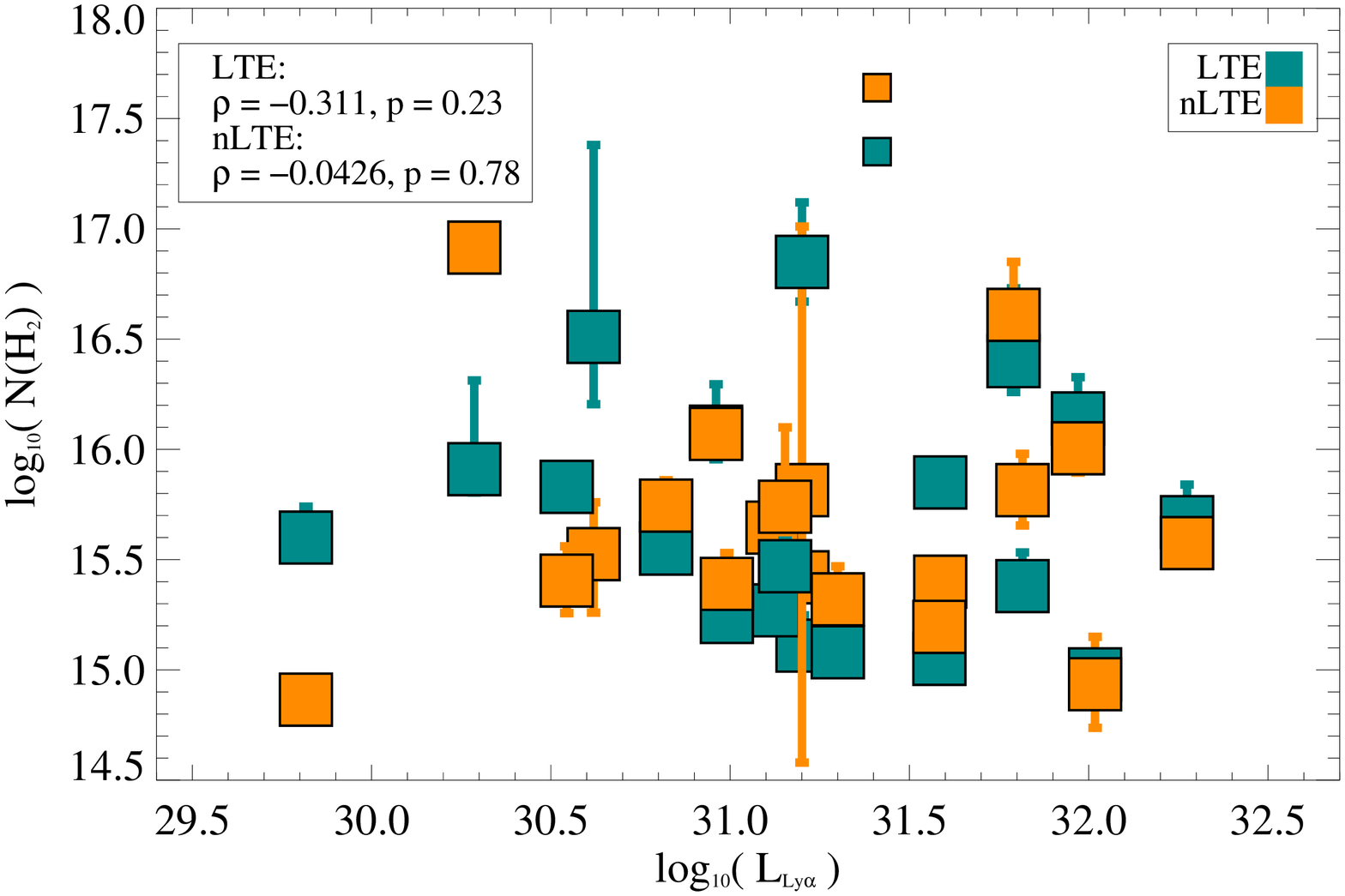}}
	\hfill
    \subfloat{\includegraphics[angle=90,width=0.465\textwidth]{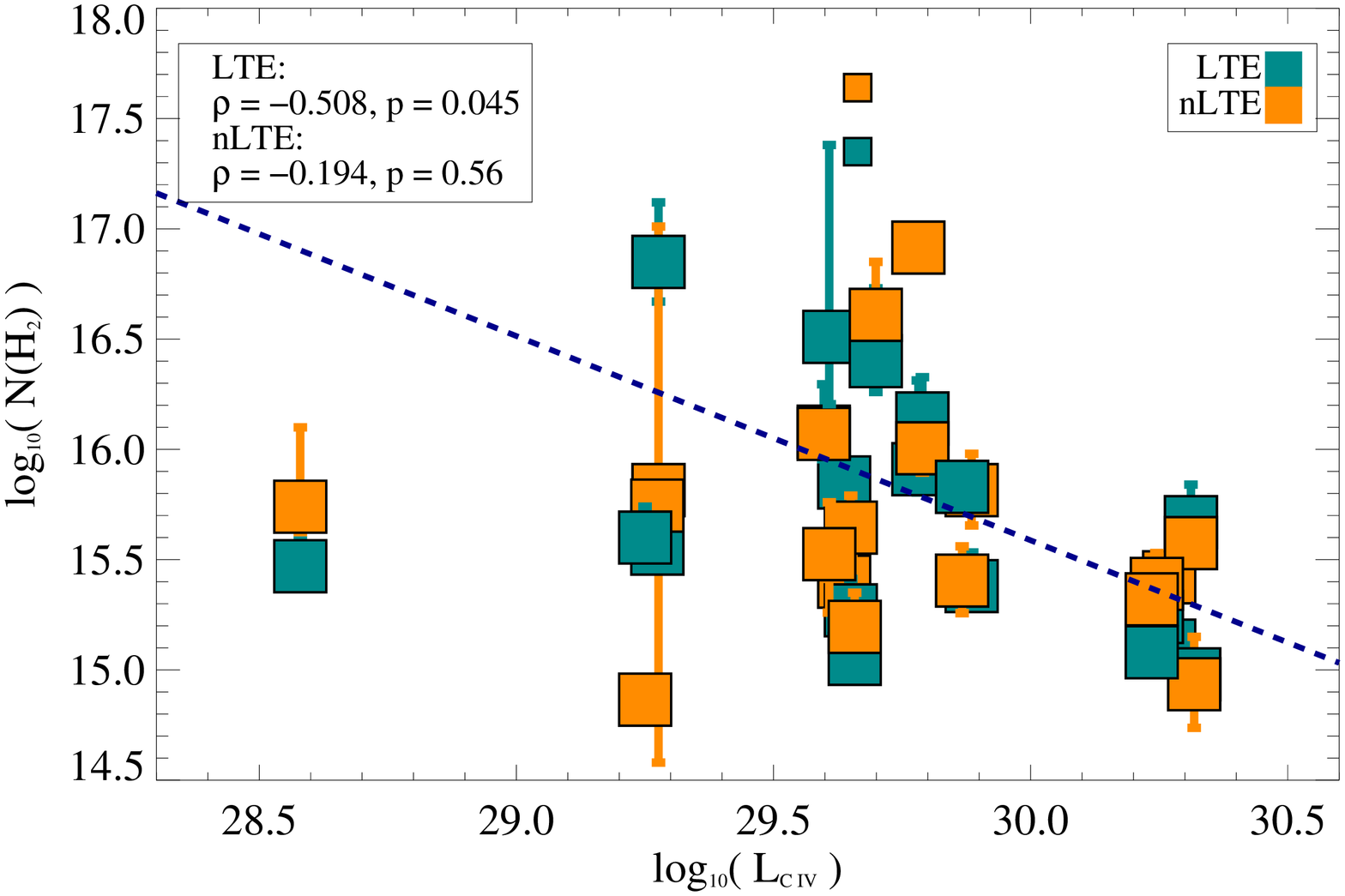}} \\
		\subfloat{\includegraphics[angle=90,width=0.465\textwidth]{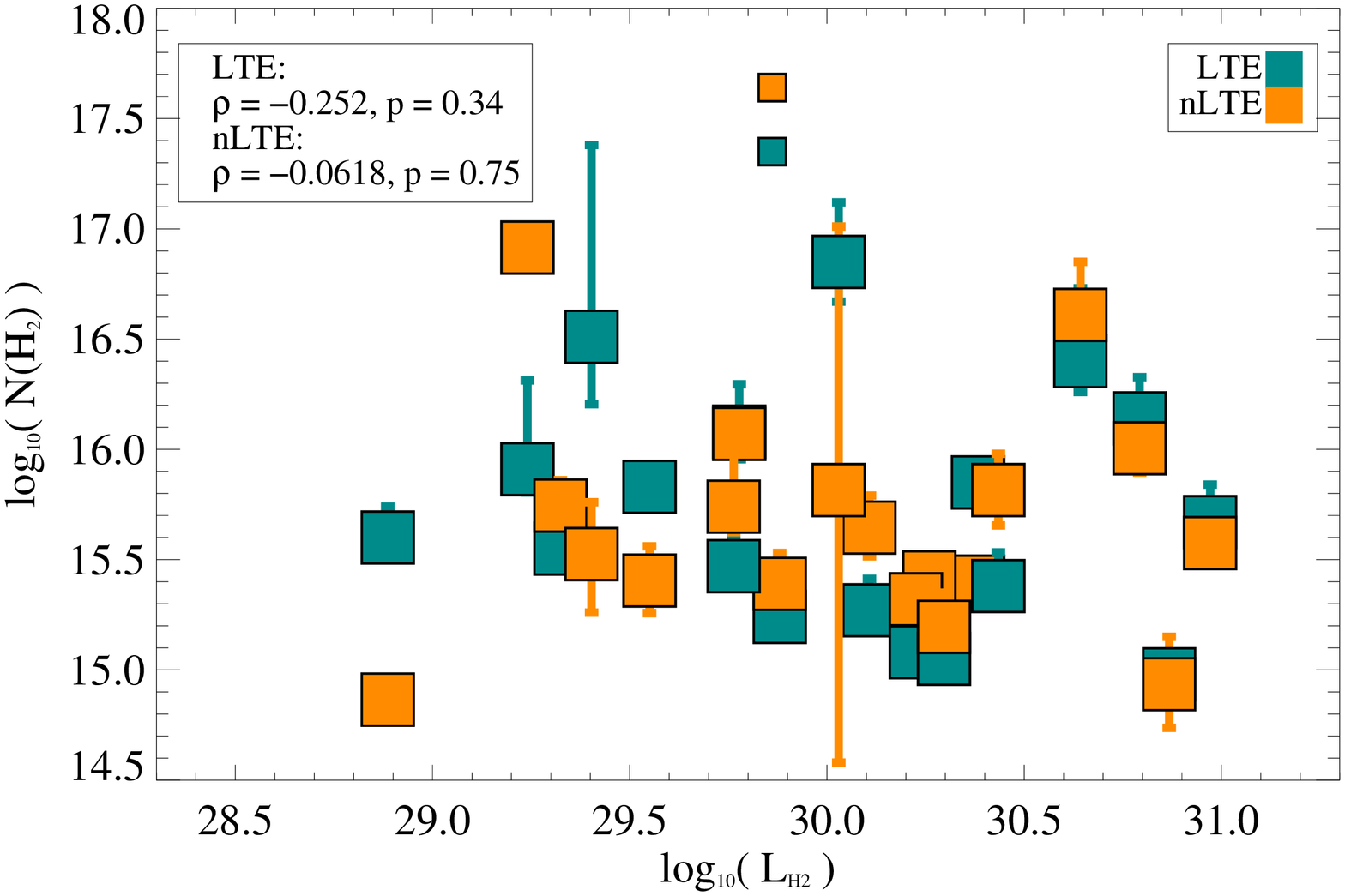}}
	\hfill
    \subfloat{\includegraphics[angle=90,width=0.48\textwidth]{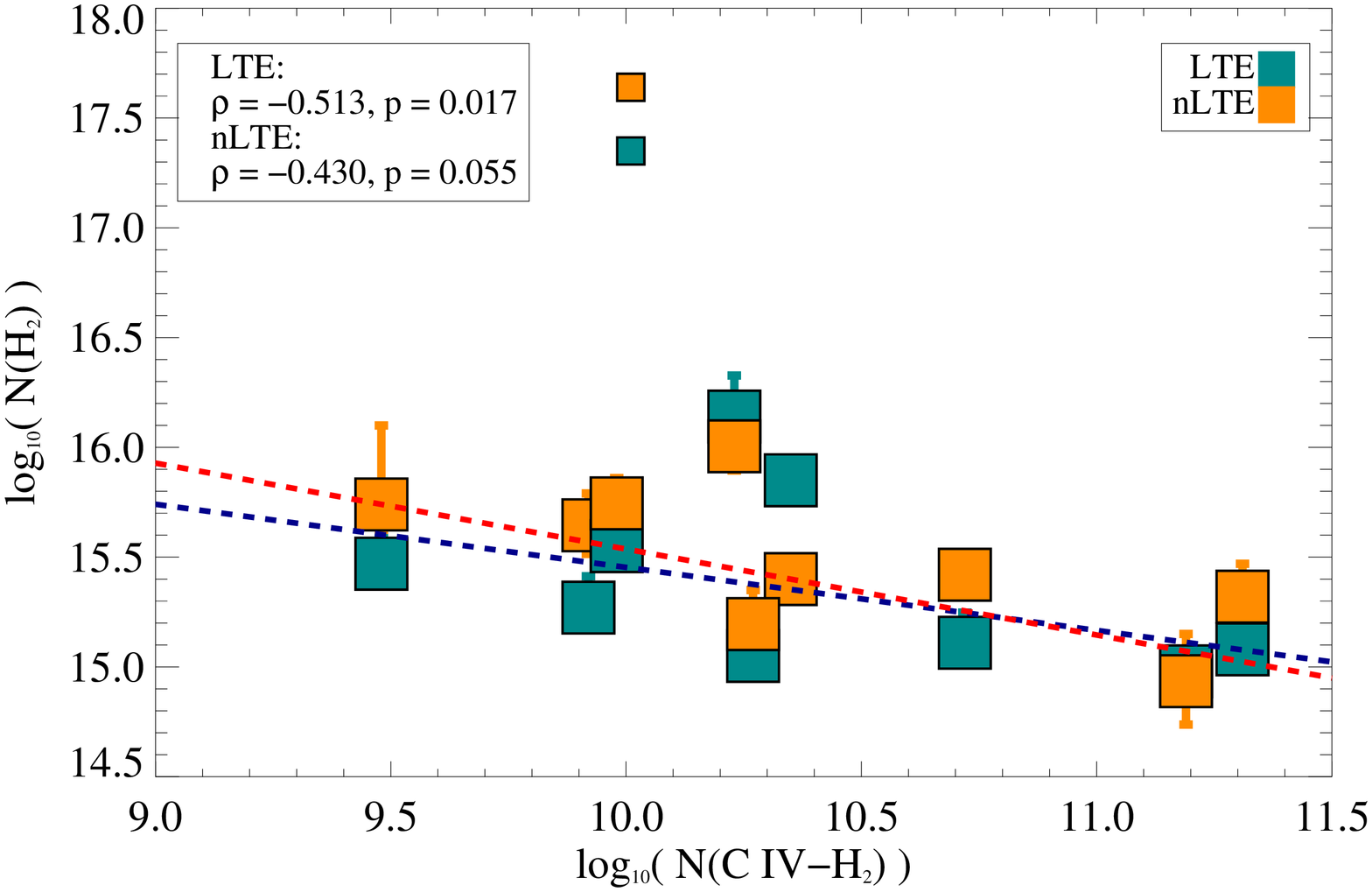}}
	\caption{We compare the total column densities of thermal and non-thermal H$_2$ to the total Ly$\alpha$ luminosity (\emph{top left}), the total \ion{C}{4} luminosity (\emph{top right}), the total H$_2$ fluorescence luminosity (\emph{bottom left}), and the total column density of H$_2$ found in H$_2$[5,18] (\emph{bottom right}). N(H$_2$) shows confident trends with L$_{C IV}$ and N(\ion{C}{4}-H$_2$), while N(H$_2$)$_{\textnormal{nLTE}}$ only displays a loose trend with N(\ion{C}{4}-H$_2$). We find no correlations between the modeled H$_2$ column densities and L$_{Ly\alpha}$ and L$_{H_2}$. \added{LkCa 15 is included in each plot as smaller square symbols.} Outside of log space, the column density variables have units of cm$^{-2}$ and the luminosity variables have units of erg s$^{-1}$.
	\label{fig8b}}
\end{figure*}

Next, we look at how discrete emission line features (from the protostar and accretion shock regions) and disk fluorescence processes may play a role on the total column densities of H$_2$ in PPD sightlines. 
We split the circumstellar parameters into the following categories: the total luminosity from stellar+shock-generated Ly$\alpha$ emission (L$_{Ly \alpha}$; ; \citealt{France+14}), the total luminosity from stellar+shock-generated \ion{C}{4} emission (L$_{C IV}$; \citealt{France+14}), the total H$_2$ fluorescence luminosity from Ly$\alpha$-pumped H$_2$ predominantly produced in the disk atmosphere (L$_{H_2}$; \citealt{France+14}), and the estimated total column density of H$_2$[5,18], derived from the statistically-determined \ion{C}{4}-pumped fluorescence features (N(\ion{C}{4}-H$_2$), derived in Section~\ref{sec:civh2}).
Figure~\ref{fig8b} shows the comparison of N(H$_2$) and N(H$_2$)$_{\textnormal{nLTE}}$ to these circumstellar variables. We find no trends between the modeled column densities of H$_2$ and L$_{Ly \alpha}$ (N(H$_2$): $\rho$ = -0.31, p = 2.34$\times$10$^{-1}$; N(H$_2$)$_{\textnormal{nLTE}}$: $\rho$ = -0.04, p = 7.86$\times$10$^{-1}$), as well as L$_{H_2}$ (N(H$_2$): $\rho$ = -0.25, p = 3.45$\times$10$^{-1}$; N(H$_2$)$_{\textnormal{nLTE}}$: $\rho$ = -0.06, p = 7.54$\times$10$^{-1}$). We do see a suggestive anti-correlation between L$_{C IV}$ and N(H$_2$) ($\rho$ = -0.51, p = 4.52$\times$10$^{-2}$), but no trend between N(H$_2$)$_{\textnormal{nLTE}}$ and L$_{C IV}$ ($\rho$ = -0.19, p = 5.62$\times$10$^{-1}$). 
Finally, we find anti-correlated behavior between both N(H$_2$) and N(H$_2$)$_{\textnormal{nLTE}}$ with N(\ion{C}{4}-H$_2$) (N(H$_2$): $\rho$ = -0.51, p = 1.71$\times$10$^{-2}$; N(H$_2$)$_{\textnormal{nLTE}}$: $\rho$ = -0.43, p = 5.50$\times$10$^{-2}$).

\subsection{The Behavior of Hot H$_2$}

We find that the \deleted{non-thermal }column densities of H$_2$ are correlated to many non-thermal diagnostics of the circumstellar environment, such as internal radiation and H$_2$ dissociation tracers. 
\explain{Text regarding interpretations of correlations have been omitted. These explanations do not add to the final interpretation of the H$_2$ populations.}
The\added{refore, the} observed distribution of H$_2$ absorption populations appear to be located somewhere in the disk environment where 1) \deleted{the H$_2$ can interact with an abundance of non-thermal electrons, and }1) the H$_2$ have access to protostellar radiation with $\lambda$ $<$ 1110 {\AA}, and 2) the H$_2$ populations are optically-thin to Ly$\alpha$ radiation. 
Piecing \deleted{all of }our results together, we suspect that the observed H$_2$ populations against the protostellar Ly$\alpha$ wing provide are not associated with the H$_2$ that fluoresces in the disk and may, instead, arise from a hot, \replaced{nebulous origin}{tenuous disk halo}. 
\citet{Adamkovics+16} explore the effects of FUV, X-ray, and Ly$\alpha$ radiation on stratified layers of molecular PPD atmospheres. In the presence of all three, FUV continuum and X-ray radiation create a hot, atomic layer along the uppermost disk surface, and Ly$\alpha$ radiation penetrates deeper into the disk via \ion{H}{1} scattering. The penetration of Ly$\alpha$ into the molecular disk is found to photodissociate trace molecules like H$_2$O and OH, which, along with H$_2$ formation on dust grains, heat this region of the disk and create a warm molecular layer (T$_{gas}$ $>$ 1500 K). This warm layer is found to have an appreciable column of warm H$_2$ (N(H$_2$) $>$ 10$^{19}$ cm$^{-2}$) in the appropriate temperature regime to reproduce fluorescent emission signatures in PPDs, though \deleted{\citet{Adamkovics+16} acknowledges that }the distribution of H$_2$ rovibrational levels is not computed \replaced{with their}{in the \citet{Adamkovics+16}} models. 

\begin{figure*}[htp]
\centering
\includegraphics[angle=0,width=0.95\textwidth]{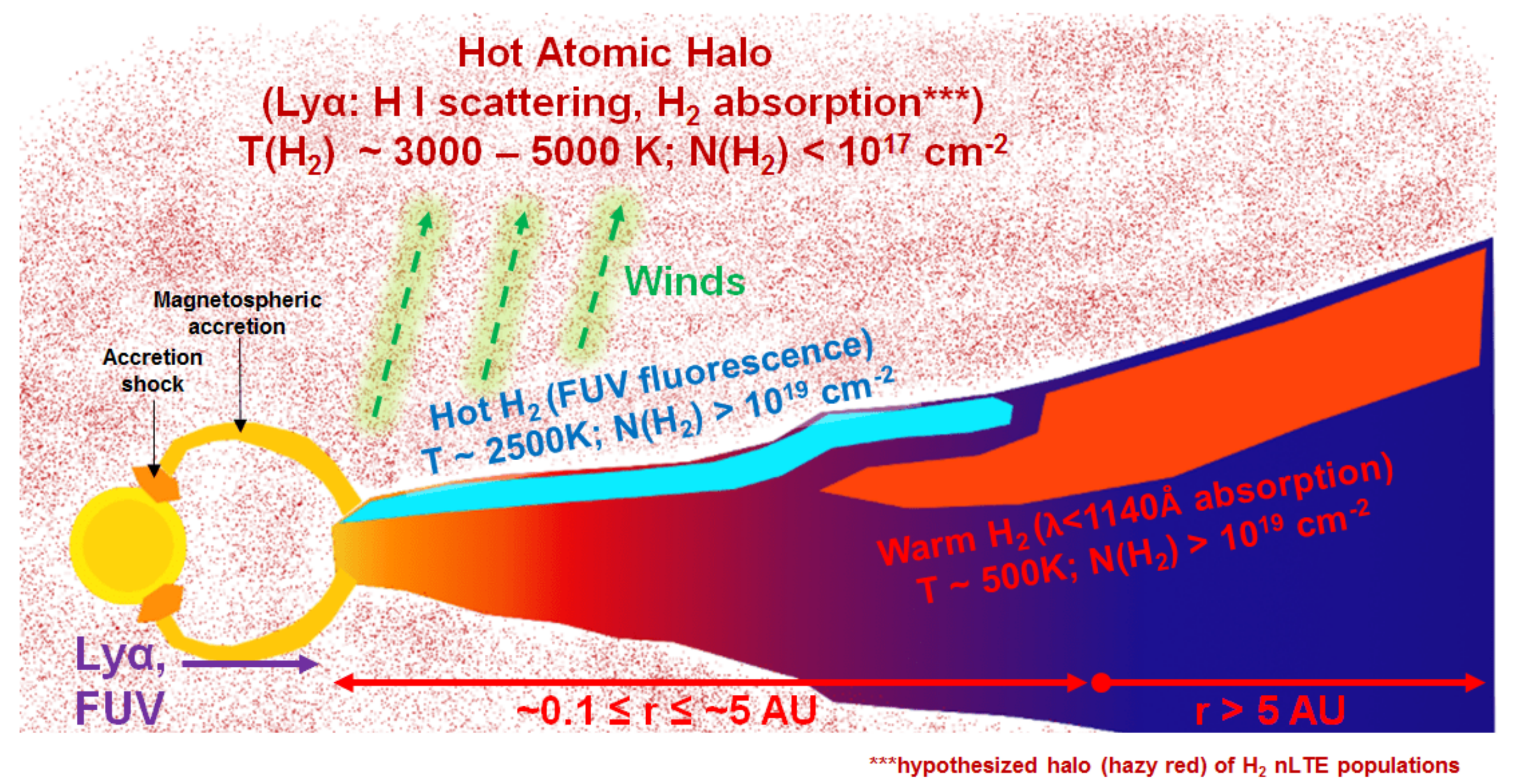}
	\caption{\added{A schematic showing the inner region of the warm PPD and environment. Regions where H$_2$ is observed as various features are labeled: hot H$_2$ fluorescence (emission - light blue) appear to come from a layer in the inner disk atmosphere where FUV/X-ray/Ly$\alpha$ radiation heat the gas \citep{Adamkovics+16}, whereas the H$_2$ populations observed by \citet{France+14b} in absorption (orange-red) are cooler, indicating a warm layer of molecules probed either further out in the disk than the H$_2$ fluorescent populations, closer to the disk midplane, or a combination of both. The hot H$_2$ populations probed against the protostellar Ly$\alpha$ wings, however, do not appear related to either of the other observed H$_2$ populations. Instead, we speculate that these hot H$_2$ populations are probed within a ``halo'' of hot atomic gas surrounding the protostar+disk (hazy dark red). Since it is difficult to pinpoint a specific region where this gas resides, we include all regions where these H$_2$ populations may reside, including close to the protostar and/or magnetosphereic accretion front from the disk to the star, to regions surrounding the disk, such as hot gas near the disk atmosphere or gas flowing away from PPD as thermal/photoevaporative winds. }
	\label{fig_diskhalo} }
\end{figure*}

The \citet{Adamkovics+16} study produces a hot (T $\sim$ 5000 K) atomic layer in the uppermost disk atmosphere, which is similar in nature to a photodissociation region (PDR; \citealt{Hollenbach+99} and references therein). This hot atomic layer is produced in all of their model parameter space, \replaced{inconsequential}{independent} of stellar Ly$\alpha$ luminosity or dust grain distributions. This layer of hot atomic gas contains a minute abundance of H$_2$ ($x$(H$_2$) $\lesssim$ 10$^{-5}$) with total column densities of hot H$_2$ similar to those found in this study (N(H$_2$)$_{hotlayer}$ $\sim$ 10$^{15}$ cm$^{-2}$; $\langle$N(H$_2$)$_{H_{2}abs} \rangle$ $\sim$ 10$^{15.5}$ cm$^{-2}$). This hot atomic layer is modeled above the warm molecular layer (where H$_2$ fluorescence may arise) and extends substantially further away from the disk midplane \citep{Adamkovics+16}. What their study finds is that Ly$\alpha$ radiation is key to producing the warm molecular regions that may be associated with warm H$_2$ and CO populations, but the hot, atomic layer is driven by the FUV continuum and X-ray luminosities, which cannot penetrate into the cooler disk like Ly$\alpha$ can.

Connecting the findings from this work and the \citet{Adamkovics+16} models, we suggest that the observed H$_2$ absorption populations, probed in the wings of protostellar Ly$\alpha$ profiles, reside in this tenuous, hot atomic region of the circumstellar environment. 
We argue that the behavior of the Ly$\alpha$ transition, being by nature a powerful resonance line, 
allows Ly$\alpha$ radiation to scatter through both the PPD and the surrounding PDR-like environment. Rather than probing a discrete line source coming straight from the accretion shock near the protostellar surface, we probe Ly$\alpha$ that has scattered through the circumstellar environment by \ion{H}{1} atoms before reaching the observer.  
\replaced{The scattering of Ly$\alpha$ radiation by neutral hydrogen causes Doppler shifts away from the rest wavelength of Ly$\alpha$, which is observed as a broadening of the emission line profile to several hundred km s$^{-1}$ before leaving the hot atomic environment around the PPD.}{The scattering of Ly$\alpha$ radiation by \ion{H}{1}, which occurs when a Ly$\alpha$ photon is absorbed and emitted many times into many different directions and results in changes in the frequency of Ly$\alpha$ away from rest wavelength, causes significant broadening of the Ly$\alpha$ profile on order of several hundred km s$^{-1}$ before escaping the star (see \citealt{McJunkin+14} for a more complete overview of this process).} It appears that the H$_2$ probed in absorption against these observed Ly$\alpha$ wings may be tied to this optically-thin, hot halo surrounding the PPD, where optically-thin densities of H$_2$ absorb Ly$\alpha$ before it can exit the system. \added{Figure~\ref{fig_diskhalo} presents a cartoon disk showing the possible locations of H$_2$ fluorescence populations (blue), warm H$_2$ in the disk (red), and the hot halo of gas where H$_2$ is probed against the protostellar Ly$\alpha$ profile (diffuse red haze). For now, this haze is assumed to be anywhere surrounding the protostar and protoplanetary disk. Additional work is being conducted to constrain the spatial origins of these hot H$_2$ populations.}

\subsubsection{H$_2$ ``Multiple Pumping'' Versus Cooling}	\label{sec:calc}

The scattering of Ly$\alpha$ radiation through the hot atomic regions surrounding PPDs may help explain the non-thermal behavior of H$_2$ associated with these environments. The \deleted{odd }behavior of the \replaced{absorption}{observed} rovibrational levels may be the result of ``multiple pumping'' happening with the hot H$_2$, meaning that the excitation rate by UV photon absorption (in this case, specifically Ly$\alpha$ photons) is faster than the molecules can decay (cool) via rovibrational emission lines or collisions. 

We perform a back-of-the-envelope comparison of the H$_2$ rovibrational emission and total collision rates required to \replaced{counteract}{balance} H$_2$ photo-excitation (``Ly$\alpha$-pumping''), assuming the H$_2$ species are located in a hot atomic layer above the PPD. The hot atomic region is assumed to be a plane-parallel slab above the inner disk ($r$ $<$ 1 AU; \citealt{Adamkovics+16}) with a thickness $a$ $\sim$ 1 AU. We assume the average Ly$\alpha$ luminosity for a typical PPD system $\langle$L$_{Ly\alpha}\rangle$ $\sim$ 10$^{31}$ erg s$^{-1}$ \citep{Schindhelm+12b,France+14}, which translates into an average photon rate $\langle \Gamma_{Ly\alpha}\rangle$ = $\langle$L$_{Ly\alpha}\rangle$ / E$_{Ly\alpha}$ $\sim$ 10$^{42}$ photons s$^{-1}$ incident on the hot H$_2$. 
Since H$_2$ is expected to only be a trace species in this region ($x$(H$_2$) $\sim$ 10$^{-5}$; \citealt{Adamkovics+16}), we include a ``coverage factor'' for the total Ly$\alpha$ luminosity on the H$_2$ populations.
This leads to an estimation of the total photo-excitation rate of H$_2$ in the hot atomic layer, $\langle \Gamma_{Ly\alpha}\rangle$ $\sim$ $x$(H$_2$)$\times$10$^{42}$ photons s$^{-1}$ $\sim$ 10$^{37}$ photons s$^{-1}$. We calculate the average rate of incident Ly$\alpha$ photons on the H$_2$ populations in the PDR slab to be $\gamma_{Ly\alpha}$ $\sim$ $\langle \Gamma_{Ly\alpha}\rangle$ / ($\sigma$(H$_2$)$\times a^2$) $\approx$ 10$^{-3}$ photon s$^{-1}$, where $\sigma$(H$_2$) is the average Ly$\alpha$ line absorption cross-section of an individual molecules, given by 
\begin{equation}
	\sigma(H_2) = \frac{\sqrt{\pi}e^2}{m_e c b_{H_2}} \lambda_i f_i
\end{equation}
\citep{McCandliss+03,Cartwright+70}, where $\lambda_i$ is the absorption wavelength for a given transition in the Ly$\alpha$ profile (taken as 1215.67 {\AA} for this example), $f_i$ is the oscillator strength (the average assumed as $\approx$0.01), and $b_{H_2}$ is the b-value of the line, assumed to match our models ($b_{H_2}$ = 5 km s$^{-1}$), producing an average cross section for Ly$\alpha$ photon absorption $\sigma$(H$_2$) $\sim$ 10$^{-14}$ cm$^{2}$.

We do not include additional losses of Ly$\alpha$ flux due to absorption from other atomic species, as it is assumed that the dominant constituent of the disk PDR is neutral hydrogen at an average $T_{gas}$ $\sim$ 3500 - 5000 K, which will scatter Ly$\alpha$ around the region. 
We can quantify the ratio of the UV photo-excitation rate to the average transition probability for quadrupolar H$_2$ IR emission lines ($A_{quad}$ $\sim$ 10$^{-7}$ s$^{-1}$; \citealt{Wolniewicz+98}), $\gamma_{Ly\alpha}$ / $A_{quad}$ $\sim$ 10$^4$ photons, meaning that of order 10,000 Ly$\alpha$ photons are absorbed for every one quadrupolar photon emitted. Therefore, quadrupole emission is not an effective means of cooling the photo-excited H$_2$ populations in these regions.

Next, we explore what the expected collisional rate between H$_2$ and other particles in the hot atomic slab must be to balance with the UV photo-excitation rate. First, we set the total collisional rate of all particle interactions with H$_2$ in this region to match the photo-excitation rate of H$_2$ in the hot atomic region, such that $\Sigma \alpha_{H_2,i}$ = $\gamma_{Ly\alpha}$ $\sim$ 10$^{-3}$ collisions s$^{-1}$. 
Given $\langle$N(H$_2$)$\rangle$ from our empirical models, we estimate the total number density of H$_2$ in the hot atomic layer to be $n$(H$_2$) $\sim$ 10$^3$ cm$^{-3}$. Finally, we estimate the total collisional rate with H$_2$ needed to match the photo-excitation rate of H$_2$ via Ly$\alpha$-pumping, $\Sigma C_{H_2,i}$ $\sim$ $\Sigma \alpha_{H_2,i}$ / $n$(H$_2$) $\sim$ 10$^{-6}$ cm$^3$ s$^{-1}$. 

This result suggests that, at T$_{gas}$ $\approx$ 3500 - 5000 K, interactions between H$_2$ and dominant particles in the hot atomic environment, like \ion{H}{1}, protons (p$^{+}$), and electrons (e$^{-}$), are expected to occur at a total rate of $\sim$10$^{-6}$ cm$^3$ s$^{-1}$. \citet{Mandy+93} and \citet{Roberge+82} find collisional rates between H$_2$ + \ion{H}{1} to be of order $C_{H_2,HI}$ $\sim$ 10$^{-10}$ cm$^3$ s$^{-1}$ for gas with T$_{gas}$ $\approx$ 2000 - 4500 K (which is similar to interactions between H$_2$ + p$^{+}$; \citealt{Black+77,Smith+82}). The rate of collisions between H$_2$ + e$^{-}$, for gas with T$_{gas}$ $\sim$ 3500 K, is found to be $C_{H_2,e^-}$ $\sim$ 10$^{-11}$ cm$^3$ s${-1}$ \citep{Prasad+Huntress+80a}. Additionally, interactions between H$_2$ + H$_2$ are expected to occur much less frequently, with $C_{H_2,H_2}$ $\sim$ 10$^{-16}$ cm$^3$ s${-1}$ for $T_{gas}$ $\sim$ 3500 K \citep{Mandy+16}. 

We find that the integrated collision rate of H$_2$ in these environments, derived from literature values, is $\sim$ 4 dex lower than the photo-excitation rate of H$_2$ by Ly$\alpha$ radiation alone. When we quantify the ratio of the UV photo-excitation rate to the total collisional rate of particles with H$_2$ in this exercise (optimistically assuming $\Sigma C_{H_2,i}$ $\sim$ 10$^{-9}$ cm$^3$ s$^{-1}$), $\gamma_{Ly\alpha}$ / ( $\Sigma C_{H_2,i} \times n$(H$_2$) ) $\sim$ 10$^3$ photons, or that $\sim$1,000 Ly$\alpha$ photons are absorbed for every one de-excitation collision of H$_2$. 
 
\replaced{It appears viable that ``multiple pumping''}{We conclude that is it therefore plausible that ''Ly$\alpha$ multiple pumping''} may play a key role in re-distributing H$_2$ rovibrational states in this hot gas region of the circumstellar environment before collisions or rovibrational emission can cool the molecules. Indeed, our simple calculation compliments observed behaviors of H$_2$ rovibration levels in ISM PDR environments (e.g., \citealt{Draine+96,Hollenbach+99}, and references therein). The critical density of most H$_2$ rovibration levels, or the ratio of the radiative lifetime of a given state ($A_{ul}$, in s$^{-1}$) to the collision rate for de-excitation out of the same state ($C_{H_2,i}$, in cm$^3$ s$^{-1}$), is typically of order 10$^4$ cm$^{-3}$ for $T_{gas}$ $>$ 2000 K \citep{Mandy+93}. In our estimation, the density of H$_2$ is near this critical density, but is still under it, allowing ``multiple pumping'' to \deleted{re}populate H$_2$ states by UV pumping before collisions de-excite the level populations \citep{Draine+96,Hollenbach+99}.

\subsubsection{A Simple Model of Ly$\alpha$-pumped H$_2$} \label{sec:mod3}

What, then, is the expected distribution of H$_2$ rovibration levels if Ly$\alpha$-pumping plays a significant role in regulating the ground states of the molecules? We create a simple model of H$_2$ photo-excitation, in the absence of cooling routes (i.e., rovibrational emission and collisional de-excitation), which tracks the column densities of individual H$_2$ rovibrational levels in the presence of an appreciable Ly$\alpha$ radiation field. This model tracks the fluorescence cascade of H$_2$ from excited electronic levels, pumped by photo-excitation, back to the ground electronic level until the column densities of rovibration states settles to a preferential distribution, (i.e. the states no longer significantly change due to the photo-excitation process). The framework of the model, which we will refer to as Model 3, is as followings:
\begin{enumerate}
	\item We start with a thermal distribution of hot H$_2$, where rovibrational levels are statistically defined by the total column density (N(H$_2$)) and temperature (T(H$_2$)) of the bulk molecular population.
	\item A constant, uniform radiation distribution of Ly$\alpha$ photons are generated and exposed to the initially-defined thermal population of H$_2$. 
	\item H$_2$ in the correct [$v$,$J$] ground level will have some \deleted{cross-sectional }probability to absorb Ly$\alpha$ photons incident on the H$_2$ populations. If the H$_2$ molecules absorb the photons, they are pumped to an excited electronic level, either in the Lyman or Werner bands. From there, they immediately decay back to the ground state in \deleted{one of multiple routes, or in }a fluorescent cascade. The probability for a Ly$\alpha$-pumped H$_2$ to decay back to a specific ground level is defined by the branching ratios (transition probabilities) from the excited electronic level [$v^{\prime}$,$J^{\prime}$] to the ground electronic level [$v^{\prime \prime}$,$J^{\prime \prime}$].
	\item All rovibration levels of H$_2$ are followed simultaneously and allowed to redistribute themselves by transition probabilities after initially being photo-pumped out of their original ground electronic level, [$v$,$J$]. The model runs until the ground rovibration levels settle to a nearly constant distribution of levels in the presence of this unchanging Ly$\alpha$ radiation field. 
\end{enumerate}

The Ly$\alpha$ radiation distribution used in Model 3 is assumed to mimic the observed line width and shape on a target-by-target basis. The Ly$\alpha$ line shape is assumed to be Gaussian, with parameters describing the line shape adapted from \citet{McJunkin+14}. The flux in the Ly$\alpha$ line, $F_{Ly\alpha}$, is allowed to float in each model run, as are N(H$_2$) and T(H$_2$), which set the initial conditions for each model iteration. For the duration of each model, the Ly$\alpha$ line emission is assumed to neither change in shape nor in peak flux, effectively providing the H$_2$ populations with a constant, uniform distribution of Ly$\alpha$ photons until the H$_2$ ground states relax to some preferential distribution. The basic mechanics of the model take advantage of $\sim$100 H$_2$ cross sections coincident with the Ly$\alpha$ emission profiles of typical PPD targets (i.e., Classic T Tauri stars; \citealt{France+14}). These cross sections are calculated using intrinsic transition properties of H$_2$ with Ly$\alpha$ provided by \citet{Abgrall+93} and \citet{Abgrall+93b}. Based on the energy of a given Ly$\alpha$ photon, H$_2$ in a receptive rovibration level [$v$,$J$] will absorb the photon and be pumped to either the Lyman or Werner excited electronic band. The excited H$_2$ molecules will decay back to one of many potential ground electronic rovibration levels via branching ratio probabilities, again inferred from intrinsic molecular properties provided by \citet{Abgrall+93} and \citet{Abgrall+93b}. This process is repeated until the rovibration levels of H$_2$ relax to some distribution of states under the constant Ly$\alpha$ flux (i.e., no significant change in the column densities of rovibration levels is detected, to within $\delta$log$_{10}$N(H$_2$[$v$,$J$]) $\lesssim$ 0.1 for all rovibration levels). See Appendix~\ref{model3} for more details about the models, including the iteration process used for Ly$\alpha$-pumping, H$_2$ electronic fluorescence and further details regarding the MCMC and statistics of the process.

\begin{deluxetable}{l c c c c}
\tabletypesize{\small}
\tablecaption{Ly$\alpha$-pumped H$_2$ Column Density \& Temperature Results
	\label{tab5}}
\tablewidth{0pt}
\tablehead{
	    & \multicolumn{3}{c}{Model 3}  &	 \\
												\cmidrule(r{1em}){2-4}                   
	    \colhead{Target}  &  {N(H$_2$)\tablenotemark{a}} & {T(H$_2$)\tablenotemark{b}} & {F$_{Ly\alpha}$\tablenotemark{c}} & {$\Delta$N(H$_2$)\tablenotemark{a,d}} 	}              
\startdata
	AA Tau & 16.28$^{+0.52}_{-0.33}$ & 3214$^{+570}_{-810}$ & -10.4$^{+0.8}_{-0.7}$ & 14.13 \\
	AB Aur & 15.60$^{+0.29}_{-0.16}$ & 3437$^{+410}_{-691}$ & -10.5$^{+0.5}_{-0.6}$ & 13.44 \\
	AK Sco & 15.65$^{+0.50}_{-0.27}$ & 3601$^{+290}_{-522}$ & -10.2$^{+0.7}_{-0.3}$ & 13.35 \\
	BP Tau & 17.09$^{+0.94}_{-0.58}$ & 2557$^{+1113}_{-1339}$ & -6.7$^{+0.6}_{-0.5}$ & 13.09 \\
	CS Cha & 18.36$^{+0.68}_{-1.17}$ & 1596$^{+1700}_{-340}$ & -6.9$^{+0.3}_{-0.4}$ & 13.61 \\
	DE Tau & 16.21$^{+0.45}_{-0.34}$ & 2982$^{+693}_{-812}$ & -9.4$^{+0.7}_{-0.7}$ & 13.94 \\
	DF Tau A & 16.48$^{+1.23}_{-1.38}$ & 2678$^{+940}_{-1258}$ & -6.9$^{+0.4}_{-4.2}$ & 12.63 \\
	DM Tau & 16.30$^{+0.80}_{-0.28}$ & 3670$^{+232}_{-888}$ & -8.7$^{+1.0}_{-0.6}$ & 13.90 \\
	GM Aur & 16.15$^{+0.32}_{-0.22}$ & 3469$^{+376}_{-650}$ & -7.5$^{+0.3}_{-0.4}$ & 13.55 \\
	HD 104237 & 17.87$^{+0.57}_{-0.52}$ & 2200$^{+1060}_{-766}$ & -5.7$^{+0.2}_{-0.3}$ & 13.25 \\
	HD 135344 B & 16.78$^{+0.46}_{-0.31}$ & 3185$^{+517}_{-1128}$ & -6.4$^{+0.3}_{-0.3}$ & 13.26 \\
	HN Tau A & 16.95$^{+0.99}_{-0.64}$ & 2140$^{+1088}_{-998}$ & -8.9$^{+1.5}_{-1.3}$ & 12.24 \\
	LkCa15 & 18.09$^{+1.00}_{-0.53}$ & 3456$^{+394}_{-858}$ & -8.9$^{+2.1}_{-1.3}$ & 13.81 \\
	RECX 11 & 16.72$^{+0.32}_{-0.25}$ & 3087$^{+593}_{-411}$ & -6.7$^{+0.3}_{-0.7}$ & 13.60 \\
	RECX 15 & 17.13$^{+0.55}_{-0.53}$ & 2679$^{+933}_{-798}$ & -5.9$^{+0.2}_{-0.5}$ & 13.85 \\
	RU Lupi & 17.26$^{+0.46}_{-0.47}$ & 2735$^{+621}_{-976}$ & -5.7$^{+0.1}_{-0.4}$ & 13.95 \\
	RW Aur A & 18.03$^{+0.68}_{-0.71}$ & 2504$^{+1489}_{-627}$ & -5.6$^{+0.1}_{-0.2}$ & 14.25 \\
	SU Aur & 17.59$^{+1.31}_{-1.20}$ & 2739$^{+857}_{-1631}$ & -6.1$^{+0.4}_{-4.0}$ & 12.85 \\
	SZ 102 & 16.97$^{+1.31}_{-0.89}$ & 2662$^{+940}_{-1435}$ & -6.9$^{+1.0}_{-2.3}$ & 13.16 \\
	TW Hya & 17.19$^{+1.22}_{-0.61}$ & 1910$^{+1514}_{-1029}$ & -6.6$^{+0.6}_{-0.3}$ & 13.03 \\
	UX Tau A & 17.54$^{+1.40}_{-0.49}$ & 2734$^{+880}_{-1789}$ & -6.5$^{+0.8}_{-1.2}$ & 13.75 \\
	V4046 Sgr & 16.24$^{+1.05}_{-0.32}$ & 2803$^{+771}_{-1309}$ & -6.6$^{+0.7}_{-1.1}$ & 12.76 \\
	\hline 
	Avg. Results &  16.93$^{+1.40}_{-1.33}$ & 2820$^{+850}_{-1224}$ & -7.4$^{+1.8}_{-3.1}$ & 13.43 \\
\enddata
\tablenotetext{a}{All column densities are to the power of 10 (log$_{10}$N(H$_2$)).}
\tablenotetext{b}{Temperatures of H$_2$ (T(H$_2$)) are in Kelvin.}
\tablenotetext{c}{The integrated Ly$\alpha$ fluxes that pump H$_2$ populations out of thermal equilibrium are described by the sum of a narrow and broad Gaussian component, with FWHMs of each component adapted from \citet{McJunkin+14}. Flux are to the power of 10 (log$_{10}$F(Ly$\alpha$)). F(Ly$\alpha$) has units of ergs cm$^{-2}$ s$^{-1}$.}
\tablenotetext{d}{The integrated residual between the observed column densities of H$_2$ in states [$v$,$J$] to the model prediction of column density in the same rovibrational levels, $\Sigma | \textnormal{N(H}_2[v,J])_{data} - \textnormal{N(H}_2[v,J])_{model}| $.	}
\end{deluxetable}

\begin{figure}[htp]
\centering
\includegraphics[angle=270,width=0.475\textwidth]{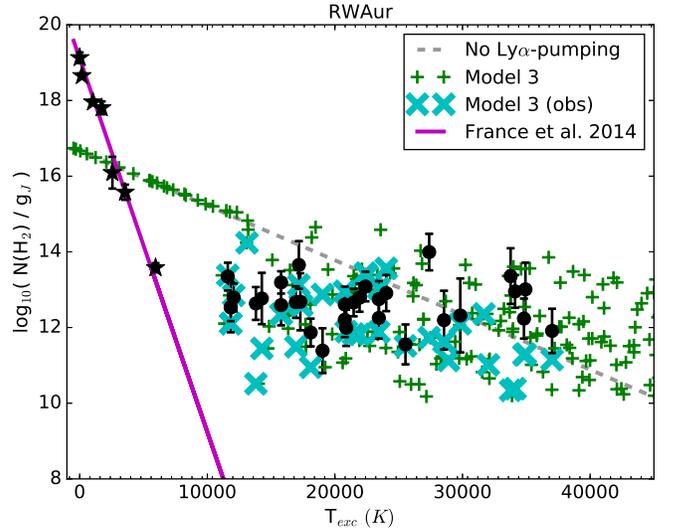}
	\caption{The rotation diagram for RW Aur, with rovibrational column densities derived in this study (black circles) and lower energy states calculated by \citet{France+14b} (black stars; $\lambda \lambda$ 1092.5 $-$ 1117 {\AA}). The magenta solid line shows the thermal distribution H$_2$ levels examined by \citet{France+14b}, with log$_{10}$( N(H$_2$) ) $=$ 19.90 cm$^{-2}$ and T(H$_2$) $=$ 440 K. The green plus symbols represent the H$_2$ rovibrational levels output by the Ly$\alpha$-pumping models (Model 3). The cyan ``X''s mark rovibrational levels from Model 3 which match the observed $H_2$ levels, so the reader can directly compare the the data with the modeled states. The gray dashed line presents the initial thermal distribution of H$_2$ in the models (i.e., without Ly$\alpha$ pumping), which is described by log$_{10}$( N(H$_2$) ) $=$ 18.03 and T(H$_2$) = 2504 K.
	\label{fig10} }
\end{figure}

We present Model 3 results in Table~\ref{tab5}. Figure~\ref{fig10} shows the observed rotation diagram of RW Aur A and the resulting modeled distribution of H$_2$ rovibration levels produced by Model 3. The Ly$\alpha$ photo-excitation models for all targets are presented in Appendix~\ref{app:fig4}. Green plus symbols represent all H$_2$ rovibrational states for $v$ $\leq$ 15, $J$ $\leq$ 25, while cyan ``X'' symbols represent modeled rovibration levels with the same rovibration level as those empirically measured in the stellar Ly$\alpha$ wings of the target. Model 3 for RW Aur finds a total column density of H$_2$, log$_{10}$( N(H$_2$) ) $\approx$ 18.0, which is $\sim$2 dex lower than results from \citet{France+14}, at a temperature T(H$_2$) $\approx$ 2500 K (in \citet{France+14}, T(H$_2$)$_{warm}$ = 440 K). 

The total column density of thermal H$_2$ for RW Aur is slightly larger than the average best-fit N(H$_2$) for all targets ($\langle$log$_{10}$N(H$_2$)$\rangle$ $\sim$ 17.0), with the smallest total column density log$_{10}$N(H$_2$) $\approx$ 15.5. Interestingly, for almost all samples in our survey, the derived total column density of thermal H$_2$ distributions is larger than those estimated by our thermal models (i.e., Models 1 and 2).   
For all targets, the derived thermal temperatures of H$_2$ from the Ly$\alpha$-pumping model range from 1500 - 4000 K ($\langle$T(H$_2$)$\rangle$ $\sim$ 2800 K). Overall, the final results from the Ly$\alpha$-pumping models \deleted{slightly }overestimate the total column density of H$_2$ for a hot atomic layer origin by $\sim$1-2 dex and underestimate the total column density of H$_2$ for a warm molecular layer origin by the same amount \citep{Adamkovics+16}. Additionally, the temperature of thermal H$_2$ is found somewhere between the two layers.

One of the first things we notice about the Model 3 results is that the H$_2$ rovibrational levels are redistributed in such a way that more highly \replaced{thermal}{excited} H$_2$ \replaced{populations}{levels} ($T_{exc}$ $\gtrsim$ 30,000 K) can be pumped to higher column densities than they are expected to be in thermal distributions. Rovibrational levels of H$_2$ most affected by the flux of Ly$\alpha$ (i.e., $v$ $\geq$ 2; $T_{exc}$ $\sim$ 10,000 K) first appear diminished in column density, relative to the native thermal distributions, but for rovibrational levels with $T_{exc}$ $\gtrsim$ 30,000 K, the relative column densities of highly energetic states appears to return back towards the level of the thermal distribution, with many states being pumped by $\gtrsim$ 1 dex more than they would otherwise be in thermally-distributed states. 

Additionally, the re-distributed H$_2$ rovibrational levels appear scattered, with the behavior of the scattered distributions appearing roughly consistent for rovibrational levels with $T_{exc}$ $\gtrsim$ 10,000 K and with a spread of $\sim$1 dex. We note that this behavior matches the characteristic distributions of empirically-derived H$_2$ rovibration levels measured against Ly$\alpha$ for most, if not all, of our PPD sightlines. The Ly$\alpha$ redistribution appears to scatter most H$_2$ states out of thermal equilibrium at T$_{exc}$ $\gtrsim$ 10,000 K, suggesting that the H$_2$ absorption coincident on the Ly$\alpha$ wings do not probe thermal populations of H$_2$ in these sightlines. The fact that we see this same peculiar H$_2$ population behavior for all disks in our survey, regardless of orientation of the disk in the line of sight (i.e., i$_{disk}$), suggests that the sampling of H$_2$ may not be co-spatial with the same H$_2$ populations observed in fluorescence from each disk. The models also suggest that, for rovibrational levels insensitive to Ly$\alpha$ radiation (i.e., $v$ $<$ 2), H$_2$ may still be thermally populated. Theoretically, if we could observe rovibrational levels of H$_2$ not pumped by Ly$\alpha$ radiation, we could test this hypothesis.

We do have one case study - RW Aur - where this test is currently possible. The sightline to RW Aur probes both hot H$_2$ embedded in the Ly$\alpha$ profile of the protostar and warm H$_2$ in the FUV continuum ($\lambda \lambda$ 1090 - 1120 {\AA}; \citealt{France+14}). If the warm disk H$_2$ populations and the hot Ly$\alpha$ H$_2$ populations were co-spatial with one another, we would expect to find signatures in the FUV-continuum probing the same hot H$_2$ population (specifically for $v$ = 0, $J$ = 4, 5, 6; $\lambda$ = 1100.2, 1104.1, 1104.5, 1109.3, 1109.9, 1115.5, 1116.0 {\AA}, where the distributions of warm and hot H$_2$ populations overlap). From the Ly$\alpha$-pumping model results for RW Aur, we expect to find appreciable thermal columns of hot H$_2$ in the sightline, which is several dex denser than the warm H$_2$ probed by the FUV continuum. The FUV continuum is much less likely to scatter through the gas disk than Ly$\alpha$, and therefore is expected to provide a better probe of the geometry through the disk material. The fact that the \citet{France+14} study does not see clear deviations to the larger column density found by Model 3 for hot H$_2$[$v$ = 0, $J$ = 4, 5, 6] in the FUV continuum is further evidence supporting our original hypothesis - that the resonance nature of Ly$\alpha$ allows the radiation to scatter through a hot atomic \replaced{haze or layer}{halo} above the PPD, and the observed H$_2$ signatures observed in the protostellar Ly$\alpha$ wings probe residual H$_2$ in these environments, rather than in the disk.

\deleted{However, we find that with Model 3, while the general behavior of rovibration levels matches the characteristic behavior of empirically-derived Ly$\alpha$-H$_2$ absorption species, we do not perfectly replicate the observed level populations of H$_2$. We state the total residual H$_2$ densities unaccounted for by the models in Table~\ref{tab5}, which are $\sim$4 dex lower than residuals found between purely thermal models (Models 1 and 2) and highly energetic H$_2$ states (T$_{exc}$ $>$ 20,000 K). The models we put forth for this experiment were simplified, and perhaps including more physical mechanisms to the simulations, such as additional UV-pumping throughout the FUV, dissociation and formation routes, rovibration emission, and collisional de-excitation, will better estimate the distribution of H$_2$ levels under all these processes simultaneously. Still, the Ly$\alpha$-pumping models were successful in replicating the general behavior of hot H$_2$ rovibrational levels observed in PPD sightlines against the protostellar Ly$\alpha$ features. This result suggests that photo-excitation from \ion{H}{1}-Ly$\alpha$ may be an effective process for re-distributing the ground electronic levels of the hot H$_2$ probed in the Ly$\alpha$ profiles. This has its limits, as demonstrated at the end of Section~\ref{sec:calc}: Ly$\alpha$ photo-excitation may only be an effective means of re-distributing H$_2$ ground levels in optically-thin regimes (e.g., PDRs), or where rovibrational cooling and collisional de-excitation of H$_2$ occur much less frequently than photo-excitation, but otherwise likely does not have as strong a role in other astrophysical environments where these conditions do not hold true.}

%
%
%
\section{Conclusions}

We perform the first empirical survey of H$_2$ \deleted{rovibrational }absorption observed against the stellar Ly$\alpha$ emission profiles of 22 PPD hosts. The aim of this study was to identify thermal and non-thermal H$_2$ species in each sightline and investigate excitation mechanisms responsible for the distributions of non-thermal H$_2$ populations.
We normalize each Ly$\alpha$ profile \deleted{with a smoothing kernel }and create optical depth models to \deleted{simultaneously }synthesize H$_2$ absorption features observed across the normalized Ly$\alpha$ spectra. Each optical depth model estimates the column density of H$_2$ in ground states [$v$,$J$] from the absorption depth in the Ly$\alpha$ wings, and we present the H$_2$ rotation diagrams of all samples in our survey to examine the behavior of the H$_2$ rovibrational populations in all sightlines. 
Below, we highlight our findings and conclusions:
\begin{itemize}
	\item Thermally-distributed H$_2$ models alone cannot reproduce observed rovibration levels. \replaced{When we look at the general behavior of all PPD hosts, we see there is a repeating pattern, where h}{H}ighly-energetic states are ``pumped'' when compared with lower energy rovibrational states. This appears to happen at ``knee'' junctures, which are consistently found at T$_{exc}$ = 20,000 K, 25,000-26,000 K, and 31,000-32,000 K. 
	\item We find roughly-equivalent total column densities of thermal and non-thermal H$_2$ populations in transitional disk samples and samples with detectable \ion{C}{4}-pumped H$_2$ fluorescence. \replaced{Interestingly, p}{P}rimordial disk targets have more spread in this relation, and show more samples with larger total column densities of thermal H$_2$ than non-thermal H$_2$ populations.
	\item High energy continuum radiation, produced primarily by accretion processes onto the host protostar, appears to play an important role in regulating the total density of non-thermal H$_2$ in the circumstellar environment. \replaced{High energy FUV photons (912 {\AA} $<$ $\lambda$ $<$ 1110 {\AA}) and X-rays are effective at both grain heating and creating of free electrons, which can excite molecules to non-thermal states \citep{Bergin+04}, and w}{W}e find correlations between the X-ray and FUV luminosities and N(H$_2$)$_{\textnormal{nLTE}}$\replaced{. We find}{ and} little evidence that line emission from protostellar accretion processes plays a significant role in regulating the total column densities of thermal and non-thermal H$_2$ states, except \ion{C}{4}, which appears to be anti-correlated with the total thermal column densities of H$_2$.
	\item There is a clear anti-correlation between N(H$_2$)$_{\textnormal{nLTE}}$ and H$_2$ dissociation continuum, suggesting that photo-excitation may be more effective at unbinding H$_2$ already in highly energized levels than lower energy thermal states.
	\item From one target that has access to cooler H$_2$ populations observed against the FUV continuum (RW Aur A; \citealt{France+14}), we see \replaced{a significant discrepancy between}{two populations of H$_2$:} warm H$_2$ \replaced{populations}{probing higher density material in the protoplanetary disk,} and \deleted{the} hot \replaced{Ly$\alpha$ absorption populations}{H$_2$ in an atomic halo surrounding the protostar and disk}. The total column of \replaced{cool}{warm} H$_2$ is several dex higher than the total column of \replaced{warm}{hot} H$_2$ in the Ly$\alpha$ wings. We see a crossing point, where we should begin to see warmer columns of H$_2$ in the FUV continuum (T$_{exc}$ $\approx$ 3,000 K), but, observationally, this does not appear to be the case. \ion{H}{1}-Ly$\alpha$ is a strong resonance \added{transition}, and a small amount of residual \ion{H}{1} in the protostellar environment will scatter Ly$\alpha$ \deleted{around the system many times }before it escapes. We suspect that the H$_2$ populations probed in the protostellar Ly$\alpha$ wings are not associated with the disk, but rather found in \replaced{this tenuous haze of hot}{a tenuous halo of hot, mostly atomic} gas around the disk.\deleted{, since we do not appear to see these same population columns densities in absorption against the FUV continuum (which will not scatter out of the line of sight as Ly$\alpha$ will and is a more likely tracer of the disk at a given observed geometry).} The hot H$_2$ also probes much lower column densities ($\langle$N(H$_2$)$\rangle$ $\sim$ 10$^{16}$ cm$^{-2}$) of H$_2$ than is required to produce the observe fluorescence in these same PPD samples, 
strongly suggesting that absorption and fluorescence H$_2$ populations are not co-spatial.
\end{itemize}

While this study examined the behavior of hot H$_2$ in protoplanetary disk environments, further investigation and proper implementation of non-LTE models is necessary to pinpoint the physics driving H$_2$ to higher [$v$,$J$] states. Studies have been performed that point to several mechanisms driving H$_2$ populations, including collisions with other particles and higher energy photons (FUV/EUV/X-ray; \citet{Nomura+05,Nomura+Millar+05,Adamkovics+16}) and reformation/destruction of H$_2$ by chemical evolution, especially H$_2$O dissociation in the warm disk atmosphere \citep{Du+Bergin+14,Glassgold+Najita+15}. The next step forward would be to implement radiative, collisional, and chemical processes simultaneously to simulate the PPD environmental behavior. \deleted{Unfortunately, this is more easily said than done. Protoplanetary disk environments are complicated; interstellar medium formalisms do not necessarily apply to the warm, stratified layers we expect to probe in PPDs. Furthermore, dust grain sizes and evolution throughout the PPD lifetime add complications, such as inaccuracy of H$_2$ formation rates throughout the evolution of planet-forming disks. All of these complex physical problems will become critical to address as better observations (i.e., \textit{JWST}, \textit{LUVOIR}) are made of the warm materials in PPDs.}
\deleted{Until then, laboratory astrophysical experiments may help clarify our understanding of the role of various radiative processes on the physical state of H$_2$ in circumstellar environments. While it is difficult to replicate the exact environments of PPDs in a laboratory setting, we can take a step back and understand how H$_2$ states respond to the input of different radiative mechanisms, such as X-ray flux, continuum emission through the FUV, and discrete line emission from \ion{H}{1}-Ly$\alpha$, \ion{H}{1}-Ly$\beta$, and \ion{H}{1}-LyC. These experiments may also help in our understanding of the role of other molecular and gaseous materials in the circumstellar environment that are not directly probed in this study, but may have meaningful impacts on the final ``equilibrium'' state of H$_2$ we observe in PPDs.}
Paper II will address the spatial origins of this H$_2$ absorption, based on results from this study and empirical evidence from the absorption features themselves. \deleted{However, further research is needed to help relate these H$_2$ species to other atomic and molecular tracers in and around the disk atmosphere, which will provide an important stepping stone towards tying together disk processes and creating a more complete picture of the evolutionary mechanisms that drive the dispersal of gas from these systems over, on average, a few Myr.}

%
%
%
\acknowledgments

This research was funded by the NASA Astrophysics Research and Analysis (APRA) grant NNX13AF55G, \textit{HST} GO program 12876, \textit{HST} GO program 13372,and \textit{HST} AR program 13267, and uses archival NASA/ESA \textit{Hubble Space Telescope} observations, obtained through the Barbara A. Mikulski Archive for Space Telescopes at the Space Telescope Science Institute. The authors thank the anonymous referee for their helpful feedback, which ultimately improved the quality of this paper. KH would like to thank Allison Youngblood, Ilaria Pascucci, Andrea Banzatti, and Klaus Pontoppidan for enjoyable discussion and constructive suggestions.

\clearpage

%
%
%
\appendix

\section{Additional Details on H$_2$ Absorption Line Analysis}\label{app:h2abs}

\figsetstart
\figsetnum{1}
\figsettitle{The Ly$\alpha$ profiles of each PPD host}

\figsetgrpstart
\figsetgrpnum{1.1}
\figsetgrptitle{AA Tau
}
\figsetplot{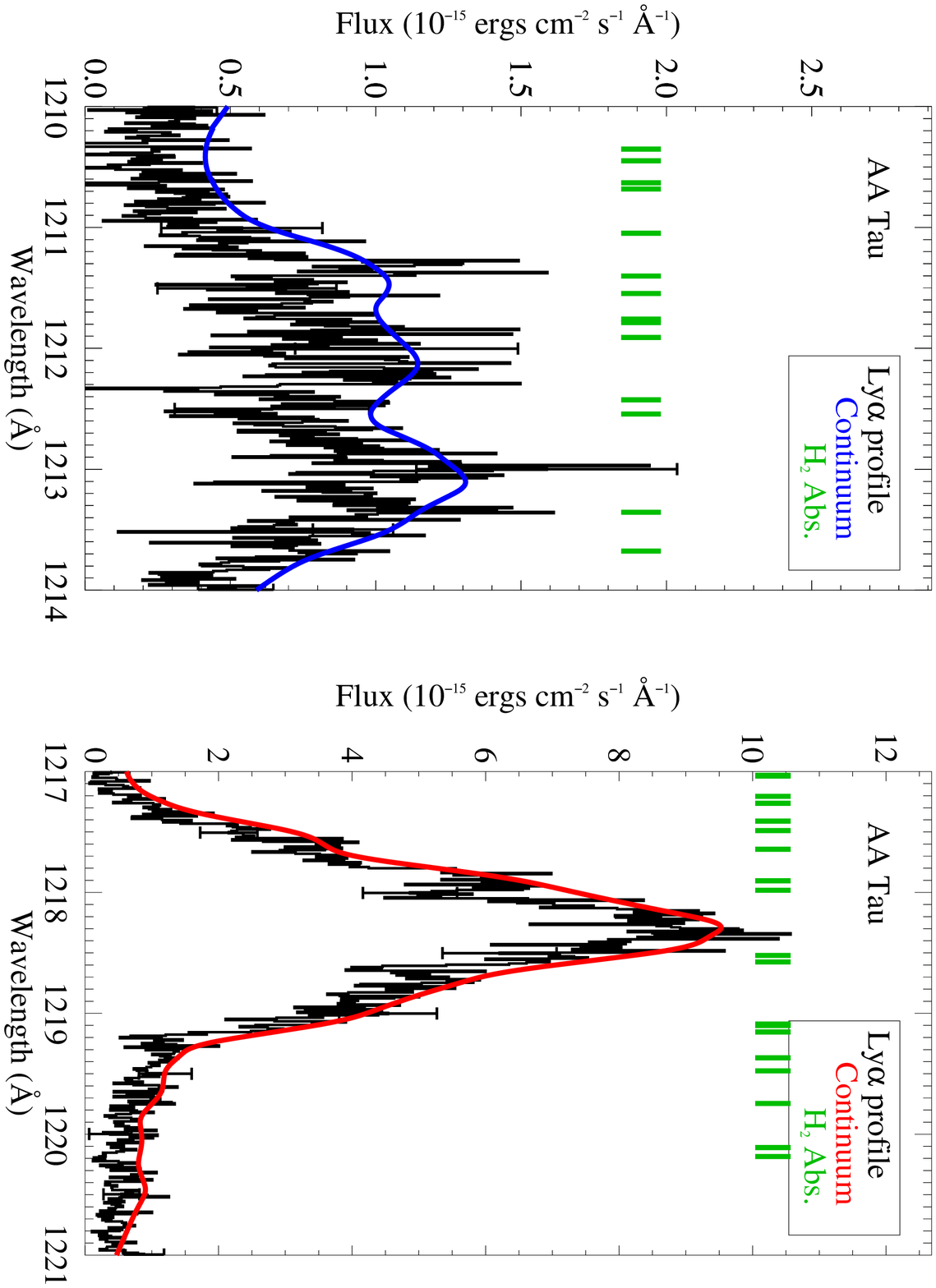}
\figsetgrpnote{}
\figsetgrpend

\figsetend

\begin{figure}
\figurenum{1}
\centering
\includegraphics[angle=90,width=0.9\textwidth]{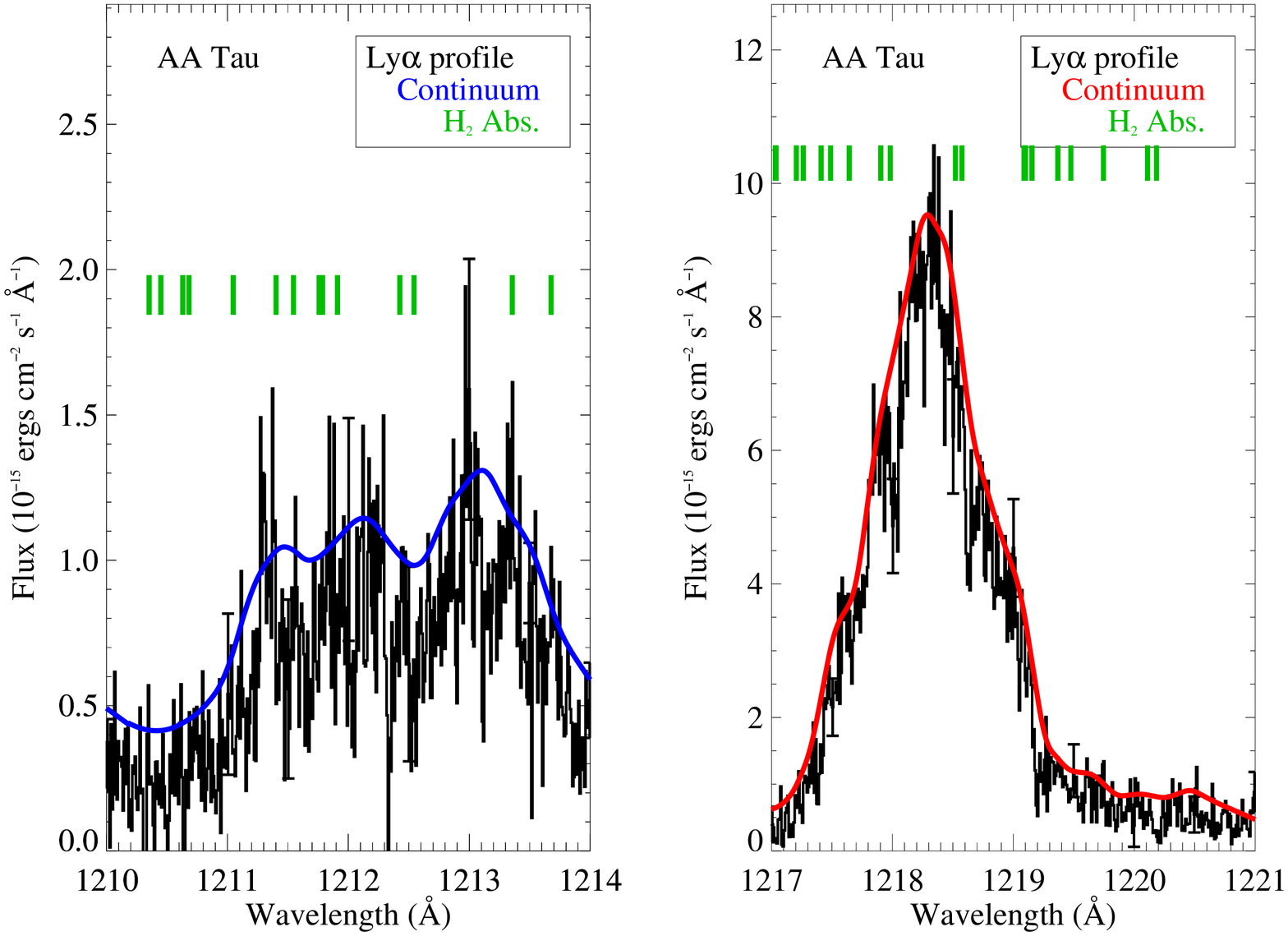}
\caption{The Ly$\alpha$ profiles of each target, overlaid with the Ly$\alpha$ ``continuum'' fit determined from our functional processes (Section~\ref{sec:lya_norm}). Two targets are shown per row, where the blue and red Ly$\alpha$ profiles are presented. The Ly$\alpha$ mean flux arrays are over-plotted in blue (over the blue-wing Ly$\alpha$ component) and red (over the red-wing Ly$\alpha$ component). We mark the location of H$_2$ absorption transitions in the Ly$\alpha$ profiles with green hashes. The continuum fit is determined to normalize the Ly$\alpha$ emission profile, which is achieved by dividing the mean flux continuum through the emission line, creating a normalized spectral region across the Ly$\alpha$ wing.} \label{app:fig2}
\end{figure}

Information about each H$_2$ absorption transition was found either in \citet{Abgrall+93} or \citet{Abgrall+93b}, specifically the Einstein A-coefficient, describing the rate of spontaneous decay from state $u$ $\rightarrow$ $l$ ($A_{ul}$), and the wavenumber. All H$_2$ transitions were selected from \citet{Roncin+95} between 1210 - 1221 {\AA}, with transitions preferentially considered from those previously called out by \citet{Herczeg+02} and \citet{France+12a}. Other H$_2$ transitions included in the line-fitting analysis met a minimum (A$_{ul}$) $\geq$ 3.0$\times$10$^{7}$ s$^{-1}$, to ensure that the absorption transition probabilities were large enough for detection, assuming a warm thermal population of H$_2$. The energy levels of ground state H$_2$ in vibration and rotation levels [$v$,$J$] (E$_{gr}$) were derived from equations outlined in \texttt{H$_{2}$ools} \citep{McCandliss+03}, with physical constants taken from \citet{Herzberg+50}, \citet{Jennings+84}, and \citet{Draine+11}. The physical properties of the H$_2$ transition were derived from intrinsic properties of the molecule:
\begin{equation}	
	\sigma(\lambda) = \left(\frac{\lambda_{\lambda}^{3}}{8\pi c} \right) \left(\frac{g_{u}}{g_{l}} \right) A_{ul} 
    \label{eq1}
\end{equation}
\begin{equation}	
	f_{lu} = \left( \frac{m_{e} c}{8 \pi^2 e^{2}} \right) \left( \frac{g_u}{g_l} \right) \lambda_{lu}^2 A_{ul}
    \label{eq2}
\end{equation}
where $\lambda_{\lambda}$ is the photo-excitation wavelength, Ly$\alpha$, of H$_2$ in ground state [$v$,$J$]; $g_{u}$ and $g_{l}$ are the statistical weights of the electronically-excited [$v^{\prime}$,$J^{\prime}$] and ground [$v$,$J$] states, respectively; and ($\pi e^{2}/m_{e} c$) is the definition of the classical cross section, expressed as 0.6670 cm$^{2}$ s$^{-1}$ in cgs units.
Table~\ref{tab3} shows all transitions used in our H$_2$ synthetic absorption model, including physical properties (E$_{gr}$, $f_{lu}$, $A_{ul}$) and level transition information. Not all transitions were implemented for every target. Depending on the effective range of the stellar Ly$\alpha$ wing in wavelength space, many of the transitions found on the edges of the wings (1210 $-$ 1212: 1213.5 $-$ 1215.2 {\AA} for the blue wing; 1216 $-$ 1218: 1219.5 $-$ 1221 {\AA} for the red wing) were omitted. 

The modeled b-value is fixed in all synthetic absorption spectra to replicate the thermal width of a warm bulk population of H$_2$ (T(H$_2$) $\geq$ 2500 K) in the absence of turbulent velocity broadening. If the b-value were larger, the broadening acts to widen the absorption feature and diminish the depth of the line center, which causes degeneracy between the estimated rovibrational [$v$,$J$] level column densities and the thermal/turbulent parameters of the models. When we increased $b_{H_2}$ $=$ 10 km s$^{-1}$, the column densities of the rovibrational [$v$,$J$] levels were systematically reduced by 0.1-0.7 dex for all survey samples.

The multi-component fit of H$_2$ absorption was mostly insensitive to initial conditions. Initially, we set the same initial conditions for the start of the run ($v_r$ = 0 km s$^{-1}$; T(H$_2$) = 2500 K; log$_{10}$ N(H$_2$;$v$,$J$) varied by transition properties) and allowed the parameters float. Once an effective range of values was determined for all targets, T(H$_2$) and $b_{H_2}$ were fixed, and only $v_r$ was allowed to float. This produces column density estimates that are relatively comparable for all targets in our survey.

\begin{figure}
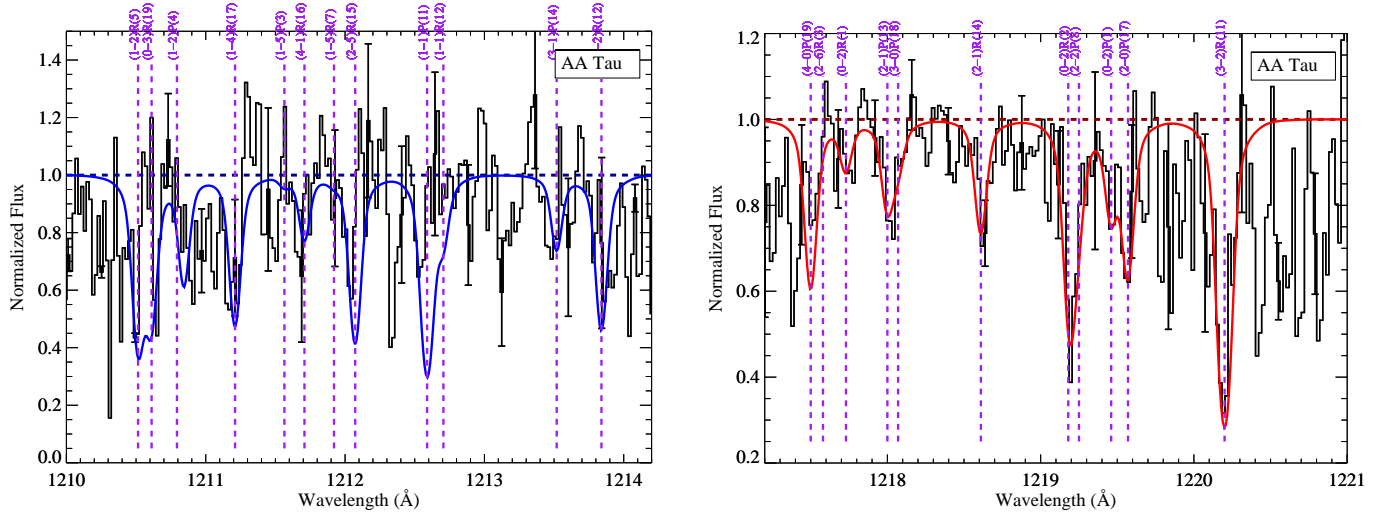

\figurenum{2}
\centering
	\subfloat{\includegraphics[angle=90,width=0.475\textwidth]{figb1a.eps}}
	\hfill
    \subfloat{\includegraphics[angle=90,width=0.485\textwidth]{figb1b.eps}} 
    \caption{The final, synthesized absorption spectra of warm H$_2$ against the Ly$\alpha$ transmission spectra. For targets that have transmission spectra for both blue and red Ly$\alpha$ wings, blue wing spectra are shown on the \emph{left} (blue H$_2$ absorption fit) and red wing spectra are on the \emph{right} (red H$_2$ absorption fit). For targets with only a red-wing transmission spectrum, red-wing fits are shown on the \emph{left} (red H$_2$ absorption fit).} \label{app:fig3}
\end{figure}

As discussed in \citet{France+12a}, only the (0-2)R(2) and (2-2)P(9) levels, whose wavelengths differed by $\Delta \lambda$ = 0.01 {\AA} (at 1219.09 and 1219.10 {\AA}, respectively), were sensitive to the initial conditions. The total column density at this wavelength range is robust, while the relative columns shared between the two transitions was not. To mitigate this, we weighed the individual columns by the product of their oscillator strengths and relative populations of the two levels at T(H$_2$) = 2500 K. Using the methodology laid out in \texttt{H$_2$ools} and Equation~\ref{eq2}, we calculate the oscillator strengths and relative populations of the two lines to be [$f_{R(2)}$ = 25.5 $\times$ 10$^{-3}$; $P_{R(2)}$ = 5.76 $\times$ 10$^{-4}$] and [$f_{R(9)}$ = 31.8 $\times$ 10$^{-3}$; $P_{R(9)}$ = 6.24 $\times$ 10$^{-4}$], respectively. Therefore, N(2,2) contributes 0.425 of the total column density determined at 1219.10 {\AA}, while N(2,9) contributes 0.575 of the total column.
Column 2 of Figure~\ref{app:fig3} show the minimized multi-component synthetic spectra plotted over the normalized Ly$\alpha$ wings for the red-ward and blue-ward profile components, respectively.

\figsetstart
\figsetnum{2}
\figsettitle{The Relative Absorption Spectra and H$_2$ Optical Depth Models of each PPD host Ly$\alpha$ emission wing.}

\figsetgrpstart
\figsetgrpnum{2.1}
\figsetgrptitle{AA Tau
}
\figsetplot{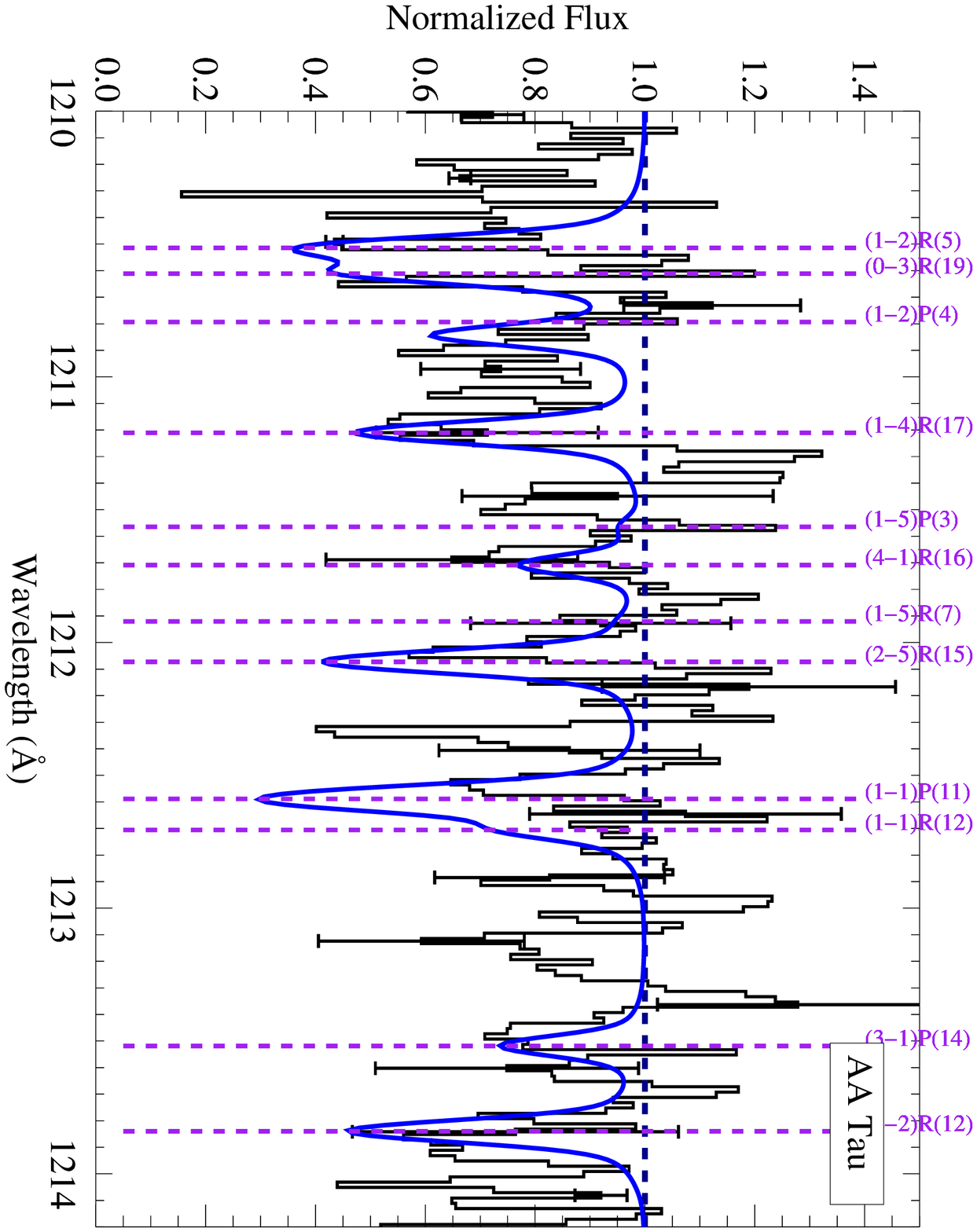}
\figsetplot{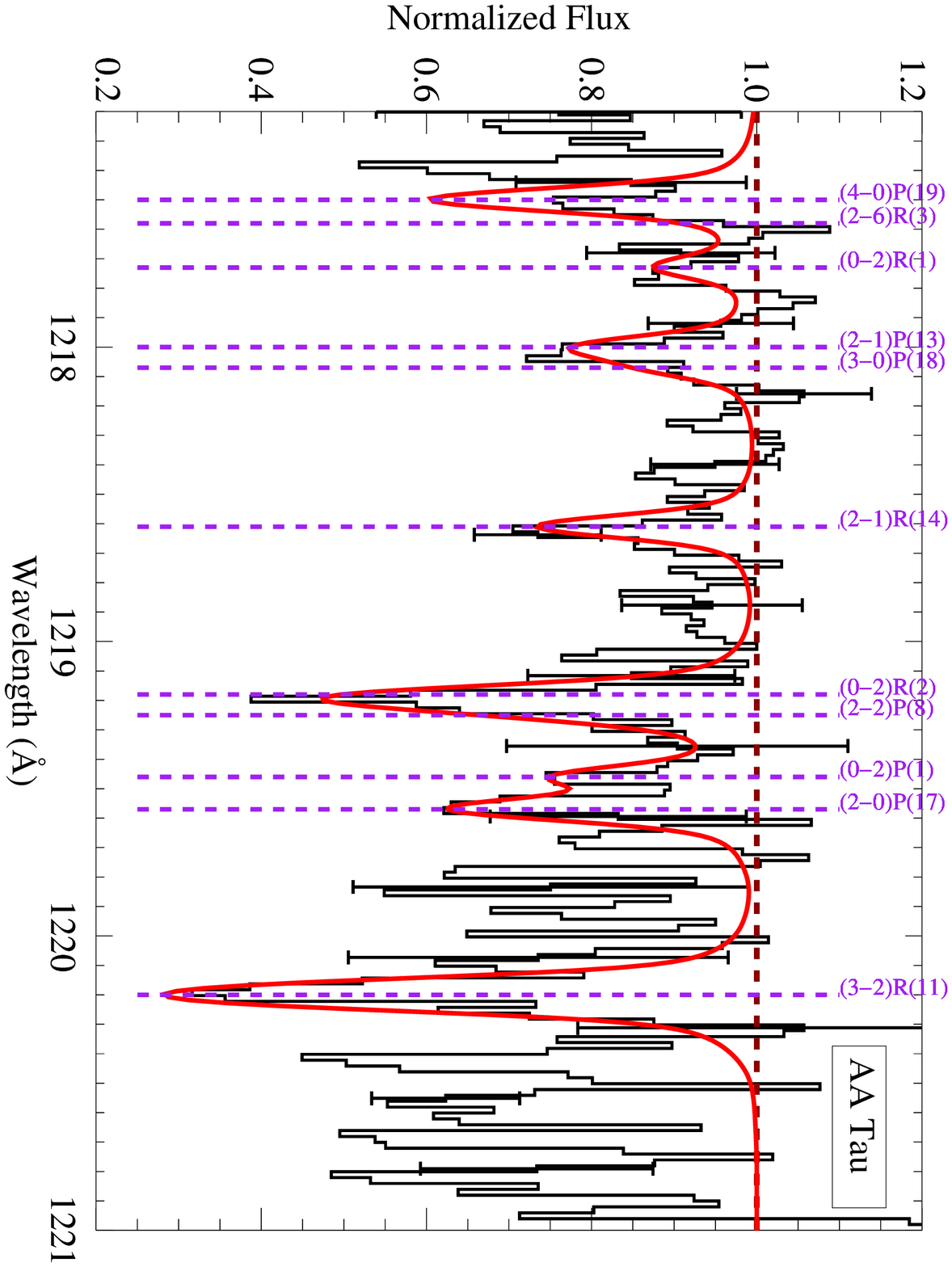}
\figsetgrpnote{}
\figsetgrpend

\figsetend

\section{H$_2$ Model Details and Monte Carlo Simulations}\label{app:models}

\subsection{Models 1 \& 2: Thermal H$_2$ Populations only}\label{model1+2}
Models 1 and 2 are simple models that follow the \texttt{H$_2$ools} layout: Given the derived column densities for observed H$_2$ ground states against the stellar Ly$\alpha$ wing N(H$_2$;$v$,$J$), we use first principles molecular physics to determine the theoretical population column densities of a bulk H$_2$ population N(H$_2$) described by a shared thermal profile T(H$_2$). The level column densities are calculated using Boltzmann populations, assuming LTE conditions, and each ground state energy level is determined by calculating the electronic, vibrational, and rotational energy levels for a ground state [$v$,$J$], as described in \citet{McCandliss+03}. 

Model 1 assumes that all data points extracted from the absorption features of each target are thermally-populated. Model 2 assumes only H$_2$ populations with ground state energies E$_{gr}$ $<$ 1.5 eV (T$_{exc}$ $\lesssim$ 17500 K) are thermally-populated, with the possibility that H$_2$ in ground states with E$_{gr}$ $>$ 1.5 eV are pumped additionally by some unknown non-thermal process(es), and so are not considered in the model-data comparison. We use Model 2 as a baseline of the minimum N(H$_2$) and T(H$_2$) of thermal H$_2$ in the disk atmosphere for each target, assuming any of the observed, absorbing H$_2$ against the Ly$\alpha$ wing is purely thermally excited. 

\begin{figure}
\centering
\includegraphics[angle=270,width=0.9\textwidth]{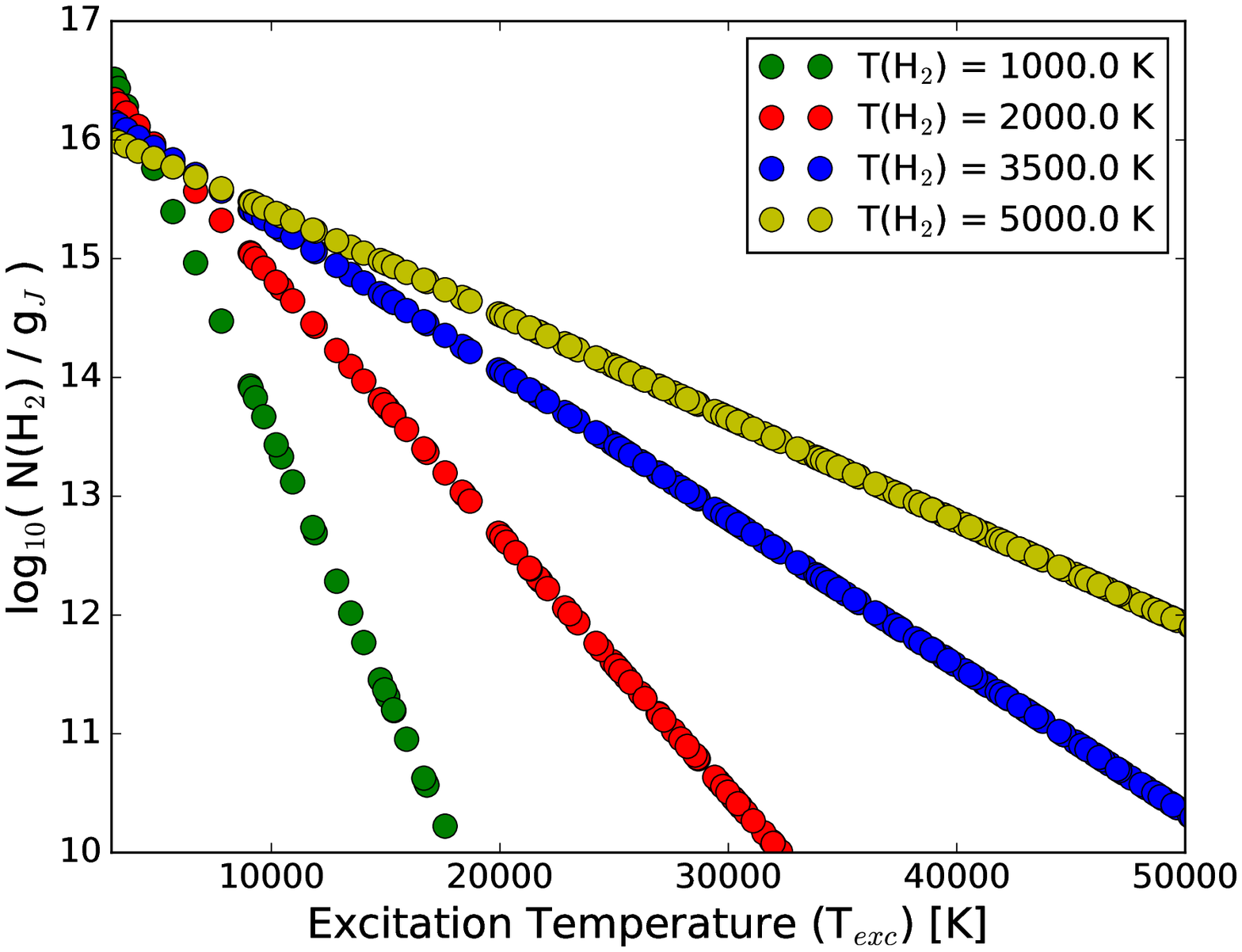}
\caption{Modeled rotation diagram of H$_2$ populations found in thermal equilibrium with a set N(H$_2$) = 10$^{17}$ cm$^{-2}$ and varying thermal descriptions T(H$_2$). As T(H$_2$) increases, more H$_2$ populations with higher excitation temperatures, T$_{exc}$, become populated, increasing the relative ratio of higher T$_{exc}$ state to lower T$_{exc}$ states, which decreases the slope of the distribution towards zero. This model is used to compare the observed rotation profiles of H$_2$ to thermally-populated states of H$_2$ for Models 1 and 2. \label{h2pop}}
\end{figure}

Figure~\ref{h2pop} shows an example of how the relative [$v$,$J$] states are populated by the thermal distribution of H$_2$. While the total column density of H$_2$ regulates the column densities of H$_2$ found in ground state [$v$,$J$], T(H$_2$) determines the relative abundances of each [$v$,$J$] to others in the ground state. For example, a lower T(H$_2$) means that, statistically, more H$_2$ is found in ground states with low [$v$,$J$] because the overall excess energy in the H$_2$ populations is low. However, as T(H$_2$) increases, the ratio of the abundances of H$_2$ found in higher [$v$,$J$] states to those in low [$v$,$J$] states increases. This appears as a ``flattening'' of the slope of H$_2$ populations in Figure~\ref{h2pop}.

\figsetstart
\figsetnum{3}
\figsettitle{Fitting Thermal Models to Each H$_2$ Rotation Diagram}

\figsetgrpstart
\figsetgrpnum{3.1}
\figsetgrptitle{AA Tau
}
\figsetplot{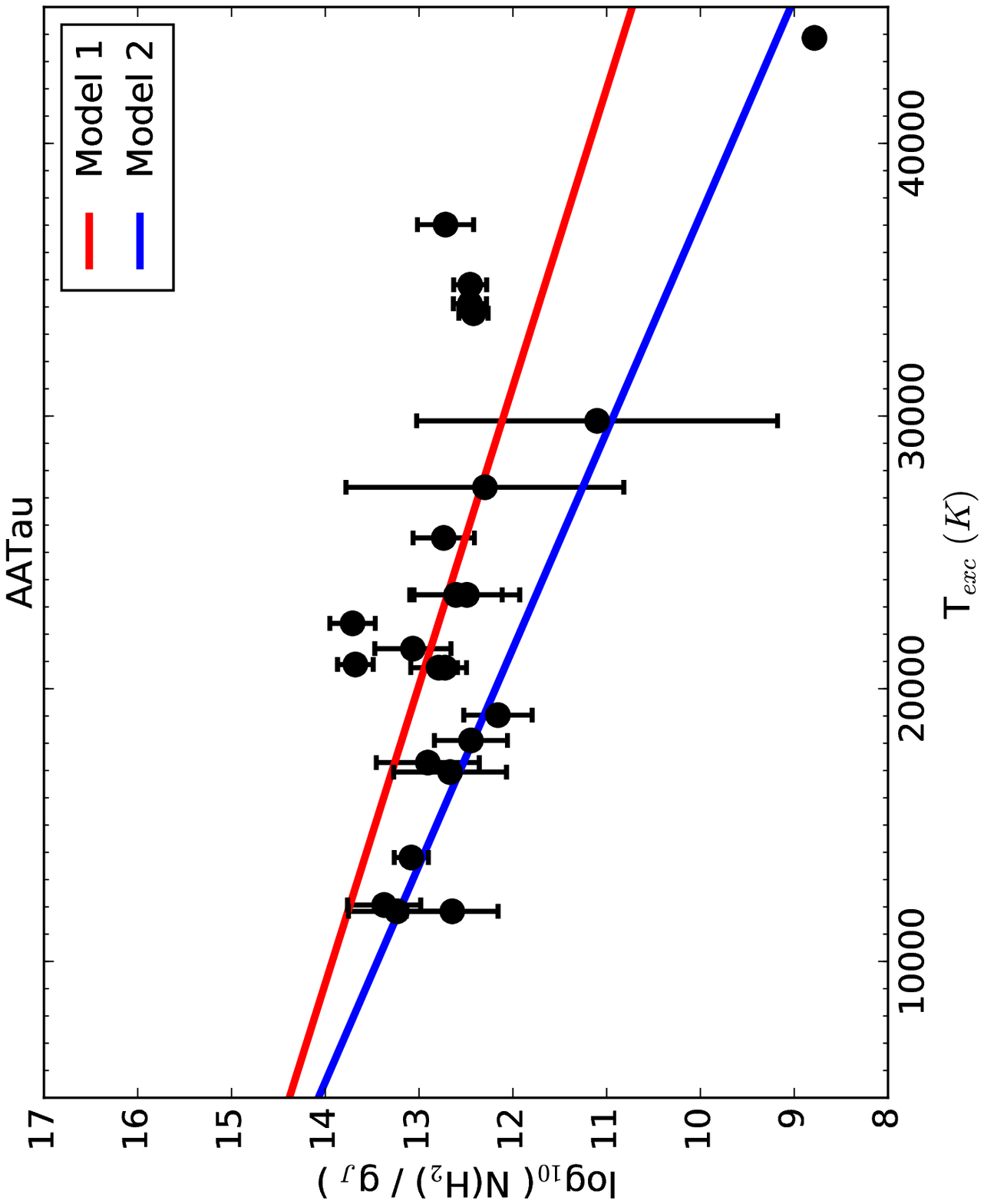}
\figsetgrpnote{}
\figsetgrpend

\figsetend

\begin{figure}
\figurenum{3}
\centering
\includegraphics[angle=270,width=0.9\textwidth]{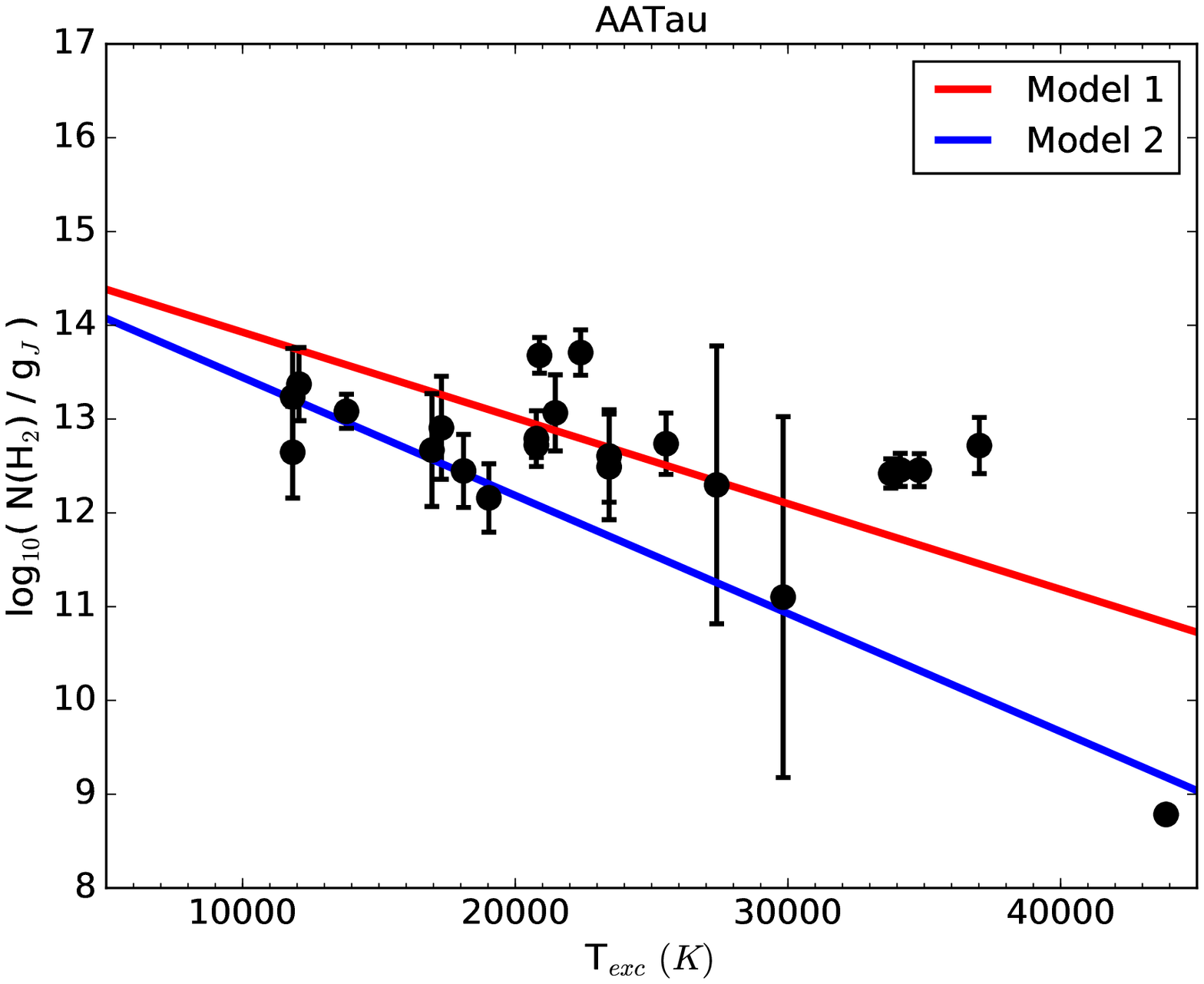}
\caption{All rotation diagrams are presented here. Each H$_2$ ground state column density is weighed by its statistical weight, g$_J$. Model 1 attempts to fit one thermally-populated bulk H$_2$ population through all data points extracted from the \textit{HST} data sets. Model 2 does the same as Model 1, but only for H$_2$ states with lower energy ground states (E$_{gr}$ $\leq$ 1.5 eV; T$_{exc}$ $\leq$ 17,500 K). } \label{app:fig4}
\end{figure}

\section{MCMC Simulations}\label{app:mcmc}
Each model is compared to the resulting rotation diagrams derived from the relative H$_2$ absorption column densities derived as explained in Section~\ref{sec:lya_norm}. This is done using a MCMC routine, which randomly-generates initial parameter conditions and minimizes the likelihood function ($\ln \Lagr$(x,$\theta$)) between the H$_2$ rovibration column densities and model parameters. We define $\ln \Lagr$(x,$\theta$) as a $\chi^2$ statistic, with an additional term to explore the weight of standard deviations on each rovibrational column density:
\begin{equation}
	\ln \Lagr(\textnormal{x},\theta, f) = \frac{(y(\textnormal{x}) - y(\textnormal{x},\theta))^2}{\sigma^2} - \ln ( (\sigma^2 + y(\textnormal{x},\theta)^2 \exp(2f))^{-1} )
    \label{eq_lklihd}
\end{equation}
In Equation~\ref{eq_lklihd}, x represents the ground state energy of H$_2$ in rovibration level [$v$,$J$], $y$(x) is the observed column density of H$_2$, $y$(x,$\theta$) is the modeled column density of H$_2$ derived from the thermal model, $\sigma^2$ is the variance in the column densities, and $f$ is an estimation on the accuracy of the column density standard deviations. For parameters shared between all thermal model runs (N(H$_2$), T(H$_2$), $\ln f$), we set prior information about each to keep the model outputs physically viable. We let the total thermal H$_2$ column density range from N(H$_2$) = 12.0 $-$ 25.0 cm$^{-2}$. Below N(H$_2$) = 12.0 cm$^{-2}$, there is not enough column in individual rovibrational levels to produce measurable absorption features in the data. Additionally, N(H$_2$) $\geq$ 25.0 cm$^{-2}$ will significantly saturate the features in the absorption spectra, which we do not see for any target in our survey. The thermal populations of H$_2$ are allowed to range from T(H$_2$) = 100 $-$ 5000 K. The H$_2$ populations must be warm enough to populate the correct rovibrational levels that absorb Ly$\alpha$ photons, while simultaneously cooler than the dissociation temperature of H$_2$ (T(H$_2$)$_{diss}$ $\approx$ 5000 K). 

For Models 1 and 2, MCMC simulations were run with 300 independent initial randomly-generated parameter realizations (walkers) and allowed to vary over 1000 steps to converge on the best representation of the observations.

\subsection{Model 3: Thermal H$_2$ Populations Photo-excited by HI-Ly$\alpha$}\label{model3}
Model 3 uses the same thermal populations of Models 1 and 2 and adds an additional photo-pumping mechanism to show how thermal populations reach an equilibrium state in the presence of an external radiation field. First, because we observe H$_2$ absorption against the Ly$\alpha$ wings of these targets and Ly$\alpha$ radiation makes up the vast majority of the FUV radiation that photo-excited H$_2$ to fluorescence, we assume the radiation pumping the thermal states to new equilibrium populations is dominated by Ly$\alpha$. To describe the amount of radiation being absorbed by H$_2$, we add two additional parameters that describe the flux input to the system, $F_n(\lambda)$ and $F_b(\lambda)$, which represent a narrow and broad flux component from the stellar Ly$\alpha$ radiation incident on the H$_2$ populations. Following the results and analysis from \citet{McJunkin+14}, we assume the Ly$\alpha$ radiation profile incident with the H$_2$ on the disk surface is described by two Gaussian components - a narrow component, where the bilk of the flux is located, and a broad component, which describes the shape of the observed outer wings. \citet{McJunkin+14} find full width at half maximum (FWHM) fits for both the narrow and broad Gaussian components of the radiation distribution, and we use those results to describe the width of our input flux. We allow the peak fluxes of both the narrow and broad flux distributions to vary and have final input Ly$\alpha$ flux distributions described by:
\begin{equation}
\begin{split}
	F_{\textnormal{Ly}\alpha}(\lambda) &= F_n \left( \lambda \right) + F_b \left( \lambda \right) \\
	 &=  F_n \exp\left( \frac{-\Delta\lambda ^2}{2\sigma_n^2} \right) + F_b \exp\left( \frac{-\Delta\lambda ^2}{2\sigma_b^2} \right),
	\label{eq:mod3}
\end{split}
\end{equation}
where $F_n$ and $F_b$ are free parameters in the models, and $\sigma_n$ and $\sigma_b$ are derived from the FWHM found in \citet{McJunkin+14}, and $\Delta\lambda$ = $\lambda - \lambda_{lab}$, where $\lambda_{lab}$ is the rest wavelength of HI-Ly$\alpha$ (1215.67 {\AA}). 
Each flux distribution is kept constant throughout the model run, assuming the output radiation from the star over the time it takes to equalize the photo-pumped populations of H$_2$ is isotropic. We assume $F_n(\lambda)$ and $F_b(\lambda)$ are observed flux rates, and we therefore infer the flux back to the star by reddening the flux with ISM extinction values determined by \citet{McJunkin+14}. The allowed ranges of observed total Ly$\alpha$ flux are outlined in \citet{Schindhelm+12b} for reconstructed Ly$\alpha$ profiles seen in at the PPDs. We constrain the Ly$\alpha$ flux to log$_{10}$($F_n(\lambda)$) = -13 to -5 ergs cm$^{-2}$ s$^{-1}$ and log$_{10}$($F_b(\lambda)$) = -16 to -5 ergs cm$^{-2}$ s$^{-1}$, which are integrated over each Gaussian function in Equation~\ref{eq:mod3}. 

\begin{figure}
\centering
\includegraphics[angle=270,width=0.9\textwidth]{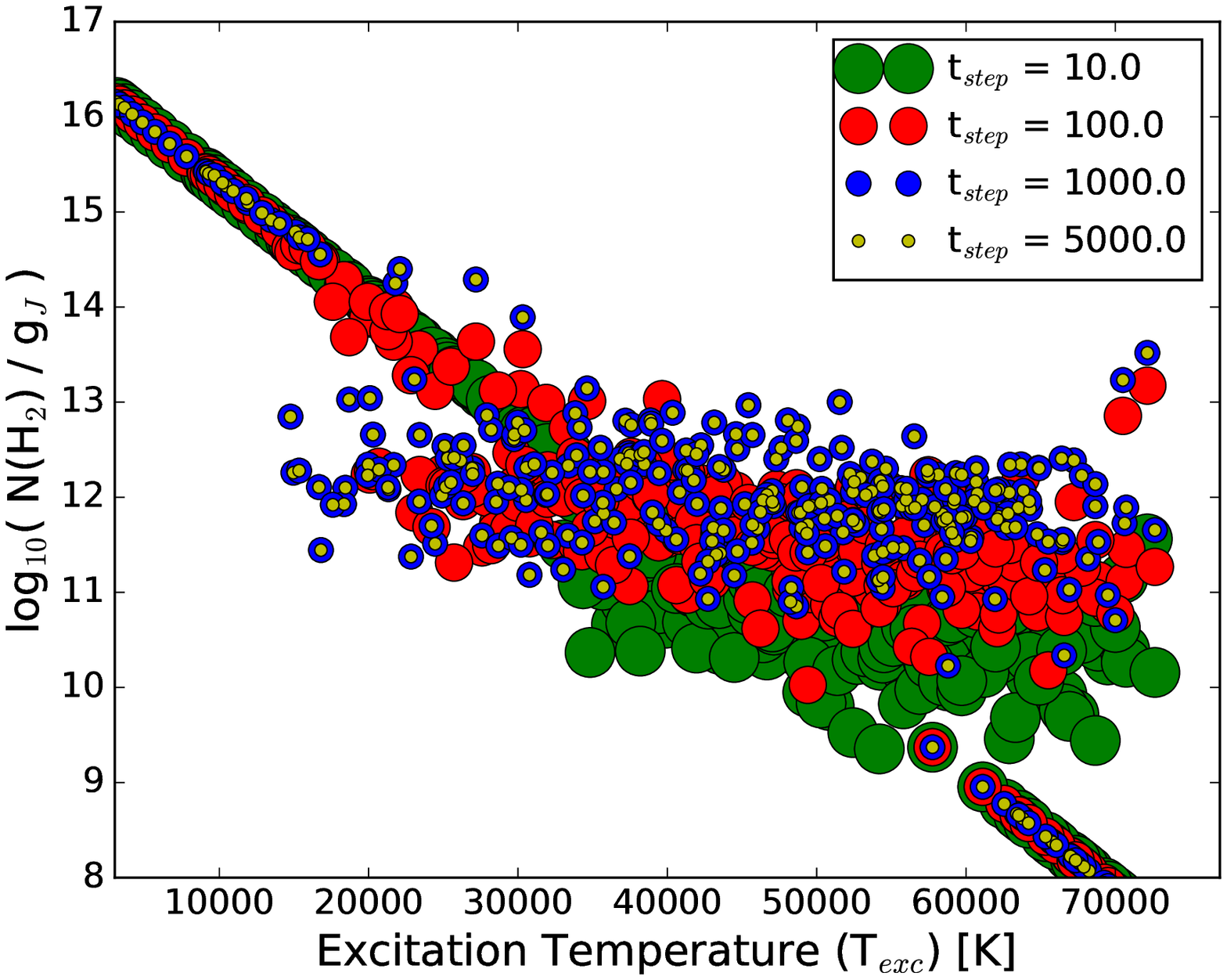}
\caption{Modeled thermal distributions of H$_2$ with N(H$_2$) = 10$^{17}$ cm$^{-2}$ and T(H$_2$) = 3500 K, assuming thermal populations of H$_2$ are constantly photo-excited by an external HI-Ly$\alpha$ radiation field to an equilibrium state. The Ly$\alpha$ radiation field is assumed to be Gaussian in shape when interacting with H$_2$ molecules, with a peak flux of 10$^{-9}$ ergs cm$^{-2}$ s$^{-1}$ at 1215.67 {\AA}. We show how the number of iterations of time the H$_2$ is exposed to the Ly$\alpha$ flux affects the distribution of H$_2$ ground states. We find that iterations t$_{step}$ = 1000 reaches a final equilibrium state. The thermal distribution + Ly$\alpha$ pumping mechanics are used to calculate theoretical H$_2$ populations in Model 3.     \label{h2pop+fluorescence}}
\end{figure}

Once flux and thermal H$_2$ population parameters are chosen, we follow the change in thermal populations in states [$v$,$J$] of H$_2$ being exposed to the pumping radiation in time iterations of the pumping process, t$_{step}$, where each t$_{step}$ is considered over some arbitrary $\Delta$t. First, we find how much H$_2$ in state [$v$,$J$] is lost to be pumped to some electronic excited state [$v^{\prime}$,$J^{\prime}$] because of the interaction with a discrete Ly$\alpha$ photon with wavelength $\lambda$. We determine how much H$_2$ is photo-excited by $\lambda$ by calculating the cross section for absorption of photon $\lambda$, given the transition probabilities of the H$_2$ rovibration levels. Once all [$v$,$J$] state losses via $\lambda$ absorption have been determined, we allow the excited state H$_2$ to fluoresce back to the ground state via the branching ratios, or transition probabilities, to some final ground state [$v^{\prime\prime}$,$J^{\prime\prime}$]. 

For this simple model, we assume that dissociation of H$_2$ molecules by Ly$\alpha$-pumping is negligible. As \citet{Dalgarno+70} describe, nearly all Ly$\alpha$-pumped excited states have bound de-exciation levels, such that transitions from the Lyman band are expected to have very low probabilities of dissociation. While there exist a handful of Werner-band transitions, which likely prose the greatest probability for molecular dissociation upon decay, this simple model does not contain a source term of H$_2$, such that we cannot control the formation of H$_2$ at any point in the model. To keep the modeled distributions of H$_2$ constant throughout the simulated experiment, therefore, we assume that all H$_2$ transitions result only in the decay of H$_2$ to arbitrary ground states, with no probability that H$_2$ dissociates via these fluorescence routes.

This process is repeated until the H$_2$ populations reach a steady-state equilibrium, such that the absorption out of state [$v$,$J$] equalizes with the cascade back to [$v^{\prime\prime}$,$J^{\prime\prime}$]. For T(H$_2$) $\leq$ 5000 K, this equilibrium is reached by t$_{step}$ = 1000. With higher N(H$_2$), we find that it takes more t$_{step}$ to reach equilibrium, but for N(H$_2$) $<$ 10$^{20}$ cm$^{-2}$ and high T(H$_2$), an equilibrium state is reached after t$_{step}$ $\approx$ 3000 steps. Figure~\ref{h2pop+fluorescence} shows how the distributions of thermal H$_2$ populations change when exposed to a constant Ly$\alpha$ flux from the host star, as a function of time steps from first exposure. If we assume the distributions of H$_2$ ground states are primarily affected by photo-pumping via Ly$\alpha$ photons and no other physical mechanisms to drive the populations to non-LTE states (collisions with other species, chemical evolution, etc), then equilibrium of H$_2$ states is reached fairly quickly and does not change from the final equilibrium state of populations.

We perform the same MCMC data-model reduction for Model 3 and the observed rotation diagrams. Model 3 required time iterations and, therefore, took longer to run. We ran two separate iteration of Model 3, the first MCMC simulation having 100 independent walkers varying over the parameter space iterating over 2000 steps with 1000 time iterations of the Ly$\alpha$-pumping. We determined that, after about 100 converging steps for each walker, we were able to settle into the best realization of the data. We also determined that longer time iterations were necessary to settle the Ly$\alpha$-pumping mechanism into equilibrium for larger column densities (N(H$_2$) $>$ 10$^{20}$ cm$^{-2}$) and temperatures (T(H$_2$) $>$ 4500 K). We ran a second iteration of MCMC simulations for Model 3 using t$_{step}$ = 5000 per model realization, with 100 independent MCMC walkers iterating over 500 steps to convergence. Because of the extensive computation time of Model 3 with 5000 time steps per model realization, we chose to cut the total number of convergence steps to keep the same number of walker realizations in the MCMC. 

Table~\ref{tab5} presents parameter results for all modeled H$_2$ thermal distributions. For all Model 3 realizations, $F_b(\lambda)$ has been excluded, since the majority of the integrated flux of Ly$\alpha$ is dominated by $F_n(\lambda)$ ($F_n(\lambda)$ $>>$ $F_b(\lambda)$). Figure~\ref{app:fig5} shows the best-fit, median model parameters for Model 3 with observed rovibration H$_2$ levels overplotted. We mark each modeled rovibration level with a green plus symbol, and we mark modeled rovibration levels which are probed in the observed Ly$\alpha$ wing(s) of the target with cyan crosses.

\figsetstart
\figsetnum{4}
\figsettitle{Fitting Ly$\alpha$-Pumping Models to Each H$_2$ Rotation Diagram}

\figsetgrpstart
\figsetgrpnum{4.1}
\figsetgrptitle{AA Tau
}
\figsetplot{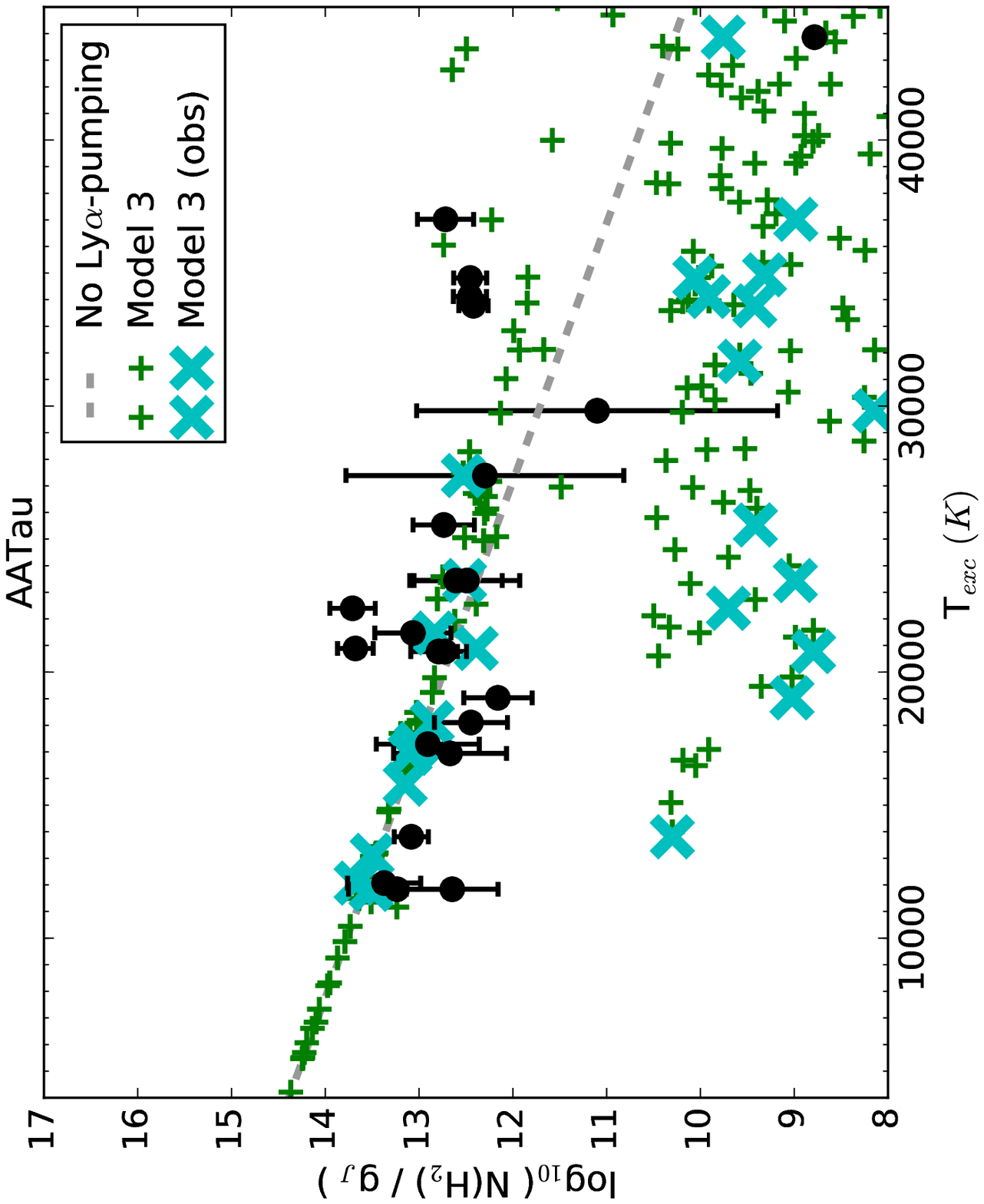}
\figsetgrpnote{}
\figsetgrpend

\figsetend

\begin{figure}
\figurenum{4}
\centering
\includegraphics[angle=270,width=0.9\textwidth]{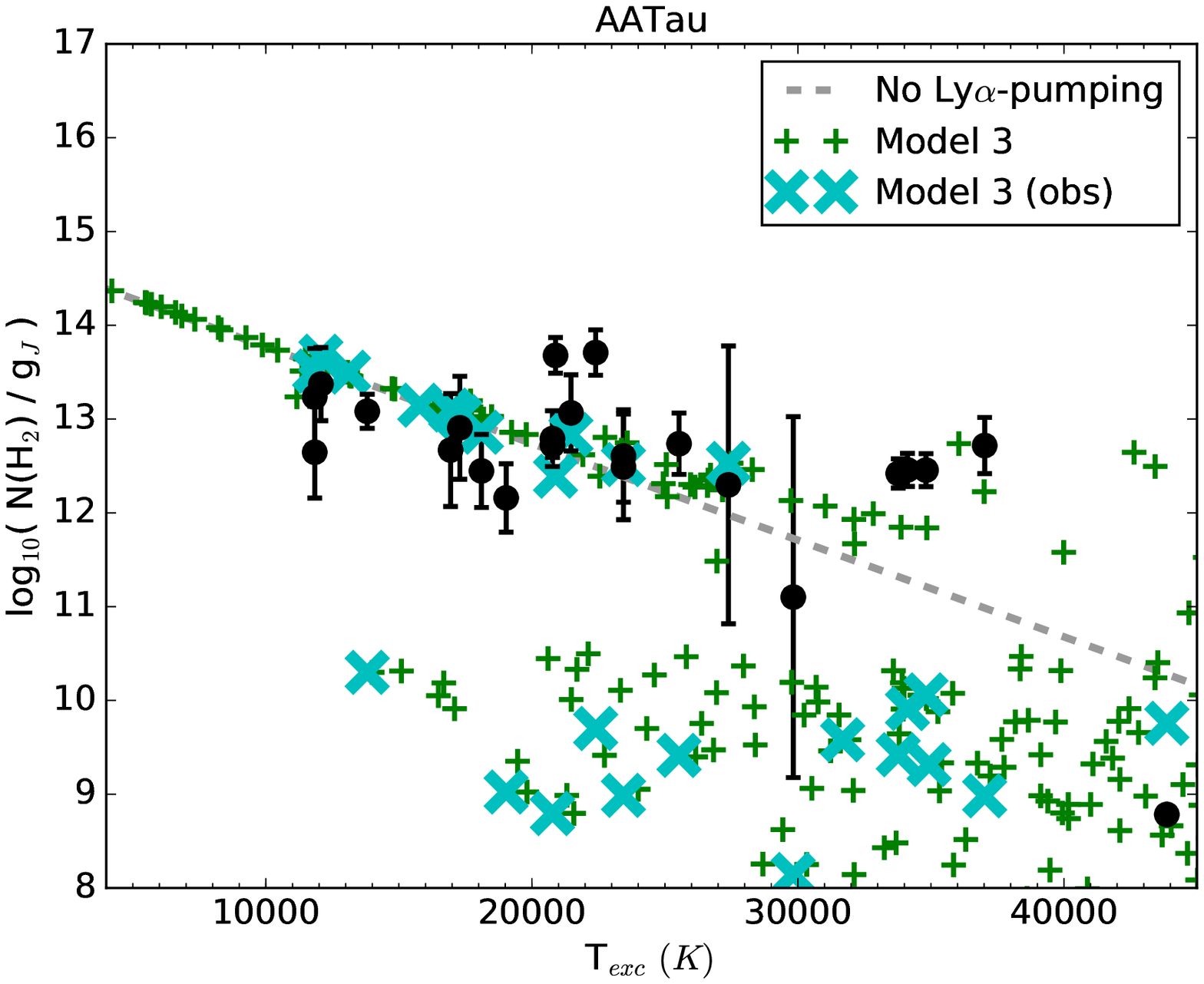}
\caption{All rotation diagrams are presented here. Each H$_2$ ground state column density is weighed by its statistical weight, g$_J$. Model 1 attempts to fit one thermally-populated bulk H$_2$ population through all data points extracted from the \textit{HST} data sets. Model 2 does the same as Model 1, but only for H$_2$ states with lower energy ground states (E$_{gr}$ $\leq$ 1.5 eV; T$_{exc}$ $\leq$ 17,500 K). } \label{app:fig4}
\end{figure}


\begin{thebibliography}{172}

\bibitem[{{Abgrall} {et~al.}(1993{\natexlab{a}}){Abgrall}, {Roueff}, {Launay},
  {Roncin}, \& {Subtil}}]{Abgrall+93}
{Abgrall}, H., {Roueff}, E., {Launay}, F., {Roncin}, J.~Y., \& {Subtil}, J.~L.
  1993{\natexlab{a}}, \aaps, 101, 273

\bibitem[{{Abgrall} {et~al.}(1993{\natexlab{b}}){Abgrall}, {Roueff}, {Launay},
  {Roncin}, \& {Subtil}}]{Abgrall+93b}
---. 1993{\natexlab{b}}, \aaps, 101, 323

\bibitem[{{{\'A}d{\'a}mkovics} {et~al.}(2016){{\'A}d{\'a}mkovics}, {Najita}, \&
  {Glassgold}}]{Adamkovics+16}
{{\'A}d{\'a}mkovics}, M., {Najita}, J.~R., \& {Glassgold}, A.~E. 2016, \apj,
  817, 82

\bibitem[{{Aikawa} \& {Nomura}(2006)}]{Aikawa+Nomura+06}
{Aikawa}, Y. \& {Nomura}, H. 2006, \apj, 642, 1152

\bibitem[{{Akeson} {et~al.}(2002){Akeson}, {Ciardi}, {van Belle}, \&
  {Creech-Eakman}}]{Akeson+02}
{Akeson}, R.~L., {Ciardi}, D.~R., {van Belle}, G.~T., \& {Creech-Eakman}, M.~J.
  2002, \apj, 566, 1124

\bibitem[{{Alencar} {et~al.}(2003){Alencar}, {Melo}, {Dullemond}, {Andersen},
  {Batalha}, {Vaz}, \& {Mathieu}}]{Alencar+03}
{Alencar}, S.~H.~P., {Melo}, C.~H.~F., {Dullemond}, C.~P., {Andersen}, J.,
  {Batalha}, C., {Vaz}, L.~P.~R., \& {Mathieu}, R.~D. 2003, \aap, 409, 1037

\bibitem[{{Alexander} {et~al.}(2014){Alexander}, {Pascucci}, {Andrews},
  {Armitage}, \& {Cieza}}]{Alexander+14}
{Alexander}, R., {Pascucci}, I., {Andrews}, S., {Armitage}, P., \& {Cieza}, L.
  2014, Protostars and Planets VI, 475

\bibitem[{{Alexander} \& {Armitage}(2007)}]{Alexander+Armitage+07}
{Alexander}, R.~D. \& {Armitage}, P.~J. 2007, \mnras, 375, 500

\bibitem[{{Alexander} {et~al.}(2006){Alexander}, {Clarke}, \&
  {Pringle}}]{Alexander+06}
{Alexander}, R.~D., {Clarke}, C.~J., \& {Pringle}, J.~E. 2006, \mnras, 369, 229

\bibitem[{{Andrews} \& {Williams}(2007)}]{Andrews+Williams+07}
{Andrews}, S.~M. \& {Williams}, J.~P. 2007, \apj, 671, 1800

\bibitem[{{Andrews} {et~al.}(2011){Andrews}, {Wilner}, {Espaillat}, {Hughes},
  {Dullemond}, {McClure}, {Qi}, \& {Brown}}]{Andrews+11}
{Andrews}, S.~M., {Wilner}, D.~J., {Espaillat}, C., {Hughes}, A.~M.,
  {Dullemond}, C.~P., {McClure}, M.~K., {Qi}, C., \& {Brown}, J.~M. 2011, \apj,
  732, 42

\bibitem[{{Ardila} {et~al.}(2002){Ardila}, {Basri}, {Walter}, {Valenti}, \&
  {Johns-Krull}}]{Ardila+02}
{Ardila}, D.~R., {Basri}, G., {Walter}, F.~M., {Valenti}, J.~A., \&
  {Johns-Krull}, C.~M. 2002, \apj, 566, 1100

\bibitem[{{Ardila} {et~al.}(2013){Ardila}, {Herczeg}, {Gregory}, {Ingleby},
  {France}, {Brown}, {Edwards}, {Johns-Krull}, {Linsky}, {Yang}, {Valenti},
  {Abgrall}, {Alexander}, {Bergin}, {Bethell}, {Brown}, {Calvet}, {Espaillat},
  {Hillenbrand}, {Hussain}, {Roueff}, {Schindhelm}, \& {Walter}}]{Ardila+13}
{Ardila}, D.~R., {Herczeg}, G.~J., {Gregory}, S.~G., {Ingleby}, L., {France},
  K., {Brown}, A., {Edwards}, S., {Johns-Krull}, C., {Linsky}, J.~L., {Yang},
  H., {Valenti}, J.~A., {Abgrall}, H., {Alexander}, R.~D., {Bergin}, E.,
  {Bethell}, T., {Brown}, J.~M., {Calvet}, N., {Espaillat}, C., {Hillenbrand},
  L.~A., {Hussain}, G., {Roueff}, E., {Schindhelm}, E.~R., \& {Walter}, F.~M.
  2013, \apjs, 207, 1

\bibitem[{{Armitage} {et~al.}(2003){Armitage}, {Clarke}, \&
  {Palla}}]{Armitage+03}
{Armitage}, P.~J., {Clarke}, C.~J., \& {Palla}, F. 2003, \mnras, 342, 1139

\bibitem[{{Arulanantham} {et~al.}(2016){Arulanantham}, {Herbst}, {Gilmore},
  {Cauley}, \& {Leggett}}]{Arulanantham+16}
{Arulanantham}, N.~A., {Herbst}, W., {Gilmore}, M.~S., {Cauley}, P.~W., \&
  {Leggett}, S.~K. 2016, ArXiv e-prints

\bibitem[{{Ayliffe} \& {Bate}(2010)}]{Ayliffe+Bate+10}
{Ayliffe}, B.~A. \& {Bate}, M.~R. 2010, \mnras, 408, 876

\bibitem[{{Ayres}(2010)}]{Ayres+10}
{Ayres}, T.~R. 2010, VizieR Online Data Catalog, 218

\bibitem[{{Bai}(2016)}]{Bai+16}
{Bai}, X.-N. 2016, \apj, 821, 80

\bibitem[{{Banzatti} \& {Pontoppidan}(2015)}]{Banzatti+15}
{Banzatti}, A. \& {Pontoppidan}, K.~M. 2015, \apj, 809, 167

\bibitem[{{Bary} {et~al.}(2003){Bary}, {Weintraub}, \& {Kastner}}]{Bary+03}
{Bary}, J.~S., {Weintraub}, D.~A., \& {Kastner}, J.~H. 2003, \apj, 586, 1136

\bibitem[{{Beck} {et~al.}(2012){Beck}, {Bary}, {Dutrey}, {Pi{\'e}tu},
  {Guilloteau}, {Lubow}, \& {Simon}}]{Beck+12}
{Beck}, T.~L., {Bary}, J.~S., {Dutrey}, A., {Pi{\'e}tu}, V., {Guilloteau}, S.,
  {Lubow}, S.~H., \& {Simon}, M. 2012, \apj, 754, 72

\bibitem[{{Beckwith} {et~al.}(1983){Beckwith}, {Evans}, {Gatley}, {Gull}, \&
  {Russell}}]{Beckwith+83}
{Beckwith}, S., {Evans}, II, N.~J., {Gatley}, I., {Gull}, G., \& {Russell},
  R.~W. 1983, \apj, 264, 152

\bibitem[{{Beckwith} {et~al.}(1978){Beckwith}, {Gatley}, {Matthews}, \&
  {Neugebauer}}]{Beckwith+78}
{Beckwith}, S., {Gatley}, I., {Matthews}, K., \& {Neugebauer}, G. 1978, \apjl,
  223, L41

\bibitem[{{Bergin} {et~al.}(2004){Bergin}, {Calvet}, {Sitko}, {Abgrall},
  {D'Alessio}, {Herczeg}, {Roueff}, {Qi}, {Lynch}, {Russell}, {Brafford}, \&
  {Perry}}]{Bergin+04}
{Bergin}, E., {Calvet}, N., {Sitko}, M.~L., {Abgrall}, H., {D'Alessio}, P.,
  {Herczeg}, G.~J., {Roueff}, E., {Qi}, C., {Lynch}, D.~K., {Russell}, R.~W.,
  {Brafford}, S.~M., \& {Perry}, R.~B. 2004, \apjl, 614, L133

\bibitem[{{Bertout} {et~al.}(1988){Bertout}, {Basri}, \&
  {Bouvier}}]{Bertout+88}
{Bertout}, C., {Basri}, G., \& {Bouvier}, J. 1988, \apj, 330, 350

\bibitem[{{Bertout} {et~al.}(1999){Bertout}, {Robichon}, \&
  {Arenou}}]{Bertout+99}
{Bertout}, C., {Robichon}, N., \& {Arenou}, F. 1999, \aap, 352, 574

\bibitem[{{Bethell} \& {Bergin}(2011)}]{Bethell+Bergin+11}
{Bethell}, T.~J. \& {Bergin}, E.~A. 2011, \apj, 739, 78

\bibitem[{{Black} \& {Dalgarno}(1977)}]{Black+77}
{Black}, J.~H. \& {Dalgarno}, A. 1977, \apjs, 34, 405

\bibitem[{{B{\"o}hm} {et~al.}(2004){B{\"o}hm}, {Catala}, {Balona}, \&
  {Carter}}]{Bohm+04}
{B{\"o}hm}, T., {Catala}, C., {Balona}, L., \& {Carter}, B. 2004, \aap, 427,
  907

\bibitem[{{Bouvier} {et~al.}(1999){Bouvier}, {Chelli}, {Allain}, {Carrasco},
  {Costero}, {Cruz-Gonzalez}, {Dougados}, {Fern{\'a}ndez}, {Mart{\'{\i}}n},
  {M{\'e}nard}, {Mennessier}, {Mujica}, {Recillas}, {Salas}, {Schmidt}, \&
  {Wichmann}}]{Bouvier+99}
{Bouvier}, J., {Chelli}, A., {Allain}, S., {Carrasco}, L., {Costero}, R.,
  {Cruz-Gonzalez}, I., {Dougados}, C., {Fern{\'a}ndez}, M., {Mart{\'{\i}}n},
  E.~L., {M{\'e}nard}, F., {Mennessier}, C., {Mujica}, R., {Recillas}, E.,
  {Salas}, L., {Schmidt}, G., \& {Wichmann}, R. 1999, \aap, 349, 619

\bibitem[{{Brown} {et~al.}(2009){Brown}, {Blake}, {Qi}, {Dullemond}, {Wilner},
  \& {Williams}}]{Brown+09}
{Brown}, J.~M., {Blake}, G.~A., {Qi}, C., {Dullemond}, C.~P., {Wilner}, D.~J.,
  \& {Williams}, J.~P. 2009, \apj, 704, 496

\bibitem[{{Brown} {et~al.}(2013){Brown}, {Pontoppidan}, {van Dishoeck},
  {Herczeg}, {Blake}, \& {Smette}}]{Brown+13}
{Brown}, J.~M., {Pontoppidan}, K.~M., {van Dishoeck}, E.~F., {Herczeg}, G.~J.,
  {Blake}, G.~A., \& {Smette}, A. 2013, \apj, 770, 94

\bibitem[{{Calvet} {et~al.}(2002){Calvet}, {D'Alessio}, {Hartmann}, {Wilner},
  {Walsh}, \& {Sitko}}]{Calvet+02}
{Calvet}, N., {D'Alessio}, P., {Hartmann}, L., {Wilner}, D., {Walsh}, A., \&
  {Sitko}, M. 2002, \apj, 568, 1008

\bibitem[{{Calvet} {et~al.}(2005){Calvet}, {D'Alessio}, {Watson},
  {Franco-Hern{\'a}ndez}, {Furlan}, {Green}, {Sutter}, {Forrest}, {Hartmann},
  {Uchida}, {Keller}, {Sargent}, {Najita}, {Herter}, {Barry}, \&
  {Hall}}]{Calvet+05}
{Calvet}, N., {D'Alessio}, P., {Watson}, D.~M., {Franco-Hern{\'a}ndez}, R.,
  {Furlan}, E., {Green}, J., {Sutter}, P.~M., {Forrest}, W.~J., {Hartmann}, L.,
  {Uchida}, K.~I., {Keller}, L.~D., {Sargent}, B., {Najita}, J., {Herter},
  T.~L., {Barry}, D.~J., \& {Hall}, P. 2005, \apjl, 630, L185

\bibitem[{{Cartwright} \& {Drapatz}(1970)}]{Cartwright+70}
{Cartwright}, D.~C. \& {Drapatz}, S. 1970, \aap, 4, 443

\bibitem[{{Coffey} {et~al.}(2004){Coffey}, {Bacciotti}, {Woitas}, {Ray}, \&
  {Eisl{\"o}ffel}}]{Coffey+04}
{Coffey}, D., {Bacciotti}, F., {Woitas}, J., {Ray}, T.~P., \& {Eisl{\"o}ffel},
  J. 2004, \apj, 604, 758

\bibitem[{{Comer{\'o}n} \& {Fern{\'a}ndez}(2010)}]{Comeron+Fernandez+10}
{Comer{\'o}n}, F. \& {Fern{\'a}ndez}, M. 2010, \aap, 511, A10

\bibitem[{{Comer{\'o}n} {et~al.}(2003){Comer{\'o}n}, {Fern{\'a}ndez},
  {Baraffe}, {Neuh{\"a}user}, \& {Kaas}}]{Comeron+03}
{Comer{\'o}n}, F., {Fern{\'a}ndez}, M., {Baraffe}, I., {Neuh{\"a}user}, R., \&
  {Kaas}, A.~A. 2003, \aap, 406, 1001

\bibitem[{{Correia} {et~al.}(2006){Correia}, {Zinnecker}, {Ratzka}, \&
  {Sterzik}}]{Correia+06}
{Correia}, S., {Zinnecker}, H., {Ratzka}, T., \& {Sterzik}, M.~F. 2006, \aap,
  459, 909

\bibitem[{{Dalgarno} \& {Stephens}(1970)}]{Dalgarno+70}
{Dalgarno}, A. \& {Stephens}, T.~L. 1970, \apjl, 160, L107

\bibitem[{{Danforth} {et~al.}(2010){Danforth}, {Stocke}, \&
  {Shull}}]{Danforth+10}
{Danforth}, C.~W., {Stocke}, J.~T., \& {Shull}, J.~M. 2010, \apj, 710, 613

\bibitem[{{Dodson-Robinson} \& {Salyk}(2011)}]{Dodson-Robinson+Salyk+11}
{Dodson-Robinson}, S.~E. \& {Salyk}, C. 2011, \apj, 738, 131

\bibitem[{{Donehew} \& {Brittain}(2011)}]{Donehew+Brittain+11}
{Donehew}, B. \& {Brittain}, S. 2011, \aj, 141, 46

\bibitem[{{Dong} {et~al.}(2015){Dong}, {Zhu}, \& {Whitney}}]{Dong+15}
{Dong}, R., {Zhu}, Z., \& {Whitney}, B. 2015, \apj, 809, 93

\bibitem[{{Draine}(2011)}]{Draine+11}
{Draine}, B.~T. 2011, {Physics of the Interstellar and Intergalactic Medium}

\bibitem[{{Draine} \& {Bertoldi}(1996)}]{Draine+96}
{Draine}, B.~T. \& {Bertoldi}, F. 1996, \apj, 468, 269

\bibitem[{{Du} \& {Bergin}(2014)}]{Du+Bergin+14}
{Du}, F. \& {Bergin}, E.~A. 2014, \apj, 792, 2

\bibitem[{{Dullemond} {et~al.}(2007){Dullemond}, {Hollenbach}, {Kamp}, \&
  {D'Alessio}}]{Dullemond+07}
{Dullemond}, C.~P., {Hollenbach}, D., {Kamp}, I., \& {D'Alessio}, P. 2007,
  Protostars and Planets V, 555

\bibitem[{{Dullemond} \& {Monnier}(2010)}]{Dullemond+Monnier+10}
{Dullemond}, C.~P. \& {Monnier}, J.~D. 2010, \araa, 48, 205

\bibitem[{{Eisner} {et~al.}(2007){Eisner}, {Hillenbrand}, {White}, {Bloom},
  {Akeson}, \& {Blake}}]{Eisner+07}
{Eisner}, J.~A., {Hillenbrand}, L.~A., {White}, R.~J., {Bloom}, J.~S.,
  {Akeson}, R.~L., \& {Blake}, C.~H. 2007, \apj, 669, 1072

\bibitem[{{Espaillat} {et~al.}(2007){Espaillat}, {Calvet}, {D'Alessio},
  {Bergin}, {Hartmann}, {Watson}, {Furlan}, {Najita}, {Forrest}, {McClure},
  {Sargent}, {Bohac}, \& {Harrold}}]{Espaillat+07a}
{Espaillat}, C., {Calvet}, N., {D'Alessio}, P., {Bergin}, E., {Hartmann}, L.,
  {Watson}, D., {Furlan}, E., {Najita}, J., {Forrest}, W., {McClure}, M.,
  {Sargent}, B., {Bohac}, C., \& {Harrold}, S.~T. 2007, \apjl, 664, L111

\bibitem[{{Espaillat} {et~al.}(2011){Espaillat}, {Furlan}, {D'Alessio},
  {Sargent}, {Nagel}, {Calvet}, {Watson}, \& {Muzerolle}}]{Espaillat+11}
{Espaillat}, C., {Furlan}, E., {D'Alessio}, P., {Sargent}, B., {Nagel}, E.,
  {Calvet}, N., {Watson}, D.~M., \& {Muzerolle}, J. 2011, \apj, 728, 49

\bibitem[{{Fedele} {et~al.}(2010){Fedele}, {van den Ancker}, {Henning},
  {Jayawardhana}, \& {Oliveira}}]{Fedele+10}
{Fedele}, D., {van den Ancker}, M.~E., {Henning}, T., {Jayawardhana}, R., \&
  {Oliveira}, J.~M. 2010, \aap, 510, A72

\bibitem[{{Feigelson} {et~al.}(2003){Feigelson}, {Lawson}, \&
  {Garmire}}]{Feigelson+03}
{Feigelson}, E.~D., {Lawson}, W.~A., \& {Garmire}, G.~P. 2003, \apj, 599, 1207

\bibitem[{{Ferreira} {et~al.}(2006){Ferreira}, {Dougados}, \&
  {Cabrit}}]{Ferreira+06}
{Ferreira}, J., {Dougados}, C., \& {Cabrit}, S. 2006, \aap, 453, 785

\bibitem[{{Fleming} {et~al.}(2010){Fleming}, {France}, {Lupu}, \&
  {McCandliss}}]{Fleming+10}
{Fleming}, B., {France}, K., {Lupu}, R.~E., \& {McCandliss}, S.~R. 2010, \apj,
  725, 159

\bibitem[{{Foreman-Mackey} {et~al.}(2013){Foreman-Mackey}, {Hogg}, {Lang}, \&
  {Goodman}}]{Foreman+12}
{Foreman-Mackey}, D., {Hogg}, D.~W., {Lang}, D., \& {Goodman}, J. 2013, \pasp,
  125, 306

\bibitem[{{France} {et~al.}(2012{\natexlab{a}}){France}, {Burgh}, {Herczeg},
  {Schindhelm}, {Yang}, {Abgrall}, {Roueff}, {Brown}, {Brown}, \&
  {Linsky}}]{France+12a}
{France}, K., {Burgh}, E.~B., {Herczeg}, G.~J., {Schindhelm}, E., {Yang}, H.,
  {Abgrall}, H., {Roueff}, E., {Brown}, A., {Brown}, J.~M., \& {Linsky}, J.~L.
  2012{\natexlab{a}}, \apj, 744, 22

\bibitem[{{France} {et~al.}(2014{\natexlab{a}}){France}, {Herczeg}, {McJunkin},
  \& {Penton}}]{France+14b}
{France}, K., {Herczeg}, G.~J., {McJunkin}, M., \& {Penton}, S.~V.
  2014{\natexlab{a}}, \apj, 794, 160

\bibitem[{{France} {et~al.}(2017){France}, {Roueff}, \& {Abgrall}}]{France+17}
{France}, K., {Roueff}, E., \& {Abgrall}, H. 2017, \apj, \textit{In
  Preparation}

\bibitem[{{France} {et~al.}(2014{\natexlab{b}}){France}, {Schindhelm},
  {Bergin}, {Roueff}, \& {Abgrall}}]{France+14}
{France}, K., {Schindhelm}, E., {Bergin}, E.~A., {Roueff}, E., \& {Abgrall}, H.
  2014{\natexlab{b}}, \apj, 784, 127

\bibitem[{{France} {et~al.}(2011){France}, {Schindhelm}, {Burgh}, {Herczeg},
  {Harper}, {Brown}, {Green}, {Linsky}, {Yang}, {Abgrall}, {Ardila}, {Bergin},
  {Bethell}, {Brown}, {Calvet}, {Espaillat}, {Gregory}, {Hillenbrand},
  {Hussain}, {Ingleby}, {Johns-Krull}, {Roueff}, {Valenti}, \&
  {Walter}}]{France+11}
{France}, K., {Schindhelm}, E., {Burgh}, E.~B., {Herczeg}, G.~J., {Harper},
  G.~M., {Brown}, A., {Green}, J.~C., {Linsky}, J.~L., {Yang}, H., {Abgrall},
  H., {Ardila}, D.~R., {Bergin}, E., {Bethell}, T., {Brown}, J.~M., {Calvet},
  N., {Espaillat}, C., {Gregory}, S.~G., {Hillenbrand}, L.~A., {Hussain}, G.,
  {Ingleby}, L., {Johns-Krull}, C.~M., {Roueff}, E., {Valenti}, J.~A., \&
  {Walter}, F.~M. 2011, \apj, 734, 31

\bibitem[{{France} {et~al.}(2012{\natexlab{b}}){France}, {Schindhelm},
  {Herczeg}, {Brown}, {Abgrall}, {Alexander}, {Bergin}, {Brown}, {Linsky},
  {Roueff}, \& {Yang}}]{France+12b}
{France}, K., {Schindhelm}, E., {Herczeg}, G.~J., {Brown}, A., {Abgrall}, H.,
  {Alexander}, R.~D., {Bergin}, E.~A., {Brown}, J.~M., {Linsky}, J.~L.,
  {Roueff}, E., \& {Yang}, H. 2012{\natexlab{b}}, \apj, 756, 171

\bibitem[{{Frisch} {et~al.}(1999){Frisch}, {Dorschner}, {Geiss}, {Greenberg},
  {Gr{\"u}n}, {Landgraf}, {Hoppe}, {Jones}, {Kr{\"a}tschmer}, {Linde},
  {Morfill}, {Reach}, {Slavin}, {Svestka}, {Witt}, \& {Zank}}]{Frisch+99}
{Frisch}, P.~C., {Dorschner}, J.~M., {Geiss}, J., {Greenberg}, J.~M.,
  {Gr{\"u}n}, E., {Landgraf}, M., {Hoppe}, P., {Jones}, A.~P.,
  {Kr{\"a}tschmer}, W., {Linde}, T.~J., {Morfill}, G.~E., {Reach}, W.,
  {Slavin}, J.~D., {Svestka}, J., {Witt}, A.~N., \& {Zank}, G.~P. 1999, \apj,
  525, 492

\bibitem[{{Furlan} {et~al.}(2009){Furlan}, {Watson}, {McClure}, {Manoj},
  {Espaillat}, {D'Alessio}, {Calvet}, {Kim}, {Sargent}, {Forrest}, \&
  {Hartmann}}]{Furlan+09}
{Furlan}, E., {Watson}, D.~M., {McClure}, M.~K., {Manoj}, P., {Espaillat}, C.,
  {D'Alessio}, P., {Calvet}, N., {Kim}, K.~H., {Sargent}, B.~A., {Forrest},
  W.~J., \& {Hartmann}, L. 2009, \apj, 703, 1964

\bibitem[{{Garcia Lopez} {et~al.}(2006){Garcia Lopez}, {Natta}, {Testi}, \&
  {Habart}}]{GarciaLopez+06}
{Garcia Lopez}, R., {Natta}, A., {Testi}, L., \& {Habart}, E. 2006, \aap, 459,
  837

\bibitem[{{Ghez} {et~al.}(1993){Ghez}, {Neugebauer}, \& {Matthews}}]{Ghez+93}
{Ghez}, A.~M., {Neugebauer}, G., \& {Matthews}, K. 1993, \aj, 106, 2005

\bibitem[{{Glassgold} {et~al.}(2004){Glassgold}, {Najita}, \&
  {Igea}}]{Glassgold+04}
{Glassgold}, A.~E., {Najita}, J., \& {Igea}, J. 2004, \apj, 615, 972

\bibitem[{{Glassgold} \& {Najita}(2001)}]{Glassgold+Najita+01}
{Glassgold}, A.~E. \& {Najita}, J.~R. 2001, in Astronomical Society of the
  Pacific Conference Series, Vol. 244, Young Stars Near Earth: Progress and
  Prospects, ed. R.~{Jayawardhana} \& T.~{Greene}, 251

\bibitem[{{Glassgold} \& {Najita}(2015)}]{Glassgold+Najita+15}
{Glassgold}, A.~E. \& {Najita}, J.~R. 2015, \apj, 810, 125

\bibitem[{{G{\'o}mez de Castro}(2009)}]{GomezdeCastro+09}
{G{\'o}mez de Castro}, A.~I. 2009, \apjl, 698, L108

\bibitem[{{Gorti} \& {Hollenbach}(2004)}]{Gorti+Hollenbach+04}
{Gorti}, U. \& {Hollenbach}, D. 2004, \apj, 613, 424

\bibitem[{{Gorti} \& {Hollenbach}(2009)}]{Gorti+Hollenbach+09}
---. 2009, \apj, 690, 1539

\bibitem[{{Gorti} {et~al.}(2015){Gorti}, {Hollenbach}, \&
  {Dullemond}}]{Gorti+15}
{Gorti}, U., {Hollenbach}, D., \& {Dullemond}, C.~P. 2015, \apj, 804, 29

\bibitem[{{Grady} {et~al.}(2009){Grady}, {Schneider}, {Sitko}, {Williger},
  {Hamaguchi}, {Brittain}, {Ablordeppey}, {Apai}, {Beerman}, {Carpenter},
  {Collins}, {Fukagawa}, {Hammel}, {Henning}, {Hines}, {Kimes}, {Lynch},
  {M{\'e}nard}, {Pearson}, {Russell}, {Silverstone}, {Smith}, {Troutman},
  {Wilner}, {Woodgate}, \& {Clampin}}]{Grady+09}
{Grady}, C.~A., {Schneider}, G., {Sitko}, M.~L., {Williger}, G.~M.,
  {Hamaguchi}, K., {Brittain}, S.~D., {Ablordeppey}, K., {Apai}, D., {Beerman},
  L., {Carpenter}, W.~J., {Collins}, K.~A., {Fukagawa}, M., {Hammel}, H.~B.,
  {Henning}, T., {Hines}, D., {Kimes}, R., {Lynch}, D.~K., {M{\'e}nard}, F.,
  {Pearson}, R., {Russell}, R.~W., {Silverstone}, M., {Smith}, P.~S.,
  {Troutman}, M., {Wilner}, D., {Woodgate}, B., \& {Clampin}, M. 2009, \apj,
  699, 1822

\bibitem[{{Grady} {et~al.}(2004){Grady}, {Woodgate}, {Torres}, {Henning},
  {Apai}, {Rodmann}, {Wang}, {Stecklum}, {Linz}, {Williger}, {Brown},
  {Wilkinson}, {Harper}, {Herczeg}, {Danks}, {Vieira}, {Malumuth}, {Collins},
  \& {Hill}}]{Grady+04}
{Grady}, C.~A., {Woodgate}, B., {Torres}, C.~A.~O., {Henning}, T., {Apai}, D.,
  {Rodmann}, J., {Wang}, H., {Stecklum}, B., {Linz}, H., {Williger}, G.~M.,
  {Brown}, A., {Wilkinson}, E., {Harper}, G.~M., {Herczeg}, G.~J., {Danks}, A.,
  {Vieira}, G.~L., {Malumuth}, E., {Collins}, N.~R., \& {Hill}, R.~S. 2004,
  \apj, 608, 809

\bibitem[{{Green} {et~al.}(2012){Green}, {Froning}, {Osterman}, {Ebbets},
  {Heap}, {Leitherer}, {Linsky}, {Savage}, {Sembach}, {Shull}, {Siegmund},
  {Snow}, {Spencer}, {Stern}, {Stocke}, {Welsh}, {B{\'e}land}, {Burgh},
  {Danforth}, {France}, {Keeney}, {McPhate}, {Penton}, {Andrews},
  {Brownsberger}, {Morse}, \& {Wilkinson}}]{Green+12}
{Green}, J.~C., {Froning}, C.~S., {Osterman}, S., {Ebbets}, D., {Heap}, S.~H.,
  {Leitherer}, C., {Linsky}, J.~L., {Savage}, B.~D., {Sembach}, K., {Shull},
  J.~M., {Siegmund}, O.~H.~W., {Snow}, T.~P., {Spencer}, J., {Stern}, S.~A.,
  {Stocke}, J., {Welsh}, B., {B{\'e}land}, S., {Burgh}, E.~B., {Danforth}, C.,
  {France}, K., {Keeney}, B., {McPhate}, J., {Penton}, S.~V., {Andrews}, J.,
  {Brownsberger}, K., {Morse}, J., \& {Wilkinson}, E. 2012, \apj, 744, 60

\bibitem[{{G{\"u}del} {et~al.}(2007){G{\"u}del}, {Skinner}, {Mel'Nikov},
  {Audard}, {Telleschi}, \& {Briggs}}]{Gudel+07}
{G{\"u}del}, M., {Skinner}, S.~L., {Mel'Nikov}, S.~Y., {Audard}, M.,
  {Telleschi}, A., \& {Briggs}, K.~R. 2007, \aap, 468, 529

\bibitem[{{Gullbring} {et~al.}(2000){Gullbring}, {Calvet}, {Muzerolle}, \&
  {Hartmann}}]{Gullbring+00}
{Gullbring}, E., {Calvet}, N., {Muzerolle}, J., \& {Hartmann}, L. 2000, \apj,
  544, 927

\bibitem[{{Gullbring} {et~al.}(1998){Gullbring}, {Hartmann}, {Brice{\~n}o}, \&
  {Calvet}}]{Gullbring+98}
{Gullbring}, E., {Hartmann}, L., {Brice{\~n}o}, C., \& {Calvet}, N. 1998, \apj,
  492, 323

\bibitem[{{Habart} {et~al.}(2004){Habart}, {Boulanger}, {Verstraete},
  {Walmsley}, \& {Pineau des For{\^e}ts}}]{Habart+04}
{Habart}, E., {Boulanger}, F., {Verstraete}, L., {Walmsley}, C.~M., \& {Pineau
  des For{\^e}ts}, G. 2004, \aap, 414, 531

\bibitem[{{Hartigan} {et~al.}(1995){Hartigan}, {Edwards}, \&
  {Ghandour}}]{Hartigan+95}
{Hartigan}, P., {Edwards}, S., \& {Ghandour}, L. 1995, \apj, 452, 736

\bibitem[{{Hartmann} {et~al.}(1998){Hartmann}, {Calvet}, {Gullbring}, \&
  {D'Alessio}}]{Hartmann+98}
{Hartmann}, L., {Calvet}, N., {Gullbring}, E., \& {D'Alessio}, P. 1998, \apj,
  495, 385

\bibitem[{{Hashimoto} {et~al.}(2011){Hashimoto}, {Tamura}, {Muto}, {Kudo},
  {Fukagawa}, {Fukue}, {Goto}, {Grady}, {Henning}, {Hodapp}, {Honda},
  {Inutsuka}, {Kokubo}, {Knapp}, {McElwain}, {Momose}, {Ohashi}, {Okamoto},
  {Takami}, {Turner}, {Wisniewski}, {Janson}, {Abe}, {Brandner}, {Carson},
  {Egner}, {Feldt}, {Golota}, {Guyon}, {Hayano}, {Hayashi}, {Hayashi}, {Ishii},
  {Kandori}, {Kusakabe}, {Matsuo}, {Mayama}, {Miyama}, {Morino}, {Moro-Martin},
  {Nishimura}, {Pyo}, {Suto}, {Suzuki}, {Takato}, {Terada}, {Thalmann},
  {Tomono}, {Watanabe}, {Yamada}, {Takami}, \& {Usuda}}]{Hashimoto+11}
{Hashimoto}, J., {Tamura}, M., {Muto}, T., {Kudo}, T., {Fukagawa}, M., {Fukue},
  T., {Goto}, M., {Grady}, C.~A., {Henning}, T., {Hodapp}, K., {Honda}, M.,
  {Inutsuka}, S., {Kokubo}, E., {Knapp}, G., {McElwain}, M.~W., {Momose}, M.,
  {Ohashi}, N., {Okamoto}, Y.~K., {Takami}, M., {Turner}, E.~L., {Wisniewski},
  J., {Janson}, M., {Abe}, L., {Brandner}, W., {Carson}, J., {Egner}, S.,
  {Feldt}, M., {Golota}, T., {Guyon}, O., {Hayano}, Y., {Hayashi}, M.,
  {Hayashi}, S., {Ishii}, M., {Kandori}, R., {Kusakabe}, N., {Matsuo}, T.,
  {Mayama}, S., {Miyama}, S., {Morino}, J.-I., {Moro-Martin}, A., {Nishimura},
  T., {Pyo}, T.-S., {Suto}, H., {Suzuki}, R., {Takato}, N., {Terada}, H.,
  {Thalmann}, C., {Tomono}, D., {Watanabe}, M., {Yamada}, T., {Takami}, H., \&
  {Usuda}, T. 2011, \apjl, 729, L17

\bibitem[{{Herczeg} \& {Hillenbrand}(2008)}]{Herczeg+Hillenbrand+08}
{Herczeg}, G.~J. \& {Hillenbrand}, L.~A. 2008, \apj, 681, 594

\bibitem[{{Herczeg} {et~al.}(2002){Herczeg}, {Linsky}, {Valenti},
  {Johns-Krull}, \& {Wood}}]{Herczeg+02}
{Herczeg}, G.~J., {Linsky}, J.~L., {Valenti}, J.~A., {Johns-Krull}, C.~M., \&
  {Wood}, B.~E. 2002, \apj, 572, 310

\bibitem[{{Herczeg} {et~al.}(2006){Herczeg}, {Linsky}, {Walter}, {Gahm}, \&
  {Johns-Krull}}]{Herczeg+06}
{Herczeg}, G.~J., {Linsky}, J.~L., {Walter}, F.~M., {Gahm}, G.~F., \&
  {Johns-Krull}, C.~M. 2006, \apjs, 165, 256

\bibitem[{{Herczeg} {et~al.}(2005){Herczeg}, {Walter}, {Linsky}, {Gahm},
  {Ardila}, {Brown}, {Johns-Krull}, {Simon}, \& {Valenti}}]{Herczeg+05}
{Herczeg}, G.~J., {Walter}, F.~M., {Linsky}, J.~L., {Gahm}, G.~F., {Ardila},
  D.~R., {Brown}, A., {Johns-Krull}, C.~M., {Simon}, M., \& {Valenti}, J.~A.
  2005, \aj, 129, 2777

\bibitem[{{Herczeg} {et~al.}(2004){Herczeg}, {Wood}, {Linsky}, {Valenti}, \&
  {Johns-Krull}}]{Herczeg+04}
{Herczeg}, G.~J., {Wood}, B.~E., {Linsky}, J.~L., {Valenti}, J.~A., \&
  {Johns-Krull}, C.~M. 2004, \apj, 607, 369

\bibitem[{{Hern{\'a}ndez} {et~al.}(2007){Hern{\'a}ndez}, {Calvet},
  {Brice{\~n}o}, {Hartmann}, {Vivas}, {Muzerolle}, {Downes}, {Allen}, \&
  {Gutermuth}}]{Hernandez+07}
{Hern{\'a}ndez}, J., {Calvet}, N., {Brice{\~n}o}, C., {Hartmann}, L., {Vivas},
  A.~K., {Muzerolle}, J., {Downes}, J., {Allen}, L., \& {Gutermuth}, R. 2007,
  \apj, 671, 1784

\bibitem[{{Herzberg}(1950)}]{Herzberg+50}
{Herzberg}, G. 1950, {Molecular spectra and molecular structure. Vol.1: Spectra
  of diatomic molecules}

\bibitem[{{Hoadley} {et~al.}(2015){Hoadley}, {France}, {Alexander}, {McJunkin},
  \& {Schneider}}]{Hoadley+15}
{Hoadley}, K., {France}, K., {Alexander}, R.~D., {McJunkin}, M., \&
  {Schneider}, P.~C. 2015, \apj, 812, 41

\bibitem[{{Hollenbach} \& {Tielens}(1999)}]{Hollenbach+99}
{Hollenbach}, D.~J. \& {Tielens}, A.~G.~G.~M. 1999, Reviews of Modern Physics,
  71, 173

\bibitem[{{Hughes} {et~al.}(1994){Hughes}, {Hartigan}, {Krautter}, \&
  {Kelemen}}]{Hughes+94}
{Hughes}, J., {Hartigan}, P., {Krautter}, J., \& {Kelemen}, J. 1994, \aj, 108,
  1071

\bibitem[{{Ingleby} {et~al.}(2011){Ingleby}, {Calvet}, {Bergin}, {Herczeg},
  {Brown}, {Alexander}, {Edwards}, {Espaillat}, {France}, {Gregory},
  {Hillenbrand}, {Roueff}, {Valenti}, {Walter}, {Johns-Krull}, {Brown},
  {Linsky}, {McClure}, {Ardila}, {Abgrall}, {Bethell}, {Hussain}, \&
  {Yang}}]{Ingleby+11}
{Ingleby}, L., {Calvet}, N., {Bergin}, E., {Herczeg}, G., {Brown}, A.,
  {Alexander}, R., {Edwards}, S., {Espaillat}, C., {France}, K., {Gregory},
  S.~G., {Hillenbrand}, L., {Roueff}, E., {Valenti}, J., {Walter}, F.,
  {Johns-Krull}, C., {Brown}, J., {Linsky}, J., {McClure}, M., {Ardila}, D.,
  {Abgrall}, H., {Bethell}, T., {Hussain}, G., \& {Yang}, H. 2011, \apj, 743,
  105

\bibitem[{{Jennings} {et~al.}(1984){Jennings}, {Bragg}, \&
  {Brault}}]{Jennings+84}
{Jennings}, D.~E., {Bragg}, S.~L., \& {Brault}, J.~W. 1984, \apjl, 282, L85

\bibitem[{{Johns-Krull} \& {Valenti}(2001)}]{Johns-Krull+Valenti+01}
{Johns-Krull}, C.~M. \& {Valenti}, J.~A. 2001, \apj, 561, 1060

\bibitem[{{Johns-Krull} {et~al.}(2000){Johns-Krull}, {Valenti}, \&
  {Linsky}}]{Johns-Krull+00}
{Johns-Krull}, C.~M., {Valenti}, J.~A., \& {Linsky}, J.~L. 2000, \apj, 539, 815

\bibitem[{{Kamp} \& {Dullemond}(2004)}]{Kamp+Dullemond+04}
{Kamp}, I. \& {Dullemond}, C.~P. 2004, \apj, 615, 991

\bibitem[{{Kamp} {et~al.}(2005){Kamp}, {Dullemond}, {Hogerheijde}, \&
  {Enriquez}}]{Kamp+05}
{Kamp}, I., {Dullemond}, C.~P., {Hogerheijde}, M., \& {Enriquez}, J.~E. 2005,
  in IAU Symposium, Vol. 231, Astrochemistry: Recent Successes and Current
  Challenges, ed. D.~C. {Lis}, G.~A. {Blake}, \& E.~{Herbst}, 377--386

\bibitem[{{Kastner} {et~al.}(2016){Kastner}, {Principe}, {Punzi}, {Stelzer},
  {Gorti}, {Pascucci}, \& {Argiroffi}}]{Kastner+16}
{Kastner}, J.~H., {Principe}, D.~A., {Punzi}, K., {Stelzer}, B., {Gorti}, U.,
  {Pascucci}, I., \& {Argiroffi}, C. 2016, \aj, 152, 3

\bibitem[{{Kimble} {et~al.}(1998){Kimble}, {Woodgate}, {Bowers}, {Kraemer},
  {Kaiser}, {Gull}, {Heap}, {Danks}, {Boggess}, {Green}, {Hutchings},
  {Jenkins}, {Joseph}, {Linsky}, {Maran}, {Moos}, {Roesler}, {Timothy},
  {Weistrop}, {Grady}, {Loiacono}, {Brown}, {Brumfield}, {Content}, {Feinberg},
  {Isaacs}, {Krebs}, {Krueger}, {Melcher}, {Rebar}, {Vitagliano}, {Yagelowich},
  {Meyer}, {Hood}, {Argabright}, {Becker}, {Bottema}, {Breyer}, {Bybee},
  {Christon}, {Delamere}, {Dorn}, {Downey}, {Driggers}, {Ebbets}, {Gallegos},
  {Garner}, {Hetlinger}, {Lettieri}, {Ludtke}, {Michika}, {Nyquist}, {Rose},
  {Stocker}, {Sullivan}, {Van Houten}, {Woodruff}, {Baum}, {Hartig}, {Balzano},
  {Biagetti}, {Blades}, {Bohlin}, {Clampin}, {Doxsey}, {Ferguson},
  {Goudfrooij}, {Hulbert}, {Kutina}, {McGrath}, {Lindler}, {Beck}, {Feggans},
  {Plait}, {Sandoval}, {Hill}, {Collins}, {Cornett}, {Fowler}, {Hill},
  {Landsman}, {Malumuth}, {Standley}, {Blouke}, {Grusczak}, {Reed}, {Robinson},
  {Valenti}, \& {Wolfe}}]{Kimble+98}
{Kimble}, R.~A., {Woodgate}, B.~E., {Bowers}, C.~W., {Kraemer}, S.~B.,
  {Kaiser}, M.~E., {Gull}, T.~R., {Heap}, S.~R., {Danks}, A.~C., {Boggess}, A.,
  {Green}, R.~F., {Hutchings}, J.~B., {Jenkins}, E.~B., {Joseph}, C.~L.,
  {Linsky}, J.~L., {Maran}, S.~P., {Moos}, H.~W., {Roesler}, F., {Timothy},
  J.~G., {Weistrop}, D.~E., {Grady}, J.~F., {Loiacono}, J.~J., {Brown}, L.~W.,
  {Brumfield}, M.~D., {Content}, D.~A., {Feinberg}, L.~D., {Isaacs}, M.~N.,
  {Krebs}, C.~A., {Krueger}, V.~L., {Melcher}, R.~W., {Rebar}, F.~J.,
  {Vitagliano}, H.~D., {Yagelowich}, J.~J., {Meyer}, W.~W., {Hood}, D.~F.,
  {Argabright}, V.~S., {Becker}, S.~I., {Bottema}, M., {Breyer}, R.~R.,
  {Bybee}, R.~L., {Christon}, P.~R., {Delamere}, A.~W., {Dorn}, D.~A.,
  {Downey}, S., {Driggers}, P.~A., {Ebbets}, D.~C., {Gallegos}, J.~S.,
  {Garner}, H., {Hetlinger}, J.~C., {Lettieri}, R.~L., {Ludtke}, C.~W.,
  {Michika}, D., {Nyquist}, R., {Rose}, D.~M., {Stocker}, R.~B., {Sullivan},
  J.~F., {Van Houten}, C.~N., {Woodruff}, R.~A., {Baum}, S.~A., {Hartig},
  G.~F., {Balzano}, V., {Biagetti}, C., {Blades}, J.~C., {Bohlin}, R.~C.,
  {Clampin}, M., {Doxsey}, R., {Ferguson}, H.~C., {Goudfrooij}, P., {Hulbert},
  S.~J., {Kutina}, R., {McGrath}, M., {Lindler}, D.~J., {Beck}, T.~L.,
  {Feggans}, J.~K., {Plait}, P.~C., {Sandoval}, J.~L., {Hill}, R.~S.,
  {Collins}, N.~R., {Cornett}, R.~H., {Fowler}, W.~B., {Hill}, R.~J.,
  {Landsman}, W.~B., {Malumuth}, E.~M., {Standley}, C., {Blouke}, M.,
  {Grusczak}, A., {Reed}, R., {Robinson}, R.~D., {Valenti}, J.~A., \& {Wolfe},
  T. 1998, \apjl, 492, L83

\bibitem[{{Kraus} \& {Hillenbrand}(2009)}]{Kraus+Hillenbrand+09}
{Kraus}, A.~L. \& {Hillenbrand}, L.~A. 2009, \apj, 704, 531

\bibitem[{{Kriss}(2011)}]{Kriss+11}
{Kriss}, G.~A. 2011, {Improved Medium Resolution Line Spread Functions for COS
  FUV Spectra}, Tech. rep.

\bibitem[{{Lawson} {et~al.}(2001){Lawson}, {Crause}, {Mamajek}, \&
  {Feigelson}}]{Lawson+01}
{Lawson}, W.~A., {Crause}, L.~A., {Mamajek}, E.~E., \& {Feigelson}, E.~D. 2001,
  \mnras, 321, 57

\bibitem[{{Lawson} {et~al.}(1996){Lawson}, {Feigelson}, \&
  {Huenemoerder}}]{Lawson+96}
{Lawson}, W.~A., {Feigelson}, E.~D., \& {Huenemoerder}, D.~P. 1996, \mnras,
  280, 1071

\bibitem[{{Lawson} {et~al.}(2004){Lawson}, {Lyo}, \& {Muzerolle}}]{Lawson+04}
{Lawson}, W.~A., {Lyo}, A.-R., \& {Muzerolle}, J. 2004, \mnras, 351, L39

\bibitem[{{Levison} {et~al.}(2015){Levison}, {Kretke}, \&
  {Duncan}}]{Levison+15}
{Levison}, H.~F., {Kretke}, K.~A., \& {Duncan}, M.~J. 2015, \nat, 524, 322

\bibitem[{{Lin} \& {Papaloizou}(1986)}]{Lin+Papaloizou+86}
{Lin}, D.~N.~C. \& {Papaloizou}, J. 1986, \apj, 309, 846

\bibitem[{{Loinard} {et~al.}(2007){Loinard}, {Torres}, {Mioduszewski},
  {Rodr{\'{\i}}guez}, {Gonz{\'a}lez-L{\'o}pezlira}, {Lachaume}, {V{\'a}zquez},
  \& {Gonz{\'a}lez}}]{Loinard+07}
{Loinard}, L., {Torres}, R.~M., {Mioduszewski}, A.~J., {Rodr{\'{\i}}guez},
  L.~F., {Gonz{\'a}lez-L{\'o}pezlira}, R.~A., {Lachaume}, R., {V{\'a}zquez},
  V., \& {Gonz{\'a}lez}, E. 2007, \apj, 671, 546

\bibitem[{{Lubow} \& {D'Angelo}(2006)}]{Lubow+D'Angelo+06}
{Lubow}, S.~H. \& {D'Angelo}, G. 2006, \apj, 641, 526

\bibitem[{{Lubow} {et~al.}(1999){Lubow}, {Seibert}, \& {Artymowicz}}]{Lubow+99}
{Lubow}, S.~H., {Seibert}, M., \& {Artymowicz}, P. 1999, \apj, 526, 1001

\bibitem[{{Luhman}(2004)}]{Luhman+04}
{Luhman}, K.~L. 2004, \apj, 617, 1216

\bibitem[{{Lyo} {et~al.}(2011){Lyo}, {Ohashi}, {Qi}, {Wilner}, \&
  {Su}}]{Lyo+11}
{Lyo}, A.-R., {Ohashi}, N., {Qi}, C., {Wilner}, D.~J., \& {Su}, Y.-N. 2011,
  \aj, 142, 151

\bibitem[{{Mamajek} {et~al.}(1999){Mamajek}, {Lawson}, \&
  {Feigelson}}]{Mamajek+99}
{Mamajek}, E.~E., {Lawson}, W.~A., \& {Feigelson}, E.~D. 1999, \pasa, 16, 257

\bibitem[{{Mandy}(2016)}]{Mandy+16}
{Mandy}, M.~E. 2016, \apj, 827, 62

\bibitem[{{Mandy} \& {Martin}(1993)}]{Mandy+93}
{Mandy}, M.~E. \& {Martin}, P.~G. 1993, \apjs, 86, 199

\bibitem[{{Markwardt}(2009)}]{Markwardt+09}
{Markwardt}, C.~B. 2009, in Astronomical Society of the Pacific Conference
  Series, Vol. 411, Astronomical Data Analysis Software and Systems XVIII, ed.
  D.~A. {Bohlender}, D.~{Durand}, \& P.~{Dowler}, 251

\bibitem[{{McCandliss}(2003)}]{McCandliss+03}
{McCandliss}, S.~R. 2003, \pasp, 115, 651

\bibitem[{{McClure} {et~al.}(2016){McClure}, {Bergin}, {Cleeves},
  {van Dishoeck}, {Blake}, {Evans}, {Green}, {Henning}, 
	{{\"O}berg}, {Pontoppidan}, \& {Salyk}}]{McClure+16}
{McClure}, M.~K. and {Bergin}, E.~A. and {Cleeves}, L.~I. and 
	{van Dishoeck}, E.~F. and {Blake}, G.~A. and {Evans}, II, N.~J. and 
	{Green}, J.~D. and {Henning}, T. and {{\"O}berg}, K.~I. and 
	{Pontoppidan}, K.~M. and {Salyk}, C. A. 2016, \apj, 831, 167

\bibitem[{{McJunkin} {et~al.}(2016){McJunkin}, {France}, {Schindhelm},
  {Herczeg}, {Schneider}, \& {Brown}}]{McJunkin+16}
{McJunkin}, M., {France}, K., {Schindhelm}, E., {Herczeg}, G., {Schneider},
  P.~C., \& {Brown}, A. 2016, \apj, 828, 69

\bibitem[{{McJunkin} {et~al.}(2014){McJunkin}, {France}, {Schneider},
  {Herczeg}, {Brown}, {Hillenbrand}, {Schindhelm}, \& {Edwards}}]{McJunkin+14}
{McJunkin}, M., {France}, K., {Schneider}, P.~C., {Herczeg}, G.~J., {Brown},
  A., {Hillenbrand}, L., {Schindhelm}, E., \& {Edwards}, S. 2014, \apj, 780,
  150

\bibitem[{{Nguyen} {et~al.}(2012){Nguyen}, {Brandeker}, {van Kerkwijk}, \&
  {Jayawardhana}}]{Nguyen+12}
{Nguyen}, D.~C., {Brandeker}, A., {van Kerkwijk}, M.~H., \& {Jayawardhana}, R.
  2012, \apj, 745, 119

\bibitem[{{Nomura}(2004)}]{Nomura+04}
{Nomura}, H. 2004, \apss, 292, 435

\bibitem[{{Nomura} {et~al.}(2005){Nomura}, {Aikawa}, {Nakagawa}, \&
  {Millar}}]{Nomura+05}
{Nomura}, H., {Aikawa}, Y., {Nakagawa}, Y., \& {Millar}, T.~J. 2005, in
  Protostars and Planets V Posters, 8157

\bibitem[{{Nomura} {et~al.}(2007){Nomura}, {Aikawa}, {Tsujimoto}, {Nakagawa},
  \& {Millar}}]{Nomura+07}
{Nomura}, H., {Aikawa}, Y., {Tsujimoto}, M., {Nakagawa}, Y., \& {Millar}, T.~J.
  2007, \apj, 661, 334

\bibitem[{{Nomura} \& {Millar}(2005)}]{Nomura+Millar+05}
{Nomura}, H. \& {Millar}, T.~J. 2005, \aap, 438, 923

\bibitem[{{Nomura} \& {Nakagawa}(2006)}]{Nomura+06}
{Nomura}, H. \& {Nakagawa}, Y. 2006, \apj, 640, 1099

\bibitem[{{{\"O}berg} {et~al.}(2010){{\"O}berg}, {van Dishoeck}, {Linnartz}, \&
  {Andersson}}]{Oberg+10}
{{\"O}berg}, K.~I., {van Dishoeck}, E.~F., {Linnartz}, H., \& {Andersson}, S.
  2010, \apj, 718, 832

\bibitem[{{Owen}(2016)}]{Owen+16}
{Owen}, J.~E. 2016, \pasa, 33, e005

\bibitem[{{Owen} {et~al.}(2010){Owen}, {Ercolano}, {Clarke}, \&
  {Alexander}}]{Owen+10}
{Owen}, J.~E., {Ercolano}, B., {Clarke}, C.~J., \& {Alexander}, R.~D. 2010,
  \mnras, 401, 1415

\bibitem[{{Pontoppidan} {et~al.}(2008){Pontoppidan}, {Blake}, {van Dishoeck},
  {Smette}, {Ireland}, \& {Brown}}]{Pontoppidan+08}
{Pontoppidan}, K.~M., {Blake}, G.~A., {van Dishoeck}, E.~F., {Smette}, A.,
  {Ireland}, M.~J., \& {Brown}, J. 2008, \apj, 684, 1323

\bibitem[{{Prasad} \& {Huntress}(1980)}]{Prasad+Huntress+80a}
{Prasad}, S.~S. \& {Huntress}, Jr., W.~T. 1980, \apjs, 43, 1

\bibitem[{{Quast} {et~al.}(2000){Quast}, {Torres}, {de La Reza}, {da Silva}, \&
  {Mayor}}]{Quast+00}
{Quast}, G.~R., {Torres}, C.~A.~O., {de La Reza}, R., {da Silva}, L., \&
  {Mayor}, M. 2000, in IAU Symposium, Vol. 200, IAU Symposium, 28P

\bibitem[{{Ramsay Howat} \& {Greaves}(2007)}]{RamsayHowat+Greaves+07}
{Ramsay Howat}, S.~K. \& {Greaves}, J.~S. 2007, \mnras, 379, 1658

\bibitem[{{Ricci} {et~al.}(2010){Ricci}, {Testi}, {Natta}, {Neri}, {Cabrit}, \&
  {Herczeg}}]{Ricci+10}
{Ricci}, L., {Testi}, L., {Natta}, A., {Neri}, R., {Cabrit}, S., \& {Herczeg},
  G.~J. 2010, \aap, 512, A15

\bibitem[{{Roberge} \& {Dalgarno}(1982)}]{Roberge+82}
{Roberge}, W. \& {Dalgarno}, A. 1982, \apj, 255, 176

\bibitem[{{Rodriguez} {et~al.}(2010){Rodriguez}, {Kastner}, {Wilner}, \&
  {Qi}}]{Rodriguez+10}
{Rodriguez}, D.~R., {Kastner}, J.~H., {Wilner}, D., \& {Qi}, C. 2010, \apj,
  720, 1684

\bibitem[{{Roncin} \& {Launay}(1995)}]{Roncin+95}
{Roncin}, J.-Y. \& {Launay}, F. 1995, in Astronomical Society of the Pacific
  Conference Series, Vol.~81, Laboratory and Astronomical High Resolution
  Spectra, ed. A.~J. {Sauval}, R.~{Blomme}, \& N.~{Grevesse}, 310

\bibitem[{{Rosenfeld} {et~al.}(2012{\natexlab{a}}){Rosenfeld}, {Andrews},
  {Wilner}, \& {Stempels}}]{Rosenfeld+12}
{Rosenfeld}, K.~A., {Andrews}, S.~M., {Wilner}, D.~J., \& {Stempels}, H.~C.
  2012{\natexlab{a}}, \apj, 759, 119

\bibitem[{{Rosenfeld} {et~al.}(2012{\natexlab{b}}){Rosenfeld}, {Qi}, {Andrews},
  {Wilner}, {Corder}, {Dullemond}, {Lin}, {Hughes}, {D'Alessio}, \&
  {Ho}}]{Rosenfeld+12b}
{Rosenfeld}, K.~A., {Qi}, C., {Andrews}, S.~M., {Wilner}, D.~J., {Corder},
  S.~A., {Dullemond}, C.~P., {Lin}, S.-Y., {Hughes}, A.~M., {D'Alessio}, P., \&
  {Ho}, P.~T.~P. 2012{\natexlab{b}}, \apj, 757, 129

\bibitem[{{Rosenthal} {et~al.}(2000){Rosenthal}, {Bertoldi}, \&
  {Drapatz}}]{Rosenthal+00}
{Rosenthal}, D., {Bertoldi}, F., \& {Drapatz}, S. 2000, \aap, 356, 705

\bibitem[{{Salyk} {et~al.}(2011){Salyk}, {Blake}, {Boogert}, \&
  {Brown}}]{Salyk+11b}
{Salyk}, C., {Blake}, G.~A., {Boogert}, A.~C.~A., \& {Brown}, J.~M. 2011, \apj,
  743, 112

\bibitem[{{Salyk} {et~al.}(2008){Salyk}, {Pontoppidan}, {Blake}, {Lahuis}, {van
  Dishoeck}, \& {Evans}}]{Salyk+08}
{Salyk}, C., {Pontoppidan}, K.~M., {Blake}, G.~A., {Lahuis}, F., {van
  Dishoeck}, E.~F., \& {Evans}, II, N.~J. 2008, \apjl, 676, L49

\bibitem[{{Schindhelm} {et~al.}(2012){Schindhelm}, {France}, {Herczeg},
  {Bergin}, {Yang}, {Brown}, {Brown}, {Linsky}, \& {Valenti}}]{Schindhelm+12b}
{Schindhelm}, E., {France}, K., {Herczeg}, G.~J., {Bergin}, E., {Yang}, H.,
  {Brown}, A., {Brown}, J.~M., {Linsky}, J.~L., \& {Valenti}, J. 2012, \apjl,
  756, L23

\bibitem[{{Schneider} {et~al.}(2015){Schneider}, {France}, {G$\ddot{u}$nther},
  {Herczeg}, {Bouvier}, {Grankin}, {McJunkin}, {Robrade}, \&
  {Schmitt}}]{Schneider+15}
{Schneider}, P.~C., {France}, K., {G$\ddot{u}$nther}, H.~M., {Herczeg}, G.,
  {Bouvier}, J., {Grankin}, K., {McJunkin}, M., {Robrade}, J., \& {Schmitt},
  J.~H.~M.~M. 2015, \aap

\bibitem[{{Shull} \& {Beckwith}(1982)}]{Shull+82}
{Shull}, J.~M. \& {Beckwith}, S. 1982, \araa, 20, 163

\bibitem[{{Simon} {et~al.}(2000){Simon}, {Dutrey}, \& {Guilloteau}}]{Simon+00}
{Simon}, M., {Dutrey}, A., \& {Guilloteau}, S. 2000, \apj, 545, 1034

\bibitem[{{Smith} {et~al.}(1982){Smith}, {Adams}, \& {Alge}}]{Smith+82}
{Smith}, D., {Adams}, N.~G., \& {Alge}, E. 1982, \apj, 263, 123

\bibitem[{{Stecher} \& {Williams}(1967)}]{Stecher+67}
{Stecher}, T.~P. \& {Williams}, D.~A. 1967, \apjl, 149, L29

\bibitem[{{Stempels} {et~al.}(2007){Stempels}, {Gahm}, \&
  {Petrov}}]{Stempels+07}
{Stempels}, H.~C., {Gahm}, G.~F., \& {Petrov}, P.~P. 2007, \aap, 461, 253

\bibitem[{{Stempels} \& {Piskunov}(2002)}]{Stempels+12}
{Stempels}, H.~C. \& {Piskunov}, N. 2002, \aap, 391, 595

\bibitem[{{Sternberg}(1989)}]{Sternberg+89}
{Sternberg}, A. 1989, \apj, 347, 863

\bibitem[{{Strom} {et~al.}(1989){Strom}, {Strom}, {Edwards}, {Cabrit}, \&
  {Skrutskie}}]{Strom+89}
{Strom}, K.~M., {Strom}, S.~E., {Edwards}, S., {Cabrit}, S., \& {Skrutskie},
  M.~F. 1989, \aj, 97, 1451

\bibitem[{{Takahashi} {et~al.}(1999){Takahashi}, {Masuda}, \&
  {Nagaoka}}]{Takahashi+99}
{Takahashi}, J., {Masuda}, K., \& {Nagaoka}, M. 1999, \apj, 520, 724

\bibitem[{{Takeuchi} \& {Artymowicz}(2001)}]{Takeuchi+Artymowicz+01}
{Takeuchi}, T. \& {Artymowicz}, P. 2001, \apj, 557, 990

\bibitem[{{Tang} {et~al.}(2012){Tang}, {Guilloteau}, {Pi{\'e}tu}, {Dutrey},
  {Ohashi}, \& {Ho}}]{Tang+12}
{Tang}, Y.-W., {Guilloteau}, S., {Pi{\'e}tu}, V., {Dutrey}, A., {Ohashi}, N.,
  \& {Ho}, P.~T.~P. 2012, \aap, 547, A84

\bibitem[{{van Boekel} {et~al.}(2005){van Boekel}, {Min}, {Waters}, {de Koter},
  {Dominik}, {van den Ancker}, \& {Bouwman}}]{vanBoekel+05}
{van Boekel}, R., {Min}, M., {Waters}, L.~B.~F.~M., {de Koter}, A., {Dominik},
  C., {van den Ancker}, M.~E., \& {Bouwman}, J. 2005, \aap, 437, 189

\bibitem[{{van den Ancker} {et~al.}(1998){van den Ancker}, {de Winter}, \&
  {Tjin A Djie}}]{vanderAncker+98}
{van den Ancker}, M.~E., {de Winter}, D., \& {Tjin A Djie}, H.~R.~E. 1998,
  \aap, 330, 145

\bibitem[{{van Leeuwen}(2007)}]{vanLeeuwen+07}
{van Leeuwen}, F. 2007, \aap, 474, 653

\bibitem[{{Walsh} {et~al.}(2012){Walsh}, {Nomura}, {Millar}, \&
  {Aikawa}}]{Walsh+12}
{Walsh}, C., {Nomura}, H., {Millar}, T.~J., \& {Aikawa}, Y. 2012, \apj, 747,
  114

\bibitem[{{Walsh} {et~al.}(2015){Walsh}, {Nomura}, \& {van
  Dishoeck}}]{Walsh+15}
{Walsh}, C., {Nomura}, H., \& {van Dishoeck}, E. 2015, \aap, 582, A88

\bibitem[{{Webb} {et~al.}(1999){Webb}, {Zuckerman}, {Platais}, {Patience},
  {White}, {Schwartz}, \& {McCarthy}}]{Webb+99}
{Webb}, R.~A., {Zuckerman}, B., {Platais}, I., {Patience}, J., {White}, R.~J.,
  {Schwartz}, M.~J., \& {McCarthy}, C. 1999, \apjl, 512, L63

\bibitem[{{Weidenschilling}(1977)}]{Weidenschilling+77}
{Weidenschilling}, S.~J. 1977, \mnras, 180, 57

\bibitem[{{White} \& {Ghez}(2001)}]{White+Ghez+01}
{White}, R.~J. \& {Ghez}, A.~M. 2001, \apj, 556, 265

\bibitem[{{Williams} \& {Murdin}(2000)}]{Williams+00}
{Williams}, D. \& {Murdin}, P. 2000, {Physics of Molecules}

\bibitem[{{Woitke} {et~al.}(2013){Woitke}, {Dent}, {Thi}, {Menard}, {Pinte},
  {Duchene}, {Sandell}, {Lawson}, \& {Kamp}}]{Woitke+13}
{Woitke}, P., {Dent}, W.~R.~F., {Thi}, W.-F., {Menard}, F., {Pinte}, C.,
  {Duchene}, G., {Sandell}, G., {Lawson}, W., \& {Kamp}, I. 2013, in Protostars
  and Planets VI Posters, 13

\bibitem[{{Woitke} {et~al.}(2009){Woitke}, {Kamp}, \& {Thi}}]{Woitke+09b}
{Woitke}, P., {Kamp}, I., \& {Thi}, W.-F. 2009, \aap, 501, 383

\bibitem[{{Wolniewicz} {et~al.}(1998){Wolniewicz}, {Simbotin}, \&
  {Dalgarno}}]{Wolniewicz+98}
{Wolniewicz}, L., {Simbotin}, I., \& {Dalgarno}, A. 1998, \apjs, 115, 293

\bibitem[{{Woodgate} {et~al.}(1998){Woodgate}, {Kimble}, {Bowers}, {Kraemer},
  {Kaiser}, {Danks}, {Grady}, {Loiacono}, {Brumfield}, {Feinberg}, {Gull},
  {Heap}, {Maran}, {Lindler}, {Hood}, {Meyer}, {Vanhouten}, {Argabright},
  {Franka}, {Bybee}, {Dorn}, {Bottema}, {Woodruff}, {Michika}, {Sullivan},
  {Hetlinger}, {Ludtke}, {Stocker}, {Delamere}, {Rose}, {Becker}, {Garner},
  {Timothy}, {Blouke}, {Joseph}, {Hartig}, {Green}, {Jenkins}, {Linsky},
  {Hutchings}, {Moos}, {Boggess}, {Roesler}, \& {Weistrop}}]{Woodgate+98}
{Woodgate}, B.~E., {Kimble}, R.~A., {Bowers}, C.~W., {Kraemer}, S., {Kaiser},
  M.~E., {Danks}, A.~C., {Grady}, J.~F., {Loiacono}, J.~J., {Brumfield}, M.,
  {Feinberg}, L., {Gull}, T.~R., {Heap}, S.~R., {Maran}, S.~P., {Lindler}, D.,
  {Hood}, D., {Meyer}, W., {Vanhouten}, C., {Argabright}, V., {Franka}, S.,
  {Bybee}, R., {Dorn}, D., {Bottema}, M., {Woodruff}, R., {Michika}, D.,
  {Sullivan}, J., {Hetlinger}, J., {Ludtke}, C., {Stocker}, R., {Delamere}, A.,
  {Rose}, D., {Becker}, I., {Garner}, H., {Timothy}, J.~G., {Blouke}, M.,
  {Joseph}, C.~L., {Hartig}, G., {Green}, R.~F., {Jenkins}, E.~B., {Linsky},
  J.~L., {Hutchings}, J.~B., {Moos}, H.~W., {Boggess}, A., {Roesler}, F., \&
  {Weistrop}, D. 1998, \pasp, 110, 1183

\bibitem[{{Yang} {et~al.}(2011){Yang}, {Linsky}, \& {France}}]{Yang+11}
{Yang}, H., {Linsky}, J.~L., \& {France}, K. 2011, \apjl, 730, L10

\bibitem[{{Youdin}(2011)}]{Youdin+11}
{Youdin}, A.~N. 2011, \apj, 731, 99

\bibitem[{{Zhu} {et~al.}(2011){Zhu}, {Nelson}, {Hartmann}, {Espaillat}, \&
  {Calvet}}]{Zhu+11}
{Zhu}, Z., {Nelson}, R.~P., {Hartmann}, L., {Espaillat}, C., \& {Calvet}, N.
  2011, \apj, 729, 47

\end{thebibliography}

\listofchanges
\end{document}